\documentclass[aps,prx,notitlepage,superscriptaddress,nofootinbib,10pt,twocolumn, floatfix]{revtex4-2}
\pdfoutput=1

\usepackage[utf8]{inputenc}
\usepackage{amsmath}
\usepackage{url}
\usepackage{braket}
\usepackage{array}   
\usepackage[all]{nowidow} 
\usepackage[normalem]{ulem}
\usepackage{soul}

\usepackage{tikz}
\usetikzlibrary{decorations.pathreplacing, calligraphy}

\usepackage{fancyhdr}
\usepackage{fancyvrb}
\usepackage{xcolor}
\usepackage[hidelinks, colorlinks=false]{hyperref}
\usepackage{natbib}
\usepackage{cleveref}

\usepackage{qcircuit}
\usepackage{mathdots}
\usepackage{hhline}
\usepackage{graphicx}
\setkeys{Gin}{width=\linewidth,totalheight=\textheight,keepaspectratio}
\graphicspath{{figures/}}

\usepackage{amssymb, amsthm, bm}
\usepackage[ruled,linesnumbered,vlined]{algorithm2e}
\usepackage{soul}

\usepackage{empheq}
\usepackage[most]{tcolorbox}
\newtcbox{\mymath}[1][]{%
    nobeforeafter, math upper, tcbox raise base,
    enhanced, colframe=blue!30!black,
    colback=white, boxrule=1pt,
    #1}
	
\usepackage{acronym}
\newacro{QSP}{quantum signal processing}
\newacro{SOCP}{second order cone program}
\newacro{QSVT}{quantum singular value transform}
\newacro{SVD}{singular value decomposition}
\newacro{QRAM}{quantum random access memory}

\newtheorem{definition}{Definition}

\newtheorem{proposition}{Proposition}

\newcolumntype{V}{>{\centering\arraybackslash}m{9em}}
\newcolumntype{U}{>{\centering\arraybackslash}m{13em}}
\newcolumntype{F}{>{\centering\arraybackslash}m{16em}}
\newcolumntype{L}[1]{>{\raggedright\let\newline\\\arraybackslash\hspace{0pt}}m{#1}}
\newcolumntype{C}[1]{>{\centering\let\newline\\\arraybackslash\hspace{0pt}}m{#1}}
\newcolumntype{R}[1]{>{\raggedleft\let\newline\\\arraybackslash\hspace{0pt}}m{#1}}

\newcommand{\bigo}[1]{\mathcal{O}(#1)}
\newcommand{\bigoLONE}{\mathcal{O}}

\renewcommand{\>}{\rangle}

\newcommand{\polylog}{\text{polylog}}
\newcommand{\ve}[1]{\boldsymbol{#1}}

\begin{document}

\title{End-to-end resource analysis for quantum interior point methods and portfolio optimization}

\author{Alexander M. Dalzell}\email{dalzel@amazon.com}
\affiliation{AWS Center for Quantum Computing, Pasadena, CA, USA}
\affiliation{California Institute of Technology, Pasadena, CA, USA}

\author{B. David Clader}
\email{dave.clader@bqpadvisors.com (Current affiliation: BQP Advisors, LLC)}
\affiliation{Goldman Sachs, New York, NY, USA}

\author{Grant Salton}
\email{saltg@amazon.com}
\affiliation{Amazon Quantum Solutions Lab, Seattle, WA, USA}
\affiliation{AWS Center for Quantum Computing, Pasadena, CA, USA}
\affiliation{California Institute of Technology, Pasadena, CA, USA}

\author{Mario Berta}
\affiliation{AWS Center for Quantum Computing, Pasadena, CA, USA}
\affiliation{California Institute of Technology, Pasadena, CA, USA}
\affiliation{Department of Computing, Imperial College London, London, UK}
\affiliation{Institute for Quantum Information, RWTH Aachen University, Aachen, Germany}

\author{Cedric Yen-Yu Lin}
\affiliation{AWS Quantum Technologies, Seattle, WA, USA}

\author{David A. Bader}
\affiliation{Goldman Sachs, New York, NY, USA}
\affiliation{New Jersey Institute of Technology, Newark, NJ, USA}

\author{Nikitas Stamatopoulos}
\affiliation{Goldman Sachs, New York, NY, USA}

\author{Martin J. A. Schuetz}
\affiliation{Amazon Quantum Solutions Lab, Seattle, WA, USA}
\affiliation{AWS Center for Quantum Computing, Pasadena, CA, USA}

\author{Fernando G.S.L. Brand\~ao}
\affiliation{AWS Center for Quantum Computing, Pasadena, CA, USA}
\affiliation{California Institute of Technology, Pasadena, CA, USA}

\author{Helmut G. Katzgraber}
\affiliation{Amazon Quantum Solutions Lab, Seattle, WA, USA}
\affiliation{AWS Center for Quantum Computing, Pasadena, CA, USA}
\affiliation{University of Washington, Seattle, WA, USA}

\author{William J. Zeng}
\affiliation{Goldman Sachs, New York, NY, USA}

\begin{abstract}
We study quantum interior point methods (QIPMs) for second-order cone programming (SOCP), guided by the example use case of portfolio optimization (PO). We provide a complete quantum circuit-level description of the algorithm from problem input to problem output, making several improvements to the implementation of the QIPM. We report the number of logical qubits and the quantity/depth of non-Clifford $T$-gates needed to run the algorithm, including constant factors. The resource counts we find depend on instance-specific parameters, such as the condition number of certain linear systems within the problem. To determine the size of these parameters, we perform numerical simulations of small PO instances, which lead to concrete resource estimates for the PO use case. Our numerical results do not probe large enough instance sizes to make conclusive statements about the asymptotic scaling of the algorithm. However, already at small instance sizes, our analysis suggests that, due primarily to large constant pre-factors, poorly conditioned linear systems, and a fundamental reliance on costly quantum state tomography, fundamental improvements to the QIPM are required for it to lead to practical quantum advantage.
\end{abstract}

\maketitle



\section{Overview}\label{sec:overview}


\subsection{Introduction}\label{sec:overview-intro}

The practical utility of finding optimal solutions to well-posed optimization problems has been known since the days of antiquity, with Euclid considering the minimal distance between two points using a line. In the modern era, optimization algorithms for business and financial use cases continue to be ubiquitous. Partly as a result of this utility, algorithmic techniques for optimization problems have been well studied since even before the invention of the computer, including a famous dispute between Legendre and Gauss on who was responsible for the invention of least squares fitting \cite{Stigler1981Gauss}. With the advent of the quantum era, there has been great interest in developing quantum algorithms that solve optimization problems with provable speedups over classical algorithms. Some of the earliest proposals rely on quantum annealing \cite{Apolloni1989} or more recent work in variational algorithms \cite{Farhi2014qaoa, Moll2018variational} to solve combinatorial optimization problems. Quantum algorithms have also been developed that allow for more efficient convex optimization, including algorithms for semidefinite, second-order cone, and linear programs~\cite{brandao2017sdp,brandao2019sdpLarge,vanApeldoorn2020quantumsdpsolvers,vanapeldoorn2019sdpWithApplications,GSLBrandao2022fasterquantum,Kerenidis2020lpsdp,augustino2021quantum,huang2022faster,Kerenidis2021quantumalgorithms,augustino2021inexact}, as well as algorithms for solving systems of linear equations~\cite{HHL09,Childs17,subasi2019quantumLinearSystemAdiabatic,costa2021optimal,an2022}, which can be used for quantum data fitting \cite{Wiebe2012datafitting}. Using these techniques, specific financial use cases such as solving the portfolio optimization problem have been studied~\cite{rebentrost2018quantum,kerenidis2019quantum,palmer2021quantum,mugel2022dynamic}. 

Unfortunately, it can be difficult to evaluate whether these quantum algorithms will be \emph{practically} useful. In some cases, the algorithms are heuristic, and their performance can only be measured empirically once it is possible to run them on actual quantum hardware.  In other cases, the difficulty in evaluating practicality stems from the inherent complexity of combining many distinct ingredients, each with their own caveats and bottlenecks. To make an apples-to-apples comparison and quantify advantages of a quantum algorithm, a truly end-to-end resource analysis that accounts for all costs from problem input to problem output must be performed.

In this work, we perform such an end-to-end analysis for a \textit{quantum interior point method} (QIPM) for solving second-order cone programs (SOCPs), which was originally proposed in Ref.~\cite{Kerenidis2021quantumalgorithms}, based on earlier QIPMs for semidefinite and linear programs \cite{Kerenidis2020lpsdp}. In particular, we focus on a concrete use case with very broad application, but of primary interest in the financial services sector: \textit{portfolio optimization} (PO). In general, PO is the task of determining the optimal resource allocation to a collection of possible classes, so as to optimize a given objective. In finance, one seeks to determine the optimal allocation of funds across a set of possible assets that maximizes returns and minimizes risk, subject to constraints. Importantly, many variants of the PO problem can be cast as a SOCP and subsequently solved with a classical or quantum interior point method. Indeed, classical interior point methods (CIPMs) are efficient not only in theory, but also in practice; they are the method of choice within fast numerical solvers for SOCPs and other conic programs (e.g., {\cite{domahidi2013ecos}}), which encompass a large variety of optimization problems that appear in industry. Notably, QIPMs structurally mirror CIPMs, and seek improvements by replacing certain subroutines with quantum primitives. Thus, compared to other proposed quantum algorithms for conic programs not based on widely used classical techniques (e.g., solvers that leverage the multiplicative weights update method~\mbox{\cite{brandao2017sdp,brandao2019sdpLarge,vanapeldoorn2019sdpWithApplications,vanApeldoorn2020quantumsdpsolvers}}), QIPMs are uniquely positioned to provide not only a theoretical asymptotic advantage, but also a practical quantum solution for this common class of problem. 

However, the QIPM is a complex algorithm that delicately combines some purely classical steps with multiple distinct quantum subroutines. The runtime of the QIPM is stated in terms of several parameters that can only be evaluated once a particular use case has been specified; depending on how these parameters scale, an asymptotic speedup may or may not be achievable. Additionally, any speedup is contingent on access to a large quantum random access memory (QRAM), an ingredient that in prior asymptotic-focused analyses has typically been assumed to exist without much further justification or cost analysis.

Our resource analysis is detailed and takes care to study all aspects of the end-to-end pipeline, including the QRAM component. We report our results in terms of relevant problem parameters, and then we perform numerical experiments to determine the size and scaling of these parameters for actual randomly chosen instances of the PO problem, based on historical stock data. This approach allows us to estimate the exact resource cost of the QIPM for an example PO problem, including a detailed breakdown of costs by various subroutines. This estimate incorporates several optimizations to the underlying subroutines, and technical improvements to how they are integrated into the QIPM. Consequently, our analysis allows us to evaluate the prospect that the algorithm could exhibit a practical quantum advantage, and it clearly reveals the computational bottlenecks within the algorithm that are most in need of further improvement. 

While we focus on the QIPM and in particular on its application to the PO problem, our work has more general takeaways for quantum algorithms and for quantum computing applications. Firstly, our results emphasize the importance of end-to-end analysis when evaluating a proposed application. Furthermore, our modular treatment of the underlying algorithmic primitives produces quantitative and qualitative takeaways that would be relevant for end-to-end treatments of a large number of other algorithms that also rely on these subroutines, especially those in the area of machine learning, where data access via QRAM and quantum linear algebra techniques are often required {\cite{ciliberto2018quantum}}.  


\subsection{Results}\label{sec:overview-results}

Our resource analysis focuses on three central quantities that determine the overall cost of algorithms implemented on fault-tolerant quantum computers: the number of \emph{logical} qubits, the total number of $T$ gates (``$T$-count''), and the number of parallel layers of $T$ gates (``$T$-depth'') needed to construct quantum circuits for solving the problem. The $T$-depth acts as a proxy for the overall runtime of the algorithm, whereas the $T$-count and number of logical qubits are important for determining how many physical qubits would be required for a full, fault-tolerant implementation. We justify the focus on $T$ gates by pointing out that, in many prominent approaches to fault-tolerant quantum computation (such as lattice surgery \cite{Horsman2012latticesurgery,Litinski2018latticesurgery,Litinski2019gameofsurfacecodes,chamberland2022universal}), quantum circuits are decomposed into Clifford gates and $T$ gates, and the cost of implementing the circuit is dominated by the number and depth of the $T$ gates. The fault-tolerant Clifford gates can be performed transversally or even in software, whereas the $T$ gates require the expensive process of magic state distillation \cite{Knill2004magic,Bravyi2005magic}. We stop short of a full analysis of the algorithm at the physical level, as we believe the logical analysis already suffices to evaluate the overall outlook for the algorithm and identify its main bottlenecks. 

At the core of any interior point method (IPM) is the solving of a linear system of equations. The QIPM performs this step using a quantum linear system solver (QLSS) together with pure state quantum tomography. The cost of QLSS depends on a parameter $\kappa_F$,  the Frobenius condition number $\lVert G \rVert_F \lVert G^{-1} \rVert$ of the matrix $G$ that must be inverted (where $\lVert \cdot \rVert_F$ denotes the Frobenius norm, and $\lVert \cdot \rVert$ denotes the spectral norm), while the cost of tomography depends on a parameter $\xi$, a precision parameter. We evaluate these parameters empirically by simulating the QIPM on small instances of the PO problem.  

In \cref{tab:summary_scaling}, we report a summary of our overall resource calculation, in which we show the asymptotically leading term (along with its constant prefactor) in terms of parameters $\kappa_F$ and $\xi$, as well as $n$, the number of assets in the PO instance, and $\epsilon$, the desired precision to which the portfolio should be optimized.  We find (numerically) that $\kappa_F$ grows with $n$, and that $\xi$ shrinks with $n$; we estimate that, at $n=100$ and $\epsilon = 10^{-7}$, our implementation of the QIPM would require $8 \times 10^{6}$ qubits and $7 \times 10^{29}$ total $T$ gates spread out over $2 \times 10^{24}$ layers. Needless to say, these resource counts are decidedly out of reach both in the near and far term for quantum hardware, even for a problem of modest size by classical standards. Even if quantum computers one day match the gigahertz-level clock-speeds of modern classical computers, $10^{24}$ layers of $T$ gates would take millions of years to execute. By contrast, the PO problem can be easily solved in a matter of seconds on a laptop for $n=100$ stocks. 



\begin{table*}
\caption{\label{tab:summary_scaling}Asymptotic, leading-order contributions to the total quantum resources for an end-to-end portfolio optimization (including constant factors), in terms of the number of assets in the portfolio ($n$), the desired precision to which the portfolio should be optimized ($\epsilon$), the maximum Frobenius condition number of matrices encountered by the QIPM ($\kappa_F$), and the minimum tomographic precision necessary for the algorithm to succeed ($\xi$). The $T$-depth and $T$-count expressions represent the cumulative cost of $\bigo{\xi^{-2}n^{1.5}\log(n)\log(\epsilon^{-1})}$ individual quantum circuits performed serially, 
a quantity that we estimate evaluates to $6\times 10^{12}$ circuits at $n=100$; see \cref{tab:resources} for a detailed accounting. The right column uses a numerical simulation of the quantum algorithm (see \cref{sec:numerics}) to compute the instance-specific parameters in the resource expression and estimate the resource cost at $n=100$ and $\epsilon = 10^{-7}$.  }
\begin{ruledtabular}
\renewcommand{\arraystretch}{1.2}
\begin{tabular}{lll}
\textbf{Resource} & \textbf{QIPM complexity} & \textbf{Estimated at $n=100$}\\
\hline
\textbf{Number of logical  qubits}  & $800n^2 $ & $8 \times 10^{6}$ \\ 
\textbf{$T$-depth}  & $(1 \times 10^{10}) \kappa_F n^{1.5}\xi^{-2}\log_2(n)\log_2(\epsilon^{-1})\log_2(\kappa_F n^{14/27}\xi^{-1})$ & $2 \times 10^{24}$ \\
\textbf{$T$-count}  & $(5 \times 10^{11})\kappa_F n^{3.5}\xi^{-2}\log_2(n)\log_2(\epsilon^{-1})\log_2(\kappa_F\xi^{-1})$ & $7 \times 10^{29}$ \\
\end{tabular}
\end{ruledtabular}
\end{table*}


We caution that the numbers we report should not be interpreted as the final word on the cost of the QIPM for PO. We are certain that further examination of the algorithm could uncover many improvements and optimizations that would reduce the costs compared to our calculations. On the other hand, we note that our results do already incorporate several innovations we made to reduce the resource cost, including a basic attempt at preconditioning the linear system. Moreover, the pessimistic outlook our results convey is robust in the sense that the calculation would need to decrease by many orders of magnitude for the algorithm to be practical, suggesting that fundamental changes are necessary to multiple aspects of the algorithm, rather than merely superficial optimizations.

Besides the main resource calculation, we make several additional contributions and observations:
\begin{enumerate}
    \item We provide explicit quantum circuits for the important subroutines of the QIPM, namely the state-of-the-art QLSS based on the discrete adiabatic theorem \cite{costa2021optimal} and pure state tomography, which complement the explicit circuits for block-encoding (using QRAM) that a subset of the authors already reported separately in Ref.~\cite{clader2022quantum}. These circuits, and their precise resource calculations, could be useful elsewhere, as these subroutines are ubiquitous in quantum algorithms. See \cref{sec:qipm-circuits} and \cref{sec:implementation} for additional details. 

    \item We break down the resource calculation into its constituents to illustrate which parts of the algorithm are most costly. We find that many independent factors create significant challenges toward realizing quantum advantage with QIPMs, and our work underscores those aspects of the algorithm that must be improved for it to be useful. We also note that the conditions under which QIPMs would be most successful (namely, when $\kappa_F$ is small) also allow for classical IPMs based on iterative classical linear system solvers to be competitive. See \cref{sec:discussion} for additional details. 

    \item We numerically simulate several versions of the full QIPM solving the PO problem on portfolios as large as $n=120$ stocks, and we report the empirical size and scaling of the relevant parameters $\kappa_F$ and $\xi$. There is considerable variability in the trends we observe, depending on which version of the QIPM is chosen, and when the QIPM is terminated, which makes it difficult to draw robust conclusions. However, we find that both $\kappa_F$ and $\xi^{-1}$ appear to grow with $n$. Note that previous numerical experiments on a similar formulation of the PO problem \cite{kerenidis2019quantum} suggested $\kappa_F$ does not grow with problem size, but those authors scaled the number of ``time epochs'' while keeping $n$ constant. Additionally, we observe that the ``infeasible'' version of the QIPM originally proposed by \cite{Kerenidis2021quantumalgorithms} empirically performs similarly to more sophisticated ``feasible'' versions \cite{augustino2021inexact}, despite not enjoying the same theoretical guarantees of fast convergence. Finally, contrary to theoretical expectation, we observe that $\kappa_F$ and $\xi^{-1}$ do \emph{not} diverge as $\epsilon \rightarrow 0$ in our examples. See \cref{sec:numerics} for additional details. 

    \item We make various technical improvements to the underlying ingredients of QIPMs. A subset of authors previously reported \cite{clader2022quantum} a quadratic improvement in the minimum depth required for the problem of preparing an arbitrary $L$-dimensional quantum state or block-encoding an arbitrary $L \times L$ matrix, along with explicit quantum circuits and exact resource expressions. In this manuscript, we additionally contribute the following:
    \begin{itemize}
    \item Tomographic precision: Performing tomography on the output of a QLSS necessarily causes the classical estimate of the solution to the linear system to be inexact. We illustrate how the allowable amount of tomography precision can be determined adaptively rather than relying on theoretical bounds. Nonetheless, we also improve the constant prefactor in the tomographic bounds. The total number of state preparation queries needed to learn an unknown $L$-dimensional pure state to $\xi$ error using the tomography method of Ref.~\cite{Kerenidis2020lpsdp,Kerenidis2021quantumalgorithms} is to leading order at most $115 L\ln(L) / \xi^2$.\footnote{\label{foot:new-tomography}In the late stages of this project, an alternative method for pure state tomography was proposed in Ref.~\cite{apeldoorn22} with superior asymptotic query complexity, reducing $\bigo{L\ln(L)/\xi^2}$ to $\bigo{L\ln(L)/\xi}$. However, the protocol is more complicated than our approach, and it requires additional gate overhead to implement. Furthermore, for the values of $\xi$ and $L$ we consider in \cref{tab:summary_scaling}, a conservative estimate of the improvement from this method (ignoring potentially large constants) only yields about 2 orders of magnitude improvement in our final estimates of $T$-depth and $T$-count -- not enough to change our results qualitatively. Thus, we do not incorporate this method into our analysis, but we remark that we do expect a marginal improvement in our final counts.}
    \item Norm of the linear system: Since QLSSs output a normalized quantum state, tomography does not directly yield the norm of the solution to the linear system. The norm can be learned through more complicated protocols, but we observe that in the context of QIPMs, a sufficient estimate for the norm can be learned classically. 
    \item Preconditioning: We propose a simple preconditioning method that is compatible with the QIPM, while reducing the parameter $\kappa_F$. Our numerical simulations suggest the reduction is more than an order of magnitude for the portfolio optimization problem.
    \item Feasible QIPM: We implement a ``feasible'' version of the QIPM proposed by \cite{augustino2021inexact}, which relies on finding a basis for the null space of the SOCP matrix. We identified an explicit basis for the PO problem, thereby avoiding the need for a costly QR decomposition. However, we observe that finding the basis via QR decomposition leads to more stable numerical results. 
    \end{itemize}
\end{enumerate}

The outline for the remainder of the paper is as follows. In \cref{sec:portfolio} we describe and define the portfolio optimization problem in terms of Markowitz portfolio theory. In \cref{sec:socp} we describe Second Order Cone Programming (SOCP) problems, illustrate how portfolio optimization can be represented as an instance of SOCP, and discuss how IPMs can be used for solving SOCPs. In \cref{sec:qipm} we review the \emph{quantum} ingredients needed to turn an IPM into a QIPM. In particular, we review  quantum linear system solvers, block-encoding for data loading, and quantum state tomography for data read out. We also present slightly better bounds on the required tomography procedure than were previously known. In \cref{sec:implementation} we describe the full implementation of using QIPM and quantum algorithms for SOCP for the portfolio optimization problem, including a detailed resource estimate for the end-to-end problem. In \cref{sec:numerics} we show numerical results from simulations of the full problem, and in \cref{sec:discussion} we reflect on the calculation we have performed, identifying the main bottlenecks and drawing conclusions about the outlook for quantum advantage with QIPM. 

The QIPM has many moving parts requiring several mathematical symbols. While all symbols are defined as they are introduced in the text, we also provide a full list of symbols for the reader's reference in \cref{app:notation}. Throughout the paper, we denote all vectors in bold lowercase letters to contrast with scalar quantities (unbolded lowercase) and matrices (unbolded uppercase). The only exception to this rule will be the symbols $N$, $K$, and $L$, which are positive integers (despite being uppercase), and denote the number of rows or columns in certain matrices related to an SOCP instance. 



\section{Portfolio optimization (PO)}\label{sec:portfolio}

\subsection{Background}

Portfolio optimization is the process widely used by financial
analysts to assign allocations of capital across a set of assets within a portfolio, given optimization criteria such as maximizing the expected return and minimizing the financial risk.  
The creation of the
mathematical framework for modern portfolio theory (MPT) is credited to Harry Markowitz \cite{Markowitz1952,Markowitz1959},
for which he
received the 1990 Alfred Nobel Memorial Prize in Economic
Sciences \cite{Nobel1990}. 
Markowitz describes the process of selecting
a portfolio in two stages, where the first stage starts with
``observation and experience'' and ends with ``beliefs about the
future performances of available securities.'' The second stage starts
with ``the relevant beliefs about future performances'' and ends with
``the choice of portfolio.'' The theory is also known
as \emph{mean-variance analysis}. For further history, Markowitz's 1999
essay \cite{Markowitz1999} gives the early history of portfolio
theory: 1600-1960.

Typically, portfolio optimization strategies include diversification,
which is the practice of investing in a wide array of asset types and
classes as a risk mitigation strategy. Some popular asset classes are
stocks, bonds, real estate, commodities, and cash. After building a
portfolio, we expect a return (or profit) after a specific period of
time. \emph{Risk} is defined as the fluctuations of the asset value.
MPT describes how high variance assets can be combined with other
uncorrelated assets through diversification to create portfolios with
low variance on their return. Naturally, among equal-risk portfolios, investors prefer those with higher expected return, and among equal-return portfolios, they prefer those with lower risk.

\subsection{Mathematical formulation}

Within a portfolio, $w_i$ represents the amount of an asset $i$ we
are holding over some period of time. Often, this amount is given as
the asset's price in dollars at the start of the period. When the
price is positive ($w_i > 0$), we call this a \emph{long} position;
and when the price is negative ($w_i < 0$), we call this
a \emph{short} position with an obligation to buy this asset at the
end of the period.\footnote{Typically, investment banks hold long positions,
while hedge funds build portfolios with short positions that have
higher risk due to the uncertainty of the price to buy the asset at
the end of the period.} The optimization variable in our portfolio
optimization problem is the vector of $n$ assets $\ve{w} \in \mathbb{R}^n$
in our portfolio.

The price of each asset $i$ varies over time. We define $u_i$ to be the
relative change (positive or negative) during the period of
interest. Then, we define the return of the portfolio for that period as $\bar{r} = \ve{u}^\intercal \ve{w}$
dollars. The relative changes $\ve{u} \in \mathbb{R}^n$ follow a stochastic
process, and we can model this with a random vector with mean $\ve{\hat{u}}$
and covariance $\Sigma$.  The return $\bar{r}$ is then a random
variable with mean $\ve{\hat{u}}^\intercal\ve{w}$ and covariance
 $\ve{w}^\intercal \Sigma \ve{w}$.

To capture realistic problem formulations, we add one or more mathematical constraints to the optimization problem corresponding to the problem-specific considerations. For example, two common constraints in portfolio optimization problems are that we want no short positions ($w_i \geq 0$ for all $i$, denoted by $\ve{w} \geq 0$) and that the total investment budget is limited ($\ve{1}^\intercal \ve{w} = 1$, where $\ve{1}$ denotes the vector of ones). This forms the classical portfolio optimization problem from Markowitz's mean-variance theory:
\begin{equation}\label{eq:PO_min_variance}
  \begin{array}{rrclcl}
    \displaystyle \min_{\ve{w}} & \multicolumn{3}{l}{\ve{w}^\intercal \Sigma \ve{w}} \\
    \textrm{s.t.} & \ve{\hat{u}}^\intercal \ve{w} & \geq & \bar{r}_{\min} \\
    & \ve{1}^\intercal \ve{w} & = & 1\\
    & \ve{w} & \geq & 0\\
  \end{array}
\end{equation}
This formulation is a quadratic optimization problem where we minimize
the risk, while achieving a target return of at least $\bar{r}_{\min}$ with a
fixed budget and no short positions.
In practice, the portfolio optimization problem is often reformulated in other ways, for example, to maximize return subject to a fixed amount of risk, or to optimize an objective function that weighs risk against return. In our application, we follow the latter approach, formulated as follows, where $q$ is a tunable \emph{risk-aversion coefficient}:
\begin{equation}
  \begin{array}{rrclcl}
    \displaystyle \min_{\ve{w}} & \multicolumn{3}{l}{-\ve{\hat{u}}^\intercal \ve{w} + q \sqrt {\ve{w}^\intercal \Sigma \ve{w}}  } \\
    \textrm{s.t.} & \ve{1}^\intercal \ve{w}& = & 1\\
    & \ve{w} & \geq & 0\\
  \end{array}
\end{equation}
This optimization problem is no longer a QO problem, but it can be mapped to a conic problem, as described later in \cref{sec:socp-portfolio}. Depending on the problem, additional constraints can be added.\footnote{For instance, we can add constraints to allow short positions,
component-wise short sale limits, or a total short sale limit. Another variant of this is a constraint for a collateralization requirement, which limits the total of short positions to a fraction of the total long positions.  Often, buying or selling an asset results in a transaction fee that is proportional to the amount of asset that is bought or sold. Linear transaction costs or maximum transaction amounts are often included as constraints in portfolio optimization.  Diversification constraints can limit portfolio risk by limiting the exposure to individual positions and groups of assets within particular sectors.} To illustrate the flexibility of this analysis, we include a maximum transaction constraint and use the following problem formulation in our analysis in the rest of the paper:
\begin{equation}\label{eq:PO}
  \begin{array}{rrclcl}
    \displaystyle \min_{\ve{w}} & \multicolumn{3}{l}{-\ve{\hat{u}}^\intercal \ve{w} + q \sqrt {\ve{w}^\intercal \Sigma \ve{w}}  } \\
    \textrm{s.t.} & \ve{1}^\intercal \ve{w}& = & 1\\
    & \lvert \ve{w} - \ve{\bar{w}}\rvert & \leq & \ve{\zeta}\\
    & \ve{w} & \geq & 0\,,\\
  \end{array}
\end{equation}
where $\ve{\bar{w}}$ denotes the current portfolio, so that $\lvert \ve{w} - \ve{\bar{w}}\rvert$ is the vector of transaction quantities for each asset, which are constrained to be smaller than maximum values contained in the vector $\ve{\zeta}$. 
Note that Ref.~\cite{kerenidis2019quantum} chose a formulation more akin to \cref{eq:PO_min_variance} for their numerical study of the quantum interior point method for portfolio optimization.
For more information on the theory of convex optimization problems and algorithms for solving them, we direct the reader to Refs.~{\mbox{\cite{BV04,Wri97}}}. For more information about optimization methods in finance, we refer to Refs.~{\mbox{\cite{ZV2006,Cornuejols2018,MosekPortfolio2021}}}.



\section[SOCP and IPM]{Second order cone programming (SOCP) and interior point methods (IPM)}\label{sec:socp}


\subsection{Definitions}\label{sec:socp-def}

Second-order cone programming (SOCP) is a type of convex optimization that allows for a richer set of constraints than linear programming (LP), without many of the complications of semidefinite programming (SDP). Indeed, SOCP is a subset of SDP, but SOCP admits interior point methods (IPMs) that are essentially just as efficient as IPMs for LP~\cite{alizadeh2003second}. Many real-world problems can be cast as SOCP, including the portfolio optimization problem we are interested in.

For any $k$-dimensional vector $\ve{v}$, we may write $\ve{v} = (v_0; \ve{\tilde{v}})$, where $v_0$ is the first entry of $\ve{v}$, and $\ve{\tilde{v}}$ contains the remaining $k-1$ entries.

\begin{definition}A \textbf{$k$-dimensional second-order cone} (for $k\geq 2$) is the convex set
\begin{align}
\mathcal{Q}^k = \left\{ (x_0; \ve{\tilde{x}})\in \mathbb{R}^k \;| \; x_0 \ge \lVert \ve{\tilde{x}} \rVert\right\} , \label{eq:second-order-cone}
\end{align}
where $\lVert \cdot \rVert$ denotes the vector two-norm (standard Euclidean norm). For $k=1$, $\mathcal{Q}^1 = \left\{ x_0 \in \mathbb{R} \;| \; x_0 \ge 0\right \}$.
\end{definition}

\begin{definition}\label{def:SOCP}In general, a \textbf{second-order cone problem} is formulated as
\begin{equation}\label{eq:SOCP}
\begin{split}
\min_{\ve{x}} \quad &\ve{c}^\intercal \ve{x} \\
\text{s.t.} \quad &A\ve{x} = \ve{b} \\
&\ve{x} \in \mathcal{Q} ,
\end{split}
\end{equation}
where $\mathcal{Q} = \mathcal{Q}^{N_1}\times ... \times \mathcal{Q}^{N_r}$ is a
Cartesian product of $r$ second-order cones of combined dimension $N = N_1 + \ldots + N_r$, and $A$ is a full-rank $K \times N$ matrix encoding $K$ linear equality constraints, with $K \leq N$.
\end{definition}
Note that the special case of linear programming is immediately recovered if $N_i = 1$ for all $i$.  We say that a point $\ve{x}$ is \emph{primal feasible} whenever $A\ve{x} = \ve{b}$ and $\ve{x} \in \mathcal{Q}$. It is \emph{strictly primal feasible} if additionally it lies in the interior of $\mathcal{Q}$. 

The dual to problem in \cref{eq:SOCP} is a maximization problem over a variable $\ve{y} \in \mathbb{R}^K$, given as follows:

\begin{equation}\label{eq:SOCP-dual}
\begin{split}
\max_{\ve{y}} \quad &\ve{b}^\intercal \ve{y} \\
\text{s.t.} \quad &A^\intercal \ve{y} + \ve{s} = \ve{c} \\
&\ve{s} \in \mathcal{Q} .
\end{split}
\end{equation}

We say that a pair $(\ve{s}; \ve{y})$ is \emph{dual feasible} whenever $A^{\intercal}\ve{y} + \ve{s} = \ve{c}$ and $\ve{s} \in \mathcal{Q}$. For any point $(\ve{x}; \ve{y}; \ve{s})$ with $\ve{x},\ve{s}\in \mathcal{Q}$, we define the \emph{duality gap} as 
\begin{equation}\label{eq:duality_gap}
\mu(\ve{x},\ve{s}) := \frac{1}{r}\ve{x}^\intercal \ve{s} = \frac{1}{r}(\ve{c}^\intercal \ve{x} - \ve{b}^\intercal \ve{y}),
\end{equation}
where $r$ is the number of cones, as in \cref{def:SOCP}, and the second equality holds under the additional assumption that the point is primal and dual feasible. The fact that $\ve{x},\ve{s} \in \mathcal{Q}$ implies that $\mu(\ve{x},\ve{s}) \geq 0$.  Moreover, assuming that both the primal and dual problems have a strictly feasible  point, the optimal primal solution $\ve{x}^*$ and the optimal dual solution $(\ve{y}^*; \ve{s}^*)$ are guaranteed to exist and satisfy $\ve{c}^\intercal \ve{x}^* = \ve{b}^\intercal \ve{y}^*$, and hence $\mu = \frac{1}{r}\ve{x}^{*\intercal}\ve{s}^* = \ve{x}^{*\intercal}(\ve{c} - A^\intercal \ve{y}^*) = \ve{c}^\intercal \ve{x}^* - \ve{b}^\intercal \ve{y}^* = 0$ \cite{alizadeh2003second}. Thus, the primal-dual condition of optimality can be expressed by the system
\begin{equation}\label{eq:optimal_conditions}
\begin{split}
A\ve{x} &= \ve{b} \\
A^\intercal \ve{y} + \ve{s} &= \ve{c} \\
\ve{x}^\intercal \ve{s} &= 0 \\
\ve{x}\in \mathcal{Q}, \;\;& \;\;  \ve{s} \in \mathcal{Q}.
\end{split}
\end{equation}


\subsection{Portfolio optimization as SOCP}\label{sec:socp-portfolio}

The portfolio optimization problem can be solved by reduction to SOCP~\cite{MosekPortfolio2021}, and this reduction is often made in practice.  Here we describe one way of translating the portfolio optimization problem, as given in \cref{eq:PO} into a second-order cone program.

The objective function in \cref{eq:PO} has a non-linear term $q \sqrt{\ve{w}^\intercal \Sigma \ve{w}}$, which we linearize by introducing a new scalar variable $t$, and a new constraint $t \geq \sqrt{\ve{w}^\intercal \Sigma \ve{w}}$.
We obtain the equivalent optimization problem 
\begin{equation}
\begin{split}
\label{eq:main}
\min_{\ve{x}=(\ve{w};t)} \;\; [-\ve{\hat{u}}; q]^\intercal&(\ve{w};t) \\
\text{s.t.} \;\; \ve{1}^{\intercal}\ve{w} &= 1\\
|w_i - \bar{w}_i| &\le \zeta_i\\
w_i &\ge 0 \\
t^2 &\ge \ve{w}^\intercal \Sigma \ve{w}.
\end{split}
\end{equation}

Our goal now is to write the constraints in \cref{eq:main} as second-order cone constraints. Given an $m \times n$ matrix $M$ for which $\Sigma = M^\intercal M$, the constraint on $t$ can be expressed by introducing an $m$-dimensional variable $\ve{\eta}$ subject to the equality constraint $\ve{\eta} = M \ve{w}$ and the second-order cone constraint $(t; \ve{\eta}) \in \mathcal{Q}^{m+1}$. 

The matrix $M$ can be determined from $\Sigma$ via a Cholesky decomposition, although for large matrices $\Sigma$, this computation may be costly. Alternatively, if $\Sigma$ and $\ve{\hat{\mu}}$ are calculated from stock return vectors $\ve{u}^{(1)},\ldots,\ve{u}^{(m)}$ during $m$ independent \emph{time epochs} (e.g.~returns for each of $m$ days or each of $m$ months), then a valid matrix $M^\intercal$ is given by $(\ve{u}^{(1)}-\ve{\hat{u}},\ldots, \ve{u}^{(m)}-\ve{\hat{u}})$, i.e.~the columns of $M^\intercal$ are given by the deviation of the returns from the mean in each epoch. This was the approach taken in Ref.~\cite{kerenidis2019quantum}, and is also the approach we take in our numerical experiments, presented later. The downside to this approach is that the number of time epochs must grow with the number of assets. We note that, in practice, computing the matrix $\Sigma$ can be a research topic unto itself, which is beyond the scope of this paper \cite{Ledoit2020}.

The absolute value constraints are handled by introducing a pair of $n$-dimensional variables $\ve{\phi}$ and $\ve{\rho}$, subject to equality constraints $\ve{\phi} = \ve{\zeta}- (\ve{w} - \ve{\bar{w}})$ and $\ve{\rho} = \ve{\zeta} + (\ve{w} - \ve{\bar{w}})$. The absolute value constraints are then imposed as positivity constraints $\phi_i \geq 0$, $\rho_i \geq 0$, which we include as second-order cone constraints of dimension 1.\footnote{Alternatively, the absolute value constraints could be straightforwardly encoded with $n$ second-order cone constraints of dimension 2; these formulations are equivalent up to a simple coordinate change, and we opt to use 1-dimensional cones for their simplicity of presentation.}

In summary, we may write the portfolio optimization problem from \cref{eq:PO} as the following
SOCP that minimizes over the variable $\ve{x} = (\ve{w}; \ve{\phi}; \ve{\rho};t;\ve{\eta}) \in \mathbb{R}^{3n+m+1}$:
\begin{equation}\label{eq:SOCP-PO}
\begin{split}
\min_{\ve{x}} \quad & [-\ve{\hat{u}}; \ve{0}; \ve{0}; q; \ve{0}]^\intercal
(\ve{w}; \ve{\phi}; \ve{\rho};t;\ve{\eta})
 =: \ve{c}^{\intercal}\ve{x}\\
\text{s.t.} & \begin{pmatrix}
\ve{1}^\intercal  & \ve{0}^\intercal & \ve{0}^\intercal  & 0 & \ve{0}^\intercal \\
I                 & I & 0  & \ve{0} & 0              \\
I                 & 0 & -I & \ve{0} & 0              \\
M                 & 0 & 0  & \ve{0} & -I \\
\end{pmatrix}
\begin{pmatrix}
\ve{w} \\  \ve{\phi} \\ \ve{\rho} \\ t \\ \ve{\eta}
\end{pmatrix} =
\begin{pmatrix}
1 \\  \ve{\bar{w}} + \ve{\zeta} \\ \ve{\bar{w}} - \ve{\zeta} \\ \ve{0}
\end{pmatrix}
\\
& \begin{split}(\ve{w};  \ve{\phi} ;\ve{\rho};t ;\ve{\eta}) & \in \underbrace{\mathcal{Q}^1\times..
.\times \mathcal{Q}^1}_{n\,
\text{positivity constraints}} \\
& \times
\underbrace{\mathcal{Q}^1\times...\times \mathcal{Q}^1}_{2n\,
\text{budget constraints}} \\
& \times
\underbrace{\mathcal{Q}^{m+1}}_{\text{risk}},
\end{split}
\end{split}
\end{equation}
where $I$ denotes an identity block, 0 denotes a submatrix of all 0s, $\ve{0}$ is a vector of all 0s, $\ve{1}$ is a vector of all 1s, and the size of each block of $A$ can be inferred from its location in the matrix. Thus, the total number of cones is $r = 3n + 1$, and the combined dimension is $N = 3n + m + 1$.\footnote{Note that we would have had $r=2n+1$ cones if we had represented the absolute value constraints using dimension-2 cones.} The SOCP constraint matrix $A$ is a $K \times N$ matrix, with $K = 2n+m+1$. This SOCP is very similar to that considered by Kerenidis, Prakash, and Szil\'agyi \cite{kerenidis2019quantum}; however, rather than optimize a weighted combination of risk and return, they optimized risk subject to a fixed value for return, and they did not include the budget constraints.

Notice that many of the rows of the $K \times N$ matrix $A$ are sparse and contain only one or two nonzero entries. However, the final $m$ rows of the matrix $A$ will be dense and contain $n+1$ nonzero entries due to the appearance of the matrix $M$ containing historical stock data; in total a constant fraction of the matrix entries will be nonzero, so sparse matrix techniques will provide only limited benefit.

Finally, we can observe that the primal SOCP in \cref{eq:SOCP-PO} has an interior feasible point as long as $\ve{\zeta}$ has strictly positive entries. To see this, choose $\ve{w}$ to be any strictly positive vector that satisfies $|\ve{w} - \ve{\bar{w}}| <  \ve{\zeta}$, and let $\ve{\phi} = \ve{\zeta} + (\ve{\bar{w}} - \ve{w})$, $\ve{\rho} = \ve{\zeta} - (\ve{\bar{w}} - \ve{w})$, $\ve{\eta} = M\ve{w}$, and $t$ equal to any number strictly greater than $\lVert \ve{\eta}\rVert$. It can be verified that the dual program likewise has a strictly feasible point; this guarantees that the optimal primal-dual pair for the SOCP exists and satisfies \cref{eq:optimal_conditions}.


\subsection{Interior point methods for SOCP}\label{sec:scop-interior-c}


\subsubsection{Introduction}

Interior point methods (IPMs) are a class of efficient algorithms for solving convex optimization problems including LPs, SOCPs, and SDPs, where (in contrast to the simplex method) intermediate points generated by the method lie in the \emph{interior} of the convex set, and they are guaranteed to approach the optimal point after a polynomial number of iterations of the method. Each iteration involves forming a linear system of equations that depends on the current intermediate point. The solution to this linear system determines the \emph{search direction}, and the next intermediate point is formed by taking a small step in that direction. We will consider path-following primal-dual IPMs, where, if the step size is sufficiently small, the intermediate points are guaranteed to approximately follow the \emph{central path}, which ends at the optimal point for the convex optimization problem.


\subsubsection{Central path}

To define the central path, we first establish some notation related to the algebraic properties of the second-order cone. Following formulations in prior literature \cite{alizadeh2003second,Kerenidis2021quantumalgorithms}, we let the product $\ve{u} \circ \ve{v}$ of two vectors $\ve{u}=(u_0; \ve{\tilde{u}}), \;\ve{v} = (v_0; \ve{\tilde{v}}) \in \mathcal{Q}^{k}$ be defined as
\begin{equation}\label{eq:jordan_product}
\ve{u} \circ \ve{v} = (\ve{u}^\intercal \ve{v}; u_0 \ve{\tilde{v}} + v_0 \ve{\tilde{u}} ) 
\end{equation}
and we denote the identity element for this product by the vector $\ve{e} = (1; \ve{0}) \in \mathcal{Q}^k$. For the Cartesian product $\mathcal{Q} = \mathcal{Q}^{N_1}\times \ldots \times \mathcal{Q}^{N_r}$ of multiple second-order cones, the vector $\ve{e}$ is defined as the concatenation of the identity element for each cone, and the circle product of two vectors is given by the concatenation of the circle product of each constituent. A consequence of this definition is the that $\ve{e}^\intercal \ve{e}$ is equal to the number of cones $r$.

Now, for the SOCP problem of \cref{eq:SOCP}, the central path $(\ve{x}(\nu);\ve{y}(\nu);\ve{s}(\nu))$ is the one-dimensional set of \emph{central points}, parameterized by $\nu \in [0,\infty)$, which satisfies the conditions
\begin{equation}\label{eq:central_path_conditions}
\begin{split}
A\ve{x}(\nu) &= \ve{b} \\
A^\intercal \ve{y}(\nu) + \ve{s}(\nu) &= \ve{c} \\
\ve{x}(\nu) \circ \ve{s}(\nu) &= \nu \ve{e} \\
\ve{x}(\nu)\in \mathcal{Q}, \;\;& \;\;  \ve{s}(\nu) \in \mathcal{Q}.
\end{split}
\end{equation}
We can immediately see that the central path point $(\ve{x}(\nu);\ve{y}(\nu);\ve{s}(\nu))$ has a duality gap that satisfies $\mu(\ve{x}(\nu),\ve{s}(\nu))=\nu$, and that when $\nu=0$, \cref{eq:central_path_conditions} recovers \cref{eq:optimal_conditions}.


\subsubsection{Finding an initial point on the central path via self-dual embedding}\label{sec:self-dual}

Path-following primal-dual interior point methods find the optimal point by beginning at a central point with $\nu > 0$ and following the central path to a very small value of $\nu$, which is taken to be a good approximation of the optimal point. For a given SOCP, finding an initial point on the central path is non-trivial and, in general, can be just as hard as solving the SOCP itself. One solution to this problem is the homogeneous self-dual embedding \cite{ye1994nl,andersen2003implementing}, where one forms a slightly larger self-dual SOCP with the properties that (\emph{i}) the optimal point for the original SOCP can be determined from the optimal point for the self-dual SOCP and (\emph{ii}) the self-dual SOCP has a trivial central point that can be used to initialize the IPM.

To do this, we introduce new scalar variables $\tau$, $\theta$, and $\varkappa$, which are used to give more flexibility to the constraints. Previously, we required $A \ve{x} = \ve{b}$. In the larger program, we relax this constraint to read $A \ve{x} = \ve{b}\tau  - (\ve{b} - A \ve{e})\theta$, such that the original constraint is recovered when $\tau = 1$ and $\theta = 0$, but $\ve{x} = \ve{e}$ is a trivial solution when $\tau = 1$ and $\theta = 1$. Similarly, we relax the constraint $A^\intercal \ve{y} + \ve{s} = \ve{c}$ to read $A^\intercal \ve{y} + \ve{s} = \ve{c} \tau - (\ve{c} - \ve{e})\theta$, which has the trivial solution $\ve{y} = \ve{0}$, $\ve{s} = \ve{e}$ when $\tau = \theta = 1$. We complement these with two additional linear constraints to form the program
\begin{equation}\label{eq:SOCP-self-dual}
\begin{split}
&\min_{(\ve{x};\ve{y};\tau;\theta;\ve{s};\varkappa)} \qquad (r+1)\theta \\
&
\begin{pmatrix}
0                       & A^\intercal            & -\ve{c} &  \ve{\bar{c}} \\
-A                      & 0                      & \ve{b}  & -\ve{\bar{b}} \\
\ve{c}^\intercal        & -\ve{b}^\intercal      & 0       & -\bar{z} \\
-\ve{\bar{c}}^\intercal & \ve{\bar{b}}^\intercal & \bar{z} &  0       \\
\end{pmatrix}
\begin{pmatrix}
\ve{x} \\
\ve{y} \\
\tau \\
\theta
\end{pmatrix} + \begin{pmatrix}
\ve{s} \\
\ve{0} \\
\varkappa \\
0
\end{pmatrix} 
=
\begin{pmatrix}
\ve{0} \\
\ve{0} \\
0 \\
r+1
\end{pmatrix}\\
& \ve{x},\ve{s} \in \mathcal{Q}; \qquad \tau,\varkappa \geq 0; \qquad y,\theta \text{ free},
\end{split}
\end{equation}
where $\ve{\bar{b}} = \ve{b}-A\ve{e}$, $\ve{\bar{c}} = \ve{c} - \ve{e}$, $\bar{z} = \ve{c}^\intercal \ve{e} + 1$, and $r = \ve{e}^\intercal\ve{e}$ is the number of cones in the original SOCP. While \cref{eq:SOCP-self-dual} is not exactly of the form given in \cref{eq:SOCP}, we may still think of it as a primal SOCP. Since the block matrix in \cref{eq:SOCP-self-dual} is skew-symmetric and the objective function coefficients are equal to the right-hand-side of the equality constraints, when we compute the dual program (c.f.~\cref{eq:SOCP-dual}), we arrive at an equivalent program; we conclude that \cref{eq:SOCP-self-dual} is \emph{self-dual} \cite{ye1994nl}. Thus, when applying path-following primal-dual IPMs to \cref{eq:SOCP-self-dual}, we need only keep track of the primal variables, that is, $\ve{x},\ve{y},\tau,\theta,\ve{s},\varkappa$. Taking into account the addition of $\tau$ and $\varkappa$, which are effectively an extra pair of primal-dual variables, we redefine the duality gap (c.f.~\cref{eq:duality_gap}) as 
\begin{equation}\label{eq:duality_gap_self_dual}
\mu(\ve{x},\tau, \ve{s},\varkappa) := \frac{1}{r+1}(\ve{x}^\intercal \ve{s} + \varkappa \tau).
\end{equation}
Note that if the point $(\ve{x};\ve{y};\tau;\theta;\ve{s};\varkappa)$ is feasible, i.e.~if it satisfies the four linear constraints in \cref{eq:SOCP-self-dual}, then we have the identity
\begin{align}
    \mu(\ve{x},\tau, \ve{s},\varkappa) &= \frac{-\ve{x}^\intercal A^\intercal \ve{y} + \ve{x}^\intercal \ve{c} \tau  - \ve{x}^\intercal \ve{\bar{c}}\theta + \varkappa \tau}{r+1} \nonumber \\
    &=\frac{-\ve{b}^\intercal \ve{y} \tau + \ve{\bar{b}}^\intercal \ve{y}\theta + \ve{x}^\intercal \ve{c} \tau  - \ve{x}^\intercal \ve{\bar{c}}\theta + \varkappa \tau}{r+1} \nonumber \\
    &= \frac{\ve{\bar{b}}^\intercal \ve{y}\theta  - \ve{x}^\intercal \ve{\bar{c}}\theta + \bar{z}\tau  \theta }{r+1} \nonumber \\
    &= \theta,
\end{align}
where the first, second, third, and fourth rows of \cref{eq:SOCP-self-dual} are invoked above in lines one, two, three, and four, respectively. This equality justifies the redefinition in {\cref{eq:duality_gap_self_dual}}: noting that the primal objective function in {\cref{eq:SOCP-self-dual}} is $(r+1)\theta$, and (since the program is self-dual) the associated dual objective function is $-(r+1)\theta$, we see that the \emph{gap} between primal and dual objective functions, divided by the number of conic constraints ($2r+2$), is exactly equal to $\theta$.

The central path for the augmented SOCP in \cref{eq:SOCP-self-dual} is defined by the feasibility conditions for the SOCP combined with the relaxed complementarity conditions $\ve{x} \circ \ve{s} = \nu \ve{e}$ and $\varkappa \tau = \nu$. Thus, we see that the point $(\ve{x} = \ve{e}; \ve{y} = 0; \tau = 1; \theta = 1; \ve{s} = \ve{e}; \varkappa = 1)$ is not only a feasible point for the SOCP in \cref{eq:SOCP-self-dual}, but also a central point with $\nu = 1$. 

Finally, a crucial property \cite{ye1994nl} of the self-dual SOCP in \cref{eq:SOCP-self-dual} is that the optimal point for the original SOCP in \cref{eq:SOCP} can be derived from the optimal point for the SOCP in \cref{eq:SOCP-self-dual}. Specifically, let $(\ve{x}_{sd}^*;\ve{y}_{sd}^*;\tau^*;\theta^*; \ve{s}_{sd}^*;\varkappa^*)$ be the optimal point for \cref{eq:SOCP-self-dual} (it can be shown that $\theta^*=0$). Then if $\tau^* > 0$, $(\ve{x}^*;\ve{y}^*;\ve{s}^*) = (\frac{\ve{x}_{sd}^*}{\tau^*}; \frac{\ve{y}_{sd}^*}{\tau^*}; \frac{\ve{s}_{sd}^*}{ \tau^*})$ is an optimal primal-dual point for \cref{eq:SOCP,eq:SOCP-dual}. If $\tau^*=0$, then at least one of the original primal SOCP in \cref{eq:SOCP} and the original dual SOCP in \cref{eq:SOCP-dual} must be infeasible \cite{andersen2003implementing,ye1994nl}. As previously demonstrated, the specific SOCP for portfolio optimization in \cref{eq:SOCP-PO} is primal and dual feasible, so $\tau^*\neq0$  for that example. 

What if we only have a point that is \textit{approximately} optimal for the self-dual SOCP? We can still deduce an approximately optimal point for the original SOCP. Suppose we have a feasible point for which $\mu(\ve{x},\tau,\ve{s},\varkappa) = \epsilon$. The point $(\ve{x}/\tau; \ve{y}/\tau; \ve{s}/\tau)$ is $\bigo{\epsilon}$ close to feasible for the original SOCP in the sense that the equality constraints are satisfied up to $\bigo{\epsilon}$ error 
\begin{align}
    \lVert A\frac{\ve{x}}{\tau} - \ve{b} \rVert &= \frac{\epsilon}{\tau}\lVert \ve{b} - A \ve{e} \rVert   \\ 
    \lVert A^\intercal \frac{\ve{y}}{\tau} + \frac{\ve{s}}{\tau} - \ve{c} \rVert &= \frac{\epsilon}{\tau}\lVert \ve{c} - \ve{e} \rVert.
\end{align}
Moreover, since $\varkappa > 0$ and $\theta = \epsilon$, we can assert using the third row of \cref{eq:SOCP-self-dual} that the difference in objective function achieved by the primal and dual solutions is also $\bigo{\epsilon}$, that is
\begin{equation}
    \ve{c}^\intercal \frac{\ve{x}}{\tau} - \ve{b}^\intercal \frac{\ve{y}}{\tau} \leq \frac{\lvert \ve{c}^\intercal \ve{e} + 1 \rvert }{\tau}\epsilon.
\end{equation}
In summary, by using the self-dual SOCP of \cref{eq:SOCP-self-dual}, we obtain a trivial point from which to start the IPM, and given an (approximately) optimal point we obtain either an (approximately) optimal point to the original SOCP or a certificate that the original SOCP was not feasible to begin with. 


\subsubsection{Iterating the IPM}

Each iteration of the IPM takes as input an intermediate point $(\ve{x};\ve{y};\tau;\theta;\ve{s};\varkappa)$ that is feasible (or in some formulations, nearly feasible), has duality gap $\frac{1}{r+1}(\ve{x}^\intercal\ve{s} + \varkappa \tau)$ equal to $\mu$, and is close to the central path with parameter $\nu = \mu$. The output of the iteration is a new intermediate point $(\ve{x}+\ve{\Delta x};\ve{y} + \ve{\Delta y};\tau + \Delta \tau; \theta + \Delta \theta; \ve{s} + \ve{\Delta s}, \varkappa + \Delta \varkappa)$ that is also feasible and close to the central path, with a reduced value of the duality gap. Thus, many iterations leads to a solution with duality gap arbitrarily close to zero. 

One additional input is the step size, governed by a parameter $\sigma < 1$. The IPM iteration aims to bring the next intermediate point onto the central path with parameter $\nu = \sigma \mu$. This is accomplished by taking one step using Newton's method, where the vector $(\Delta \ve{x};\Delta \ve{y};\Delta \tau;\Delta \theta;\Delta \ve{s}; \Delta \varkappa)$ is uniquely determined by solving a linear system of equations called the Newton system. The first part of the Newton system is the conditions that must be met for the new point to be feasible, given in the following system of $N+K+2$ linear equations: 
\begin{equation}\label{eq:Newton-system-1}
\begin{split}
\begin{pmatrix}
0                       & A^\intercal            & -\ve{c} &  \ve{\bar{c}} \\
-A                      & 0                      & \ve{b}  & -\ve{\bar{b}} \\
\ve{c}^\intercal        & -\ve{b}^\intercal      & 0       & -\bar{z} \\
-\ve{\bar{c}}^\intercal & \ve{\bar{b}}^\intercal & \bar{z} &  0       \\
\end{pmatrix}
\begin{pmatrix}
\ve{\Delta x} \\
\ve{\Delta y} \\
\Delta \tau \\
\Delta \theta
\end{pmatrix} + \begin{pmatrix}
\ve{\Delta s} \\
\ve{0} \\
\Delta \varkappa \\
0
\end{pmatrix} 
\\
= \begin{pmatrix}
-A^\intercal \ve{y} + \ve{c}\tau - \ve{\bar{c}}\theta -\ve{s}\\
A \ve{x} - \ve{b}\tau + \ve{\bar{b}} \theta \\
-\ve{c}^\intercal \ve{x} + \ve{b}^\intercal \ve{y} + \bar{z}\theta \\
\ve{\bar{c}}^\intercal \ve{x} -\ve{\bar{b}}^\intercal \ve{y} - \bar{z}\tau
\end{pmatrix}\\
\end{split}
\end{equation}
Note that if the point is already feasible, the right-hand-side is equal to zero.

The second part of the Newton system is the linearized conditions for arriving at the point on the central path with duality gap $\sigma \mu$. That is, we aim for $(\ve{x} + \ve{\Delta x}) \circ (\ve{s} + \ve{\Delta s}) = \sigma \mu \ve{e}$ and $(\varkappa + \Delta \varkappa)(\tau + \Delta \tau) = \sigma \mu$. By ignoring second order terms (i.e.~the $\bigo{\ve{\Delta x}\circ \ve{\Delta s}}$ and $\bigo{\Delta \varkappa \Delta \tau}$ terms), these become
\begin{equation}\label{eq:Newton-intermediate}
\begin{split}
\ve{x} \circ \ve{\Delta s} + \ve{s} \circ \ve{\Delta x} &= \sigma \mu \ve{e} - \ve{x} \circ \ve{s}\\
\varkappa \Delta \tau + \tau \Delta \varkappa &= \sigma \mu  - \varkappa \tau .
\end{split}
\end{equation}

The expression above can be rewritten as a matrix equation by first defining the arrowhead matrix $U$ for a vector $\ve{u} = (u_0; \ve{\tilde{u}}) \in \mathcal{Q}^k$ as
\begin{equation}\label{eq:arrowhead}
U =
\begin{pmatrix}
u_0 & \ve{\tilde{u}}^\intercal \\
\ve{\tilde{u}} & u_0 I
\end{pmatrix} = \ve{u}\ve{e}^{\intercal} + \ve{e}\ve{u}^{\intercal} + u_0 I - 2u_0\ve{e}\ve{e}^{\intercal}.
\end{equation}
When $\ve{u} \in \mathcal{Q}$ lies in the direct product of multiple second-order cones, the arrowhead matrix is formed by placing the appropriate matrices of the above form on the block diagonal. The arrowhead matrix has the property that for any vector $\ve{v}$, $U\ve{v} = \ve{u} \circ \ve{v}$.

Using this notation, the Newton equations in \cref{eq:Newton-intermediate} can be written as
\begin{equation}\label{eq:Newton-system-2}
\begin{pmatrix}
S & 0 & 0      & 0 & X & 0 \\
0 & 0 & \varkappa & 0 & 0 & \tau
\end{pmatrix}
\begin{pmatrix}
\ve{\Delta x} \\
\ve{\Delta y} \\
\Delta \tau \\
\Delta \theta \\
\ve{\Delta s} \\
\Delta \varkappa
\end{pmatrix}
=
\begin{pmatrix}
\sigma \mu \ve{e} - X\ve{s} \\
\sigma \mu - \varkappa \tau 
\end{pmatrix},
\end{equation}
where $X$ and $S$ are the arrowhead matrices for vectors $\ve{x}$ and $\ve{s}$.

\Cref{eq:Newton-system-1,eq:Newton-system-2} together form the Newton system. We can see that there are $2N+K+3$ constraints to match the $2N+K+3$ variables in the vector $(\Delta \ve{x};\Delta \ve{y};\Delta \tau;\Delta \theta;\Delta \ve{s}; \Delta \varkappa)$. In Ref.~\cite{monteiro2000polynomial}, it is shown that, as long as the duality gap is positive and $(\ve{x};\ve{y};\tau;\theta;\ve{s};\varkappa)$ is not too far from the central path (which will be the case as long as $\sigma$ is chosen sufficiently close to $1$ in every iteration), the Newton system has a single unique solution. Note that one can choose different \emph{search directions} than the one that arises from solving the Newton system presented here; this consists of first applying a scaling transformation to the product of second-order cones, then forming and solving the Newton system that results, and finally applying the inverse scaling transformation. Alternate search directions are explained in \cref{app:alt_search_directions}, but in the main text we stick to the basic search direction illustrated above, since in our numerical simulations the simple search direction gave equal or better results than more complex alternatives, and it enjoys the same theoretical guarantee of convergence \cite{monteiro2000polynomial}. 


\subsubsection{Solving the Newton system}\label{sec:solve_newton_system}

The Newton system formed by combining \cref{eq:Newton-system-1,eq:Newton-system-2} is an $L \times L$ linear system of the form $G\ve{u} = \ve{h}$, where $L = 2N+K+3$. Classically this can be solved \emph{exactly} a number of ways, the most straightforward being Gaussian elimination, which scales as $\bigo{L^3}$. Using Strassen-like tricks \cite{strassen1969gaussian}, this can be asymptotically accelerated to $\bigo{L^{\omega}}$ where $\omega < 2.38$ \cite{alman2021refined}, although practically the runtime is closer to $\bigo{L^3}$. Meanwhile, the linear system can be \emph{approximately} solved using a variety of iterative solvers, such as conjugate gradient descent or the randomized Kaczmarz method \cite{strohmer2009randomized}. The complexity of these approaches depends on the condition number of the Newton matrix. \Cref{sec:qipm} discusses \emph{quantum} approaches to solving the Newton system. 

It is important to distinguish methods that exactly solve the Newton system, and methods that solve it inexactly, because inexact solutions typically lead to infeasible intermediate points. As presented above, the Newton system in \cref{eq:Newton-system-1,eq:Newton-system-2} can tolerate infeasible intermediate points; the main consequence is that the right-hand-side of \cref{eq:Newton-system-1} becomes non-zero. This inexact formulation was the one pursued by Kerenidis, Prakash, and Szil\'agyi \cite{Kerenidis2021quantumalgorithms}, who first examined QIPMs for SOCP (although they did not implement the self-dual embedding as we have done). However, it was pointed out in Refs.~\cite{augustino2021quantum,augustino2021inexact} that the theoretical convergence analysis that Ref.~\cite{Kerenidis2021quantumalgorithms} relies upon requires intermediate points to be exactly feasible (i.e.~the right-hand-side of \cref{eq:Newton-system-1} is always zero), and that analyses allowing for infeasibility generally have poorer guaranteed convergence time (although in practice they can be just as fast \cite{Wri97}). As discussed in \cref{sec:qipm}, exact feasibility is difficult to maintain in quantum IPMs, since the Newton system cannot be solved exactly. 

Ref.~\cite{augustino2021inexact} proposed a workaround by which exact feasibility can be maintained despite an inexact linear system solver, which they call an \emph{inexact-feasible IPM} (IF-IPM). For the IF-IPM, we assume we have access to a basis for the null space of the feasibility constraint equations, that is, a linearly independent set of solutions to \cref{eq:Newton-system-1} when the right-hand-side is zero.  We arrange these basis vectors as the columns of a matrix $B$; since there are $N+K+2$ linear feasibility constraints and $2N+K+3$ variables, the matrix $B$ should have $N+1$ columns. In the case of portfolio optimization, a matrix $B$ satisfying this criteria can be deduced by inspection, as discussed in \cref{app:null-space}; however, this choice does not yield a $B$ with orthogonal columns. Generating a $B$ with orthonormal columns can be done by performing a QR decomposition of the matrix in \cref{eq:Newton-system-1}, which would incur a large one-time classical cost of $\bigo{(N+K)^3}$ operations\footnote{Better asymptotic scaling for QR decomposition can be accomplished using fast matrix multiplication \cite{camarero2018simple}.}. In either case, since $B$ is a basis for the null space of the constraint equations, there is a one-to-one correspondence between vectors $\ve{\Delta z} \in \mathbb{R}^{N+1}$, and vectors that satisfy \cref{eq:Newton-system-1} via the relation $(\ve{\Delta x};\ve{\Delta y};\Delta \tau;\Delta \theta;\ve{\Delta s};\Delta \varkappa) = B\ve{\Delta z}$. Thus, our Newton system can be reduced to
\begin{align} \label{eq:Newton-system-IF}
\left[\begin{pmatrix}
S & 0 & 0      & 0 & X & 0 \\
0 & 0 & \varkappa & 0 & 0 & \tau
\end{pmatrix} 
B \right] \ve{\Delta z}
&=
\begin{pmatrix}
\sigma \mu \ve{e} - Xs\\
\sigma \mu - \varkappa \tau 
\end{pmatrix} \\
(\ve{\Delta x};\ve{\Delta y};\Delta \tau;\Delta \theta;\ve{\Delta s};\Delta \varkappa) &= B \ve{\Delta z}. \label{eq:vec=Qz}
\end{align}

\sloppy The Newton system above can be solved by first computing $\ve{\Delta z}$ by inverting the quantity in brackets in the first line and applying it to the right-hand-side, and then computing $(\ve{\Delta x};\ve{\Delta y};\Delta \tau;\Delta \theta;\ve{\Delta s};\Delta \varkappa)$ by performing the multiplication $B\ve{\Delta z}$. This matrix-vector product can be accomplished classically in $\bigo{N^2}$ operations. Note that matrix-matrix products where one of the matrices is an arrowhead matrix ($S$ or $X$) can also be carried out in $\bigo{N^2}$ classical time, as the form of arrowhead matrices given in \cref{eq:arrowhead} implies that the product can be computed by summing several matrix-vector products. Finally, note that since the second and fourth block columns of the first matrix in \cref{eq:Newton-system-2} are zero, the second and fourth block rows of $B$ (e.g.~in \cref{eq:Q}) can be completely omitted from the calculation. 

Thus, we see three main choices for how to run the IPM when the solution to linear systems is inexact: first, by solving \cref{eq:Newton-system-1,eq:Newton-system-2} directly and allowing intermediate solutions to be infeasible; second, by finding a matrix $B$ by inspection as described in \cref{app:null-space} and then solving \cref{eq:Newton-system-IF,eq:vec=Qz}; third, by finding a matrix $B$ via QR decomposition and then solving \cref{eq:Newton-system-IF,eq:vec=Qz}. When the linear system is solved using a quantum algorithm, as discussed in \cref{sec:qipm}, we refer to the algorithm that results from each of these three options by II-QIPM, IF-QIPM, and IF-QIPM-QR, respectively. The pros and cons of each method are summarized in \cref{tab:QIPM_choices}.

\begin{table*}
\renewcommand{\arraystretch}{1.6}
\caption{\label{tab:QIPM_choices}Choices on which version of the Newton system to solve lead to different versions of the QIPM, even with the same underlying quantum subroutines.}
\begin{ruledtabular}
\begin{tabular}{p{14em}p{11em}p{13em}p{11.5em}}
\centering
&\hfil\textbf{II-QIPM} & \hfil \textbf{IF-QIPM} & \hfil \textbf{IF-QIPM-QR}\\
\hline
\hfil\textbf{Newton system} & \hfil \Cref{eq:Newton-system-1,eq:Newton-system-2} & \hfil \Cref{eq:Newton-system-IF,eq:vec=Qz} & \hfil \Cref{eq:Newton-system-IF,eq:vec=Qz} \\
\hfil\textbf{Size of Newton System ($L$)} & \hfil $2N+K+3$ & \hfil $N+1$ & \hfil $N+1$ \\
\textbf{Feasible intermediate points} & \hfil No & \hfil Yes & \hfil Yes \\
\hfil\textbf{\raisebox{-1.1em}{Caveats}} & Theoretical convergence guarantee requires $\bigo{r^2}$ (rather than $\bigo{\sqrt{r}}$) iterations & Ill-conditioned null-space basis leads to large condition number of Newton system & Requires classical QR decomposition, which could dominate overall runtime
\end{tabular}
\end{ruledtabular}
\end{table*}


\subsubsection{Neighborhood of the central path and polynomial convergence}\label{sec:neighborhood_central_path}

Prior literature establishes that if sufficiently small steps are taken (i.e., if $\sigma$ is sufficiently close to 1), then each intermediate point stays within a small neighborhood of the central path. We now review these conclusions. Following Ref.~\cite{monteiro2000polynomial}, for a vector $\ve{u} = (u_0;\ve{\tilde{u}}) \in \mathcal{Q}^k$, we define the matrix
\begin{equation}
T_{\ve{u}} = 
\begin{pmatrix}
u_0 & \ve{\tilde{u}}^\intercal \\
\ve{\tilde{u}} & \sqrt{u_0^2 - \lVert \ve{\tilde{u}} \rVert^2 }I + \frac{\ve{\tilde{u}}\ve{\tilde{u}}^\intercal}{u_0+\sqrt{u_0^2 - \lVert \ve{\tilde{u}} \rVert^2 }}
\end{pmatrix},
\end{equation}
which, as for the arrowhead matrix, generalizes to the product of multiple cones by forming a block diagonal of matrices of the above form. We use the distance metric defined in Ref.~\cite{monteiro2000polynomial}
\begin{equation}\label{eq:distance_to_cp}
\begin{split}
& d_{F}(\ve{x},\tau,\ve{s},\varkappa ) = \\
& \sqrt{2}\sqrt{\lVert  T_{\ve{x}}\ve{s} - \mu(\ve{x},\tau, \ve{s}, \varkappa) \ve{e} \rVert^2  + (\tau\varkappa -\mu(\ve{x},\tau, \ve{s}, \varkappa))^2 }.
\end{split}
\end{equation}
The distance metric induces a neighborhood $\mathcal{N}$, which includes both feasible and infeasible points, as well as the neighborhood $\mathcal{N}_F$, which includes only feasible points
\begin{align}
\label{eq:neighborhood}\mathcal{N}(\gamma) &= \{(\ve{x};\ve{y};\tau;\theta;\ve{s};\varkappa):  \\ \nonumber
& \qquad d_F(\ve{x},\tau,\ve{s},\varkappa )\leq \gamma \mu(\ve{x},\tau, \ve{s}, \varkappa)\}  \\
\mathcal{N}_F(\gamma) &= \mathcal{N}(\gamma) \cap \mathcal{P}_F, \label{eq:neighborhood_F}
\end{align}
where $\mathcal{P}_F$ denotes the set of feasible points for the self-dual SOCP. Note that the vector $T_{\ve{x}}s$ can be computed classically in $\bigo{N}$ time given access to the entries of $\ve{x}$ and $\ve{s}$. Thus, whether or not a point lies in $\mathcal{N}(\gamma)$ can be determined in $\bigo{N}$ time.  

Corollary 1 of Ref.~\cite{monteiro2000polynomial} then implies that, so long as $0 \leq \gamma \leq 1/3$ and $(\ve{x};\ve{y};\tau;\theta;\ve{s};\varkappa) \in \mathcal{N}_F(\gamma)$, then we have
\begin{equation}
(\ve{x}+\ve{\Delta x};\ve{y}+\ve{\Delta y};\tau+\Delta \tau;\theta+\Delta \theta;\ve{s}+\ve{\Delta s};\varkappa+\Delta \varkappa) \in \mathcal{N}_F(\Gamma),
\end{equation}
where
\begin{equation}
\Gamma = \frac{4(\gamma^2 + 2(r+1)(1-\sigma)^2)}{(1-3\gamma)^2\sigma}.
\end{equation}

Thus, if $\Gamma \leq \gamma$, and assuming the Newton system is solved exactly, every intermediate point will lie in $\mathcal{N}_F(\gamma)$. This condition is met, for example, if $\gamma = 1/10$ and $\sigma = 1-(20\sqrt{2}\sqrt{(r+1)})^{-1}$. Since each iteration reduces the duality gap by a factor $\sigma$, the duality gap can be reduced to $\epsilon$ after roughly only $20\sqrt{2(r+1)}\ln(1/\epsilon)$ iterations. If the Newton system is solved inexactly, but such that feasibility is preserved (e.g., by solving inexactly for $\ve{\Delta z}$ and then multiplying by $B$, as described above), then an error $\ve{\delta}$ on the vector $(\ve{x};\tau;\ve{s};\varkappa)$ can be tolerated, and the resulting vector can still be within the neighborhood at each iteration.

On the other hand, if the Newton system is not solved exactly, then the resulting vector may not be feasible. Since $\mathcal{N}_F(\gamma)$ is defined as a subset of the feasible space, the analysis of Ref.~\cite{monteiro2000polynomial} breaks down (as pointed out in Refs.~\cite{augustino2021quantum,augustino2021inexact}). Thus, the II-QIPM version of the QIPM does not enjoy the theoretical guarantee of convergence in $\bigo{\sqrt{r}}$ iterations that the IF-QIPM and IF-QIPM-QR versions do (see \cref{tab:QIPM_choices}). The best guarantees for the II-QIPM would imply convergence only after $\bigo{r^2}$ iterations \cite{augustino2021quantum,augustino2021inexact}. Nevertheless, it is unclear if a small amount of infeasibility makes a substantial difference in practice: we simulated multiple version of the QIPM and observed similar overall performance when intermediate solutions were allowed to be infeasible, despite an inferior theoretical guarantee of success. Thus, in \cref{sec:implementation,sec:numerics}, where we present the full QIPM implementation, resource count, and numerical analysis, we focus on the II-QIPM. We present some of the results of our numerical simulations of the IF-QIPM and IF-QIPM-QR results in the appendix.



\section{Quantum interior point methods (QIPM)}\label{sec:qipm}

\subsection{Basic idea of QIPM}\label{sec:qipm-idea}

As discussed in \cref{sec:socp}, each iteration of an IPM SOCP solver involves forming and solving a linear system of equations that depends on the intermediate point at the current iteration. For classical IPM implementations for SOCP, the linear systems of equations are typically solved exactly; for example the numerical SOCP solving package ECOS solves linear systems with a sparse LDL (Cholesky) factorization \cite{domahidi2013ecos}. For arbitrary dense systems, the runtime of solving an $L \times L$ system this way is $\bigo{L^3}$ \cite{krishnamoorthy2013matrix}, but by exploiting sparsity the actual runtime in practice could be much faster, by an amount that is hard to assess. 
Alternatively, it would, in principle, be possible to employ classical iterative approximate linear system solvers such as conjugate gradient descent or the randomized Kaczmarz method. The choice of the linear system solver thereby determines the overall complexity of the IPM SOCP solver. The idea of QIPM, as pioneered in Refs.~\cite{Kerenidis2020lpsdp,augustino2021quantum}, is to use a quantum subroutine to solve the linear system of equations \cite{HHL09}. Notably, all other steps of IPMs stay classical and remain the same as described in \cref{sec:socp}. As a quantum linear system solver (QLSS) does not solve the exact same mathematical problem as classical linear system solvers and, moreover, a QLSS needs coherent (quantum) access to the classical data as given by the entries of the relevant matrices, there are various additional tools we will discuss that  allow us to embed QLSS subroutines as a step of IPM SOCP solvers.

First, we discuss in \cref{sec:qipm-linear-solver} the input and output model of QLSSs and present the complexity of state-of-the-art QLSSs. Then, in \cref{sec:qipm-qram}, we give constructions based on quantum random access memory (QRAM) to load classical data as input into a QLSS and discuss the complexity overhead arising from that step. Subsequently, in \cref{sec:qipm-tomography}, we present so-called pure state quantum tomography that allows to convert the output of the QLSS into an estimate of the classical solution vector of the linear system of equations. Finally, in \cref{qipm-complexity}, we put all the steps together and state the overall classical and quantum complexities of using QLSSs as a subroutine in IPM SOCP solvers. As described in previous work \cite{kerenidis2019quantum}, the ultimate idea is to compare these costs to the complexities of classical IPM SOCP solvers and point out regimes where quantum methods can potentially scale better that any purely classical methods (e.g., in terms of the SOCP size $N$, the matrix condition number $\kappa$, etc.)

We note that the content of this section largely corresponds to collecting various state-of-the-art results from prior literature. These ingredients are used together with the conceptual framework of \cite{Kerenidis2020lpsdp,kerenidis2019quantum,augustino2021quantum,augustino2021inexact} to lift the QIPMs presented there to superior efficiency. In \cref{sec:implementation}, we present a few novel enhancements to the implementation of the QIPM and fully explicit, end-to-end quantum circuits with corresponding novel finite size complexities.


\subsection{Quantum linear system solvers}\label{sec:qipm-linear-solver}

For our purposes, a linear system of equations is given by a real invertible $L\times L$ matrix $G$ together with a real vector $\ve{h}=(h_1,\ldots,h_L)$, and one is looking to give an estimate of the unknown solution vector $\ve{u}=(u_1,\ldots,u_L)$ defined by $G\ve{u}=\ve{h}$. We define the (Frobenius) condition number
\begin{align}
\kappa_F(G):=\|G\|_F\left\|G^{-1}\right\|,
\end{align}
where $\|\cdot\|_F$ denotes the Frobenius norm and $\|\cdot\|$ for a matrix argument denotes the spectral norm.

For this setting, the input to a QLSS is then comprised of: \textit{(i)} a preparation unitary $U_{\ve{h}}$ that creates the $\ell:=\lceil\log L\rceil$ qubit quantum state
\begin{align}\label{eq:Ub}
\ket{\ve{h}}:=\|\ve{h}\|^{-1}\cdot\sum_{i=1}^Lh_i\ket{i}\quad\text{via $\ket{\ve{h}}=U_{\ve{h}}\ket{0}^{\otimes \ell}$,}
\end{align}
where $\|\cdot\|$ for a vector argument denotes the vector two-norm (standard Euclidean norm), \textit{(ii)} a block encoding unitary $U_G$ in the form
\begin{align}\label{eq:UA}
U_G := \begin{pmatrix} \frac{G}{\|G\|_F} & \cdot \\ \cdot & \cdot \end{pmatrix}
\end{align}
on $\ell+\ell_G$ qubits for some $\ell_G\in\mathbb{N}$, and \textit{(iii)} an approximation parameter $\varepsilon_\mathrm{QLSP}\in(0,1]$. The quantum linear system problem (QLSP) is stated as follows: For a triple $(G,\ve{h},\varepsilon_\mathrm{QLSP})$ as above, the goal is to create an $\ell$-qubit quantum state $ \ket{\tilde{\ve{v}}}$ such that\footnote{In this formulation, the quantum state $\ket{\ve{v}}$ corresponds to the normalized solution vector of the normalized linear system $G\ve{u} = \ve{h}$. Thus, the state $\ket{\ve{v}}$ does not carry information on the norm of the solution $\lVert \ve{u} \rVert$. This norm is related to $\ve{v}$ by the relationship $\lVert \ve{u} \rVert = \lVert \ve{h}\rVert / \lVert G \ve{v} \rVert$.}
\begin{align}\label{eq:QLSP_def}
\Big\|\ket{\tilde{\ve{v}}}-\ket{\ve{v}}\Big\|\leq\varepsilon_\mathrm{QLSP}\quad\text{for $\ket{\ve{v}}:=\frac{\sum_{i=1}^Lu_i\ket{i}}{\left\|\sum_{i=1}^Lu_i\ket{i}\right\|}$,}
\end{align}
defined by $G\ve{u} = \ve{h}$ with $\ve{u}=(u_1,\ldots,u_L)$,
by employing as few times as possible the unitary operators $U_G,U_{\ve{h}}$, their inverses $U_G^\dagger,U_{\ve{h}}^\dagger$, controlled versions of $U_G,U_{\ve{h}}$, and additional quantum gates on potentially additional ancilla qubits. The QLSP together with the first QLSS was introduced in \cite{HHL09} and then gradually improved in \cite{Childs17,Ambainis10,subasi2019quantumLinearSystemAdiabatic,an2022,Lin20}. The state-of-the-art QLSS \cite{costa2021optimal} using the fewest calls to $U_G,U_{\ve{h}}$ and their variants, is based on ideas from discrete adiabatic evolution \cite{Dranov98}. We note the following explicit complexities from \cite[Theorem 9]{costa2021optimal}, adapted to our setting.

\begin{proposition}\label{lem:adiabatic-discrete}
The QLSP for $(G,\ve{h},\varepsilon_1)$ can be solved with a quantum algorithm on $\lceil\log_2(L)\rceil+4$ qubits for
\begin{align}
\varepsilon_1\leq C\cdot\frac{\kappa_F(G)}{Q}+\bigoLONE\left(\frac{\sqrt{\kappa_F(G)}}{Q}\right)
\end{align}
for some constant $C \le 44864 $ using $Q\geq\kappa_F(G)$ controlled queries to each of $U_G$ and $U_G^\dagger$, and $2Q$ queries to each of $U_{\ve{h}}$ and $U_{\ve{h}}^\dagger$, and constant quantum gate overhead. 
\end{proposition}

We note that a stronger version of above proposition works with the (regular) condition number $\kappa(G):=\|G\|\|G^{-1}\|$, but it requires a block-encoding of the form \cref{eq:UA} in which the normalization factor is $\lVert G \rVert$ rather than $\lVert G \rVert_F$. For general matrices of classical data, we do not know of a method to produce such a block-encoding.  In our case, we work with the Frobenius version $\kappa_F(G)$, since we do have a straightforward method to perform $U_G$ with normalization factor $\lVert G \rVert_F$, described in \cref{sec:qipm-qram}. It is then sufficient to give upper bounds for the remaining $\kappa_F(G)$ to run the algorithm from \cref{lem:adiabatic-discrete}. In practice, we will give such upper bounds by using appropriate heuristics (cf.~\cref{sec:implementation} on implementations).

Note that \cref{lem:adiabatic-discrete} implies a solution to the QLSP in \cref{eq:QLSP_def} with an asymptotic query complexity of $\bigo{\kappa_F/\varepsilon_\mathrm{QLSP}}$ to $U_G$, $U_{\ve{h}}$ and their variants and under standard complexity-theoretic assumptions this is optimal in terms of the scaling $\bigo{\kappa}$ \cite{HHL09}, but not in terms of the scaling $\bigo{\varepsilon_\mathrm{QLSP}}$. To get to an improved $\bigo{\log(1/\varepsilon_\mathrm{QLSP})}$ scaling, the authors of \cite{costa2021optimal} further rely on the eigenstate filtering method of \cite[Sec.~3]{Lin20} that additionally invokes a quantum singular value transform based on a minimax polynomial. We note the following overall complexities from \cite[Theorem 11]{costa2021optimal}, adapted to our setting.

\begin{proposition}\label{lem:adiabatic-discrete-filtered}
The QLSP problem for $(G,\ve{h},\varepsilon_2)$ can be solved with a quantum algorithm on $\lceil\log_2(L)\rceil+5$ qubits that produces a quantum state
\begin{align}\label{eq:qls-state}
\sqrt{p} \ket{0^5} \ket{\tilde{\ve{v}}} + \sqrt{1-p} \ket{\perp}\ket{\rm{fail}}
\end{align}
with $\langle0^5 \vert \!\! \perp \rangle = 0$ and success probability $p \ge 1/2$. With that, the sought-after $\varepsilon_2$-approximate solution quantum state $\ket{\tilde{\ve{v}}}$ can be prepared using $Q+d$ controlled queries to each of $U_G$ and $U_G^\dagger$, and $2Q+2d$ queries to each of $U_{\ve{h}}$ and $U_{\ve{h}}^\dagger$, where
\begin{align}\label{eq:kappa-scaling}
Q &= \frac{1}{\sqrt{2-\sqrt{2}}}C\kappa_F(G) + \bigoLONE\left(\sqrt{\kappa_F(G)}\right) \\
d &= 2\kappa_F(G)\ln(2/\varepsilon_2)\,.
\end{align}
Here, $C \leq 44864$ is the same constant as in \cref{lem:adiabatic-discrete}. 
\end{proposition}

This version of the algorithm essentially uses \cref{lem:adiabatic-discrete} with a constant choice of $\varepsilon_1 \leq \sqrt{2-\sqrt{2}}$, which ensures that the state prepared has overlap at least $ 1/\sqrt{2}$ with the ideal state  $\ket{\ve{v}}$.  Then, it uses eigenstate filtering to measure whether the final state is the correct solution state. On average we need to repeat the algorithm no more than twice to produce the desired state $\ket{\tilde{\ve{v}}}$. The resulting scaling that \cref{lem:adiabatic-discrete-filtered} implies for the QLSP problem in \cref{eq:QLSP_def} is  $\bigo{\kappa\log(1/\varepsilon_\mathrm{QLSP})}$. Following the findings from \cite[Sec.~V]{costa2021optimal}, we note that in practice the $Q\approx 1.31C\kappa_F(G)$ dominates over $d$ and all other terms can be safely neglected for typical settings\,---\,even for finite scale analyses. Moreover, the constant $C$ is typically an order of magnitude smaller than the estimates given \cite[Sec.~IV.E]{costa2021optimal};  numerical estimates produced a smaller value of $2305$. No direct estimates for general matrices $G$ are available from \cite{costa2021optimal}, but we will henceforth assume $C=2305$ for our numerical estimates.
Additionally, note that for the eigenstate filtering step via QSVT, the minimax polynomial from \cite[Sec.~3]{Lin20} and its corresponding quantum signal processing angles have to be computed. This is done as part of classical pre-processing \cite[Sec.~III]{Dong21}.\footnote{Whereas the methods from \cite[Sec.~III]{Dong21} do not have a quantified worst case convergence guarantee, they work very well in practice by typically running in time $\mathrm{polylog}(d\delta^{-1})$ for precision $\delta\in(0,1]$ and $d=\bigo{\kappa_F(G)\log(1/\varepsilon_\mathrm{QLSP})}$ the degree of the underlying polynomial. Alternatively, one might resort to the provable methods from \cite{Haah2019product} that are known to run with complexity $\bigo{d^3\mathrm{polylog}(d\delta^{-1})}$, also see~\cite{gilyen2019} for a discussion.} 

Note that the implementation of the QLSS in each of \cref{lem:adiabatic-discrete} and \cref{lem:adiabatic-discrete-filtered} assume perfect implementation of the underlying circuits, without additional gate synthesis errors. In practice, however, these circuits will not be implemented perfectly, and hence we will later include additional sources of error (e.g., block-encoding error, imperfect rotation gates, etc.) that also contribute to $\varepsilon_\mathrm{QLSP}$. We include these additional contributions in \cref{sec:qipm-tomography}, for example.

In the following, we continue by laying out the additional classical and quantum resources needed to employ QLSS for estimating in an end-to-end fashion the classical solution vector $\ve{v}=(v_1,\ldots,v_L)$ instead of the quantum state $\ket{\ve{v}}$.


\subsection[Block-encoding via QRAM]{Block-encoding via quantum random access memory (QRAM)}\label{sec:qipm-qram}


\begin{table*}
\caption{\label{tab:min_T_depth}Logical quantum resources required to block-encode (left column) and control-block-encode (right column) an $L \times L$ matrix $G$ to precision $\varepsilon_G\in[0,1]$, where we assume that $L=2^\ell$. Here we have suppressed terms doubly and triply logarithmic in $L$ and $1/\varepsilon_{G}$ (see \cite{clader2022quantum}).}
\renewcommand{\arraystretch}{1.4}
\begin{ruledtabular}
\begin{tabular}{lcc}
\textbf{Resource} & \textbf{Block Encoding} & \textbf{Controlled Block Encoding}\\
\hline
\textbf{\# of qubits}  & $N_{Qbe}:=4 L^2 - 3 L + 2 \ell -1 $                  & $N_{Qcbe}:=N_{Qbe} + L$\\ 
\textbf{$T$-depth}  & $T_{Dbe}:=10\ell+24\log_2(1/\varepsilon_G) +44 $                           & $T_{Dcbe}:=T_{Dbe} + 4$ \\
\textbf{$T$-count}  & $T_{Cbe}:=(12\log_2(1/\varepsilon_G) + 56)L^2 - 24L -12 \log_2(1/\varepsilon_G)- 32\ell -32 $ & $T_{Ccbe}:=T_{Cbe} + 16(L-1)$\\
\end{tabular}
\end{ruledtabular}
\end{table*}



\begin{table*}
\caption{\label{tab:sp_min_T_depth}Logical quantum resources required to prepare an arbitrary $\ell$-qubit quantum state $\ket{\ve{h}}$ from classical data (left column) and a single-qubit controlled version (right column) to precision $\varepsilon_{\ve{h}}\in(0,1]$. Here we have suppressed terms doubly and triply logarithmic in $L$ and $1/\varepsilon_{\ve{h}}$ (see \cite{clader2022quantum}). For a single-qubit control, there are no additional Clifford gates required, which can be observed by examining the state-preparation procedure in \cite[Sec.~IIID]{clader2022quantum} and noting that we can prepare the state $\ket{0}\ket{0}^{\otimes \ell} + \ket{1}\ket{\psi}$ with minor modifications to the procedure that prepares $\ket{\psi}$. First, we use the ``flag'' qubits to control both the angle loading and unloading steps (rather than just the unloading steps), and second, we control every flip of the flag qubits in that procedure with the first single-qubit control, thus turning NOT gates into CNOT gates, which are also Clifford. When the control is ON, the procedure works as before, and when the control is OFF, none of the qubits leave the $\ket{0}$ state. }
\begin{ruledtabular}
\renewcommand{\arraystretch}{1.4}
\begin{tabular}{lcc}
\textbf{Resource} & \textbf{State Preparation} & \textbf{Controlled State Preparation}\\
\hline
\textbf{\# of qubits}  & $N_{Qsp}:= 4 L + \ell -6$                     & $N_{Qcsp}:=N_{Qsp}+1$\\ 
\textbf{$T$-depth}  & $T_{Dsp}:=3\ell+12\log_2(1/\varepsilon_{\ve{h}}) + 24$                        & $T_{Dcsp}:=T_{Dsp}$ \\
\textbf{$T$-count}  & $T_{Csp}:=(12\log_2(1/\varepsilon_{\ve{h}}) + 40)L -12\log_2(1/\varepsilon_{\ve{h}})- 16\ell -40  $ & $T_{Ccsp}:=T_{Csp}$\\
\end{tabular}
\end{ruledtabular}
\end{table*}


In many quantum algorithms (and in particular for our use case), one needs coherent access to classical data for use in the algorithm. \emph{Block-encodings} of matrices provide a commonly used access model for the classical data by encoding matrices into unitary operators, thereby providing oracular access to the data. As mentioned above, for a matrix $G \in \mathbb{R}^{L\times L}$, a unitary matrix $U_G$ block-encodes $G$ when the top-left block of $U_G$ is proportional to $G$, i.e.
\begin{align}\label{eq:block_encoding}
U_G = \begin{pmatrix} G/\alpha & \cdot \\ \cdot & \cdot \end{pmatrix},
\end{align}
where $\alpha \ge \|G\|$ is a normalization constant, chosen as $\alpha=\|G\|_F$ for our use case. The other blocks in $U_G$ are irrelevant, but they must be encoded such that $U_G$ is unitary. For our purposes, we focus on real matrices $G$, but the extension to complex matrices is straightforward. A block-encoding makes use of unitaries that implement (controlled) state preparation, as well as quantum random access memory (QRAM) data structures for loading the classical data. Specifically, we refer to QRAM as the quantum circuit that allows query access to classical data in superposition
\begin{align}\label{eq:qram_query}
\sum_j \psi_j \ket{j}\ket{0} \overset{\text{QRAM}}{\longrightarrow} \sum_j \psi_j \ket{j}\ket{a_j},
\end{align}
where $j$ is the address in superposition with amplitude $\psi_j$ and $\ket{a_j}$ is the classical data loaded into a quantum state. There are several models of QRAM one can use that differ in the way in which the data is loaded. The two most notable QRAM models are the select-swap (SS) model, which is particularly efficient in terms of $T$-gate utilization \cite{low2018trading}, and the bucket-brigade (BB) model \cite{giovannetti2008qram}, which has reduced susceptibility to errors when operated on potentially faulty hardware \cite{hann2021}. 

The block-encoding unitary $U_G$ acts on $\ell+\ell_G$ qubits, where $\ell = \lceil \log_2(L)\rceil$ and, in our construction, $\ell_G = \ell$.  To build it, we follow the prescription of~\cite{kerenidis2016quantum,gilyen2019,chakrabarti2020quantum}, in which one forms $U_G$ as the product of a pair of controlled-state preparation unitaries $U_L$ and $U_R$. Specifically,
\begin{align}
U_G &= U_R^{\dagger} U_L,\\
U_R:&\ket{0}^{\otimes l}\ket{j} \mapsto \ket{\psi_j}\ket{j} \\
U_L:&\ket{0}^{\otimes l}\ket{k} \mapsto \ket{k}\ket{\phi_k}
\end{align}
where the $\ell$-qubit states $\ket{\psi_j}$ and $\ket{\phi_k}$ are determined from the matrix elements $G_{jk}$ of $G$, as follows:
\begin{align}
    \ket{\psi_j}&= \sum_k \frac{G_{jk}}{\lVert G_{j,\cdot} \rVert} \ket{k}  \\
    \ket{\psi_k} &= \sum_j \frac{\lVert G_{j, \cdot} \rVert}{\lVert G \rVert_F}\,,
\end{align}
where $G_{j,\cdot}$ denotes the $j$th row of $G$.
That is, controlled on the second $\ell$-qubit register in the state $\ket{j}$, $U_R$ prepares the $\ell$-qubit state $\ket{\psi_j}$ into the first $\ell$-qubit register, and $U_L$ performs the same operation for the states $\ket{\phi_k}$ modulo a swap of the two registers. Both $U_L$ and $U_R$ utilize an additional $\ell'$ QRAM ancilla qubits that begin and end in the state $\ket{0}$. These controlled-state preparation unitaries $U_R$ and $U_L$ are implemented by combining a QRAM-like data-loading step with a protocol for state preparation of $\ell$-qubit states. There are several combinations of state preparation procedure and QRAM model one can choose with varying benefits and resource requirements. In~\cite{clader2022quantum}, a subset of the authors of the present work studied the resources required to implement these block-encodings and provided explicit circuits for their implementation. For our immediate purposes, we will simply import the relevant resource estimates from that work in \cref{tab:min_T_depth}, and we refer the interested reader to~\cite{clader2022quantum} for further details.\footnote{In our setting, the matrices to block encode are typically dense, which is why the general constructions from \cite{clader2022quantum} are sufficient. However, in case the relevant data has some structure, e.g., if it is sparse, more adapted strategies such as \cite{Camps22} can be preferable.} For our purposes, we will work with the minimum depth circuits that achieve a $T$-gate depth of $\bigo{\log L}$, at the price of using a total number of $\bigo{L^2}$ many qubits for the data structure implementing the block encoding unitary $U_G$. Finally, the $\ell$-qubit unitary $U_{\ve{h}}$ defined by $\ket{\ve{h}}=U_{\ve{h}}\ket{0}^{\otimes \ell}$ corresponds to the special case of quantum state preparation and is directly treated by the methods outlined in \cite[Sec.~III.C]{clader2022quantum}. The resources required to synthesize $U_{\ve{h}}$ up to error $\varepsilon_{\ve{h}}$ are also reported in \cref{tab:min_T_depth}.

The minimum-depth block encodings of \cite{clader2022quantum} also incur some classical costs. Specifically, the quoted depth values are only achievable assuming a number of angles have been classically pre-computed and for each angle a gate sequence of single-qubit Clifford and $T$ gates that synthesizes a single-qubit rotation by that angle up to small error. Calculating one of the angles can be done by summing a subset of the entries of $G$ and computing an arcsin. Meanwhile, circuit synthesis requires applying a version of the Solovay-Kitaev algorithm \cite{dawson2005solovay,ross2016optimal}. For the block-encoding procedure, $L(L-1)$ angles and their corresponding gate sequences must be computed, which requires a total runtime of $L^2\polylog(1/\varepsilon_G)$~\cite{ross2016optimal}, although this computation is amenable to parallelization. For the state preparation procedure, $L-1$ angles and their sequences are needed.


\subsection{Quantum state tomography}\label{sec:qipm-tomography}

We have described how we can produce a quantum state $\ket{\ve{\tilde{v}}}$ approximating the (real-valued) solution $\ket{\ve{v}}$ of a linear system up to precision $\varepsilon_{\mathrm{QLSP}}$. As mentioned in \cref{sec:qipm-linear-solver}, in the actual circuit implementation, the approximation error $\varepsilon_{\mathrm{QLSP}}$ accounts for both the inherent error from eigenstate filtering captured in \cref{lem:adiabatic-discrete-filtered} as well as additional gate synthesis error arising from imperfect implementation of block-encoding unitaries and single-qubit rotations. The next step is to approximately read out the amplitudes of $\ket{\ve{\tilde{v}}}$ into classical form. To start out, we will prove the following proposition, which tells us how many copies of a quantum state are needed to provide a good enough classical description of it, up to a phase on each amplitude. This proposition and its proof are adapted from \cite[Proposition 13]{apeldoorn22}, with somewhat sharpened constant factors.

\begin{proposition}\label{lem:bernstein}
Let $0 < \varepsilon, \delta < 1$ and $|\psi\rangle = \sum_{j \in [L]}\alpha_j |j\rangle$ be a quantum state. Then, \newline $\frac{5+\sqrt{21}}{3\varepsilon^2}\ln(2L/\delta) < 3.1942 \varepsilon^{-2} \ln(2L/\delta)$ measurements of $|\psi\rangle$ in the computational basis suffice to learn an $\varepsilon$-$\ell_{\infty}$-norm estimate $|\tilde{\ve{\alpha}}|$ of $|\ve{\alpha}|$, with success probability at least $1 - \delta$.
\end{proposition}

We give the proof in \cref{app:tomography}. Recall that \cref{lem:adiabatic-discrete-filtered} gives a unitary $U$ such that
\begin{align}
U\ket{0^5}\ket{0^\ell} = \sqrt{p} \ket{0^5} \ket{\tilde{\ve{v}}} + \sqrt{1-p} \ket{\perp} \ket{\mathrm{fail}}
\end{align}
with $\ket{\tilde{\ve{v}}} = \sum_{i=1}^N \tilde{v}_i\ket{i}$, $\langle0^5\ket{\perp} = 0$, and $p \ge 1/2$. The vector $\tilde{\ve{v}}$ may have complex coefficients, but it approximates a real vector $\ve{v}$ up to some error $\varepsilon_\mathrm{QLSP}$ in $\ell_2$ norm. Our goal is to obtain an estimate $\tilde{\ve{v}}'=(v_1',\ldots,v_N')$ such that
\begin{align}
\|\ve{v} - \tilde{\ve{v}}'\| \le \xi \quad\text{for an error parameter $\xi\in[0,1]$.}
\end{align}
where $\xi$ captures all sources of error. 
\Cref{lem:bernstein} is not quite sufficient because it only gives us an estimate of the absolute value of $\tilde{\ve{v}}$. However, the following procedure, adapted from \cite[Sec.~4]{Kerenidis2020lpsdp}, will be sufficient:

\begin{enumerate}
    \item Create $k = 57.5 L \ln(6L/\delta) / (\varepsilon^2(1-\varepsilon^2/4))$ many copies of the quantum state $U\ket{0^{5+\ell}} = \sqrt{p} \ket{0^5} \ket{\tilde{\ve{v}}} + \sqrt{1-p} \ket{\perp} \ket{\mathrm{fail}}$, and measure them all in the computational basis to give empirical estimates $\{p_i\}_{i=1}^L$ of the probabilities $p|\tilde{v}_i|^2$.
    \item Using controlled applications of $U$, create $k = 57.5 L \ln(6L/\delta) / (\varepsilon^2(1-\varepsilon^2/4))$ copies of 
    \begin{align}
    & 2^{-1/2}\ket{0^5}\ket{0} \sqrt{p} \ket{\ve{\tilde{v}}} \\ \nonumber
	& + 2^{-1/2}\ket{0^5}\ket{1} \sum_{i=1}^L \sqrt{p'_i} \ket{i} \\ \nonumber
 	& + \ket{\perp'} \ket{\mathrm{fail}'}, \label{eq:controlled-U}
    \end{align} which by applying a Hadamard can be mapped to 
    \begin{align}
    & \ket{0^5}\ket{0} \frac{ \sqrt{p}\ket{\tilde{\ve{v}}} + \sum_{i=1}^L \sqrt{p'_i} \ket{i}}{2} \\ \nonumber
	& + \ket{0^5}\ket{1} \frac{ \sqrt{p}\ket{\tilde{\ve{v}}} - \sum_{i=1}^L \sqrt{p'_i} \ket{i}}{2} \\ \nonumber
	& + \ket{\perp'} \ket{\mathrm{fail}''}.\label{eq:tomography_sign_state}
    \end{align}
    Here $\ket{\perp'}$ is an arbitrary state orthogonal to $\ket{0^5}$ and $\ket{\mathrm{fail}'}$ and  $\ket{\mathrm{fail}''} $ are arbitrary unnormalized states. The quantities $\sqrt{p'_i}$ are (possibly complex) amplitudes that satisfy $|\sqrt{p'_i}-\sqrt{p_i}| \le \varepsilon_\mathrm{tsp}$ for all $i$; they arise because the state $\sum_{i=1}^L \sqrt{p'_i} \ket{i}$ can only be prepared up to some error. Next, measure this state in the computational basis, denoting the measurement count of the result $0^6 i$ as $k^+_i$ and the result $0^51 i$ as $k^-_i$.
    \item Define 
    \begin{align}
        a_i^+ &= \min\left(\sqrt{p_i}, \frac{k^+_i-k^-_i}{\sqrt{p_i}}\right) \\ 
        a_i^- &= \max\left(-\sqrt{p_i}, \frac{k^+_i-k^-_i}{\sqrt{p_i}}\right)
    \end{align}
    and let
    \begin{align}
        \tilde{a}_i =
        \begin{cases}
            0 & \text{if }\sqrt{p_i} \le \frac{2}{3\sqrt{2L}} \varepsilon \sqrt{1-\frac{\varepsilon^2}{4}} + \varepsilon_\mathrm{tsp} \\
            a_i^+ & \text{if }\tilde{a}_i \neq 0 \text{ and } k^+_i \ge k^-_i \\ 
            a_i^- & \text{if }\tilde{a}_i \neq 0 \text{ and } k^+_i < k^-_i \\ 
        \end{cases}.\label{eq:tomography_atilde_i}
        \end{align}
    Output the estimate $\ket{\tilde{\ve{v}}'} = \sum_{i=1}^L \tilde{a}_i \ket{i} / \sqrt{\sum_{i=1}^L \tilde{a}_i^2}$.
\end{enumerate}

\begin{proposition}\label{lem:tomography}
Suppose that $\lVert \tilde{\ve{v}} - \ve{v}\rVert \leq \varepsilon_{\mathrm{QLSP}}$ and that $\ve{v}$ is a real-valued vector. Let $\varepsilon$ and $\varepsilon_{\mathrm{tsp}}$ be constants that satisfy $\varepsilon + \sqrt{2L}\varepsilon_\mathrm{tsp} + \sqrt{2}\varepsilon_\mathrm{QLSP} \le 1/2$. Then the algorithm above outputs an estimate $\tilde{\ve{v}}'$ such that $\|\tilde{\ve{v}}' - \ve{v}\| < \varepsilon + 1.58\sqrt{L}\varepsilon_\mathrm{tsp} + 1.58\varepsilon_\mathrm{QLSP}$ with probability $1-\delta$.
\end{proposition}

We give the proof in \cref{app:tomography}.  The statement is used to bound the total error parameter $\xi$ by the quantity $\varepsilon + 1.58\sqrt{L}\varepsilon_\mathrm{tsp} + 1.58\varepsilon_\mathrm{QLSP}$. We note that a similar procedure in \cite[Sec.~4]{Kerenidis2020lpsdp} has already been proven to work, with somewhat worse success probability guarantees and worse constants. Ref.~\cite[Proposition 16]{apeldoorn22} shows a similar result for complex-valued states, but we use a sharper proof for input states close to real-valued. \cref{lem:tomography}, together with \cref{lem:adiabatic-discrete-filtered}, produces with high probability an $\bigo{\varepsilon}$ good estimate $\tilde{\ve{v}}'$ of $\ve{v}$ by using $\bigo{L \ln(L) / \varepsilon^2}$ many samples.\footnote{If our goal is to resolve the initial linear system $G\ve{u}=\ve{h}$, then the vector $\tilde{\ve{v}}'$ produced as in \Cref{sec:qipm-tomography} as an estimate for the normalized vector $\ve{v}=\ve{u}/\lVert \ve{u} \rVert$, gives an estimate for $\ve{u}$ via
\begin{align*}
\tilde{\ve{u}}:=\tilde{\ve{v}}'\cdot\frac{\|\ve{h}\|}{\|G\tilde{\ve{v}}'\|}, 
\end{align*}
for which we find 
\begin{align*}
\|\ve{u}-\tilde{\ve{u}}\|\leq\|\ve{v}-\tilde{\ve{v}}'\|\cdot\big(1+\kappa(G)\big)\cdot\frac{\|\ve{h}\|}{\|G\tilde{\ve{v}}'\|}.
\end{align*}
Notice that as a worst case guarantee, this picks up an additional factor $\kappa(G)$ in error scaling. However, for our purposes it will be sufficient to directly work with the normalized estimate $\tilde{\ve{v}}'$ for $\ve{v}$, the reason being that only the direction of the solution vector is important to us and not its exact normalization.} There are other methods in the literature that allow to perform pure state quantum tomography with comparable query complexities (e.g. \cite{odonnell16}), but we favor the above method because of its computational simplicity, and the fact that it does not require us to solve any potentially costly additional optimization problems. Very recently, the sample complexity has been improved to $\bigo{L \ln(L) / \varepsilon}$, which comes at the cost of more complicated quantum circuits and higher constant overheads \cite[Theorem 23]{apeldoorn22}. It would be interesting to work out the more involved finite complexity of this result, and we further comment on the potential impact of this in \cref{sec:discussion}.


\subsection{Asymptotic quantum complexity}\label{qipm-complexity}

Putting everything together, the steps of our QLSS for given real $L\times L$ matrix $G$ and real vector $\ve{h}$ of size $L$ are:

\begin{enumerate}
    \item Construct the circuits that implement the block-encoding unitaries $U_G$ and $U_{\ve{h}}$ up to error $\varepsilon_G$ and $\varepsilon_{\ve{h}}$ via quantum state preparation and QRAM, which involves a classical pre-processing cost scaling as $L^2 \text{polylog}(1/\varepsilon_{G,\ve{h}})$. The quantum resources required are described in \cref{tab:min_T_depth}. The $T$-gate depth (what we call time complexity) is 
    $\bigo{\log L}$ and the total $T$-gate count is $\bigo{L^2}$.
    \item Employ the QLSS unitary from \cref{lem:adiabatic-discrete-filtered} to approximately solve the corresponding QLSP, leading to the quantum state $\ket{\tilde{\ve{v}}}$. The query complexity to $U_G$, $U_{\ve{h}}$, their controlled versions, and their inverses, is $\bigo{\kappa_F(G)\log(1/\varepsilon)}$. The number of qubits needed is $\lceil\log L\rceil+5$.
    \item Repeat the previous step $\bigo{L \ln(L/\delta)\varepsilon^{-2}}$ many times to implement the pure state quantum tomography scheme from \cref{sec:qipm-tomography}, which also requires the use of an $\bigo{L}$ qubit QRAM structure, and one ancilla qubit. Tomography leads to the sought-after classical vector estimate $\tilde{\ve{v}}'$ with $\|\tilde{\ve{v}}'-\ve{v}\|\leq\varepsilon$.
\end{enumerate}

The QLSS can then be used for each iteration of an IPM SOCP solver, which involves forming and solving a linear system of equations, resulting in the QIPM SOCP solver. We provide the quantum circuits needed to implement the solver in the \cref{sec:qipm-circuits}. However, we emphasize that we have not yet considered the various practical aspects and difficulties of setting up an \emph{end-to-end} QIPM SOCP solver, which is discussed further in \cref{sec:implementation}.


\subsection{Quantum circuits}\label{sec:qipm-circuits}

\begin{figure*}[ht]
    \centering
\begin{displaymath}
\Qcircuit @C=2em @R=1em {
\lstick{a_3} & \gate{H} & \gate{U_{Q_{\ve{h}}}} & \qw & \qw & \qw & \gate{U_{Q_{\ve{h}}}} & \gate{H} & \qw \\
\lstick{a_4} & \qw & \ctrl{-1} \qwx[2] & \multigate{1}{CR^0(s)} & \qw & \multigate{1}{\mathrm{CR}^1(s)} & \ctrlo{-1} \qwx[2] & \gate{X} & \qw \\
\lstick{a_2} &  \qw & \qw & \ghost{\mathrm{CR}^0(s)} & \multigate{3}{V_G} & \ghost{CR^1(s)} & \qw & \qw & \qw \\
\lstick{a_1} & \qw & \multigate{1}{U_{Q_{\ve{h}}}} & \qw & \ghost{V_G} & \qw & \multigate{1}{U_{Q_{\ve{h}}}} & \qw & \qw \\
\lstick{L} & \qw & \ghost{U_{Q_{\ve{h}}}} & \qw & \ghost{V_G} & \qw & \ghost{U_{Q_{\ve{h}}}} & \qw & \qw \\
\lstick{\ell_G} & \qw & \qw & \qw & \ghost{V_A} & \qw & \qw & \qw & \qw 
}
\end{displaymath}
\caption{\label{fig:U(s)-circuit}Main component of the quantum circuit for \cref{lem:adiabatic-discrete}, described in \cite[appendix E]{costa2021optimal}, enacting the unitary $U[s]$ on registers $a_3a_4a_2a_1L\ell_G$ of the scaled Hamiltonian $c(s)\cdot \mathcal{H}[s]$, where $\mathcal{H}[s]=(1-f(s))\mathcal{H}_0+f(s)\mathcal{H}_1$, on registers $a_4a_1L$. The necessary quantum gates and functions are defined in \cref{eq:hamiltonian-H0,eq:hamiltonian-H1,eq:CR0,eq:CR1,eq:VG,eq:block-factors,eq:f(s)} except for sub-circuit $U_{Q_{\ve{h}}}$, which is depicted in \cref{fig:UQh}. The unitary $U[s]$ is then used in \cref{eq:walk} to define the overall quantum circuit $U$ for \cref{lem:adiabatic-discrete}.}


\begin{displaymath}
\Qcircuit @C=3.47mm @R=2.0mm { 
& \gate{H}
& \targ
& \gate{\!e^{-i\phi_{1} \sigma_z}\!}    
& \targ 
& \qw   
& \targ
& \gate{\!e^{-i\phi_2 \sigma_z}\!}  
& \targ             
& \qw & \,\cdots\, &        
& \targ
& \gate{\!e^{-i\phi_d \sigma_z}\!}
& \targ             
& \gate{H}
& \qw \\
\lstick{a_3} & \multigate{5}{U[1]} 
& \ctrlo{-1}    
& \qw
& \ctrlo{-1}                    
& \multigate{5}{\!U^\dagger[1]\!\!}                 
& \ctrlo{-1}    
& \qw
& \ctrlo{-1}
& \qw & \,\cdots\, &        
& \ctrlo{-1}    
& \qw
& \ctrlo{-1}    
& \qw   
& \qw \\
\lstick{a_4} & \ghost{U[1]} & \qw & \qw   & \qw & \ghost{\!U^{\dagger}[1]\!\!} & \qw & \qw  & \qw & \qw & \,\cdots\, & & \qw & \qw & \qw & \qw & \qw \\
\lstick{a_2} & \ghost{U[1]} & \ctrlo{-2} & \qw   & \ctrlo{-2} & \ghost{\!U^{\dagger}[1]\!\!} & \ctrlo{-2} & \qw & \ctrlo{-2}  & \qw & \,\cdots\, & & \ctrlo{-2} & \qw & \ctrlo{-2} & \qw & \qw \\   
\lstick{a_1} & \ghost{U[1]} & \qw & \qw  & \qw & \ghost{\!U^{\dagger}[1]\!\!} & \qw & \qw & \qw & \qw & \,\cdots\, & & \qw & \qw & \qw & \qw & \qw \\    
\lstick{L}  & \ghost{U[1]} & \qw & \qw  & \qw & \ghost{\!U^{\dagger}[1]\!\!} & \qw & \qw & \qw & \qw & \,\cdots\, & & \qw & \qw & \qw & \qw & \qw \\
\lstick{\ell_G} & \ghost{U[1]} & \ctrlo{-3} & \qw  & \ctrlo{-3} & \ghost{\!U^{\dagger}[1]\!\!} & \ctrlo{-3}  & \qw & \ctrlo{-3}  & \qw & \,\cdots\, & & \ctrlo{-3}  & \qw & \ctrlo{-3}  & \qw & \qw 
}
\end{displaymath} 
    \caption{\label{fig:QSVT}Quantum singular value transform (QSVT) circuit, described in Ref.~\cite{gilyen2019}, acting on the block-encoding $U[1]$ of $\mathcal{H}(1) = \mathcal{H}_1/\sqrt{2}$, as defined in \cref{eq:hamiltonian-H1}. The circuit features one additional ancilla qubit and depends on the classically pre-computed rotation angles $\{\phi_1,\cdots,\phi_d\}$.}

\begin{displaymath}
\Qcircuit @C=1.2em @R=0.8em {
\lstick{c}      & \qw      & \qw                    & \ctrl{2}               & \ctrl{3}           & \ctrl{2}               & \ctrl{2} & \qw                    & \qw      & \qw & \\
\lstick{a_3}    & \gate{H} & \gate{U_{Q_{\ve{h}}}}         & \qw                    & \qw                & \qw                    & \qw      & \gate{U_{Q_{\ve{h}}}}         & \gate{H} & \qw & \\
\lstick{a_4}    & \qw      & \ctrl{-1} \qwx[2]      & \multigate{1}{\mathrm{CR}^0(s)} & \qw                & \multigate{1}{\mathrm{CR}^1(s)} & \targ    & \ctrlo{-1} \qwx[2]     & \qw      & \qw & \\
\lstick{a_2}    & \qw      & \qw                    & \ghost{CR^0(s)}        & \multigate{3}{V_G} & \ghost{CR^1(s)}        & \qw      & \qw                    & \qw      & \qw & \\
\lstick{a_1}    & \qw      & \multigate{1}{U_{Q_{\ve{h}}}} & \qw                    & \ghost{V_G}        & \qw                    & \qw      & \multigate{1}{U_{Q_{\ve{h}}}} & \qw      & \qw & \\
\lstick{L}      & \qw      & \ghost{U_{Q_{\ve{h}}}}        & \qw                    & \ghost{V_G}        & \qw                    & \qw      & \ghost{U_{Q_{\ve{h}}}}        & \qw      & \qw & \\
\lstick{\ell_G} & \qw      & \qw                    & \qw                    & \ghost{V_G}        & \qw                    & \qw      & \qw                    & \qw      & \qw & \\
}
\end{displaymath}
\caption{\label{fig:c-U(s)-circuit}Controlled version of the quantum circuit in \cref{fig:U(s)-circuit}, controlled on qubit $c$. Note that not all gates need to be controlled on $c$, as their inverses follows in the circuit.}
\end{figure*}

The following are the quantum circuits needed for the QLSS of \cref{lem:adiabatic-discrete}. The QLSS requires applying a unitary $U[s]$ for many different values of $s$, where $U[s]$ is a block-encoding of a certain Hamiltonian related to $G$ and $\ve{h}$, as specified below. The unitary acts on $4 + \ell + \ell_G$ total qubits, where the final $\ell_G$ qubits are ancillas associated with $U_G$. The four single-qubit registers are referred to with labels $a_1$, $a_2$, $a_3$, $a_4$, the $\ell$-qubit register with label $L$, and the $\ell_G$-qubit register with label $\ell_G$. These labels are used as subscripts on bras, kets, and operators to clarify the register to which they apply. The circuit for $U[s]$ is depicted in \cref{fig:U(s)-circuit}, and is described in \cite[appendix E]{costa2021optimal}.
Specifically, the unitary $U[s]$ is a block-encoding of the $(2+\ell)$-qubit Hamiltonian $c(s)\cdot  \mathcal{H}[s]:=(1-f(s))\mathcal{H}_0+f(s)\mathcal{H}_1$ on registers $a_4a_1L$, where $c(s)$ is a normalization factor (defined later in \cref{eq:cs(s)}),
\begin{align}
\mathcal{H}_0:={}&\begin{pmatrix} 0 & 0 & I_L -\ket{\ve{h}}\bra{\ve{h}}_L & 0 \\ 0 & 0 & 0 & -I_L \\ I_L -\ket{\ve{h}}\bra{\ve{h}}_L & 0 & 0 & 0 \\ 0 & -I_L & 0 & 0\end{pmatrix}
\label{eq:hamiltonian-H0}
\end{align}
and
\begin{align}
\mathcal{H}_1:={}&\begin{pmatrix} 0 & 0 & 0 & G \\ 0 & 0 & G^\dagger(I_L-\ket{\ve{h}}\bra{\ve{h}}_L) & 0 \\ 0 & (I_L-\ket{\ve{h}}\bra{\ve{h}}_L)G & 0 & 0 \\ G^\dagger & 0 & 0 & 0\end{pmatrix}\,,
\label{eq:hamiltonian-H1}\end{align}
and where $I_L$ denotes the identity operation on subsystem $L$, and the four rows and columns correspond to the sectors with qubits $a_4,a_1$ set to $(0,0),(0,1),(1,0),(1,1)$. 
\Cref{fig:U(s)-circuit} features the expressions
\begin{align}
\mathrm{CR}^0(s):={}&\ket{0}\bra{0}_{a_4}\otimes R(s)_{a_2}+\ket{1}\bra{1}_{a_4}\otimes H_{a_2}\label{eq:CR0}\\
\mathrm{CR}^1(s):={}&\ket{1}\bra{1}_{a_4}\otimes R(s)_{a_2}+\ket{0}\bra{0}_{a_4}\otimes H_{a_2}\label{eq:CR1}\\
V_G:={}&\ket{0}\bra{0}_{a_2}\otimes Z_{a_1}\otimes I_{L\ell_G} \nonumber \\
& \qquad +\; \ket{1}\bra{1}_{a_2}\otimes\begin{pmatrix} 0 & U_G \\ U_G^\dagger & 0 \end{pmatrix}_{a_1L\ell_G}\,,\label{eq:VG}
\end{align}
where $H$ denotes the single-qubit Hadamard gate, and $R(s)$ is given by
\begin{align}\label{eq:block-factors}
R(s):={}&\frac{1}{\sqrt{(1-f(s))^2 + f(s)^2}}\begin{pmatrix} 1-f(s) & f(s) \\ f(s) & -(1-f(s))\end{pmatrix}\\
f(s):={}&\frac{\kappa_F(G)}{\kappa_F(G)-1}\cdot\left(1-\left(1+s\left(\sqrt{\kappa_F(G)}-1\right)\right)^{-2}\right)\,. \label{eq:f(s)}
\end{align}
The normalization factor of $R(s)$ above combines with a factor of $1/\sqrt{2}$ introduced by the Hadamard gate to give an overall normalization factor for $\mathcal{H}(s)$ of \begin{equation}c(s) = \left(2((1-f(s))^2+f(s)^2)\right)^{-1/2} \in[2^{-1/2},1]
\label{eq:cs(s)}
\end{equation}
and scheduling function $f(s)$ with $f(0)=0$ and $f(1)=1$. 
Note that we have the self-inverse property $U[s]^2=1\;\forall s\in[0,1]$, as demonstrated in \cite[Appendix E]{costa2021optimal}. The overall quantum circuit $U$ for the quantum algorithm of \cref{lem:adiabatic-discrete} is then given as (cf.~\cite{low2019hamiltonian})
\begin{align}\label{eq:walk}
U:=\prod_{j=1}^{Q} P[1-j/Q]
\end{align}
with the walk operator $$P[s]:=WU[s],$$ where $W$ is the operator that acts as identity on registers $a_4a_1L$ (which host the Hamiltonian $\mathcal{H}[s]$) while performing the reflection $(2\ket{0}\bra{0}_{a_2a_3\ell_G}-I_{a_2a_3\ell_G})$ on the remaining qubits.
The unitary $U$ makes $Q$ controlled queries to each of $U_G$ and $U_G^\dagger$, and $2Q$ queries to each of $U_{\ve{h}}$ and $U_{\ve{h}}^\dagger$, and it has constant quantum gate overhead.

Next, we give the remaining QSVT eigenstate filtering quantum circuit for the refined quantum linear system solver of \cref{lem:adiabatic-discrete-filtered}. We are interested in the null space of $c(1)\cdot \mathcal{H}[1]$, which has ground-state energy equal to zero and spectral gap at least $c(1)\kappa^{-1}_F(G) = (\sqrt{2}\kappa_F)^{-1}$. As such, we employ the Chebyshev minimax polynomial
\begin{equation}
R_l(x,\kappa^{-1}_F(G)):=\frac{T_l\left(-1+2\frac{x^2-\kappa^{-2}_F(G)/2}{1-\kappa^{-2}_F(G)/2}\right)}{T_l\left(-1+2\frac{-\kappa^{-2}_F(G)/2}{1-\kappa^{-2}_F(G)/2}\right)},
\end{equation}
where $T_l(\cdot)$ is $l$-th Chebyshev polynomial of the first kind, as part of the corresponding QSVT quantum circuit. From \cite[Lemma 2]{Lin20}, $R_l$ has even degree $d$ equal to 
\begin{equation}\label{eq:eigenstate_filtering_degree}
d:=2l=2\left\lceil\kappa_F(G)\ln(2/\varepsilon_\mathrm{qsp})\right\rceil\quad\text{for some $\varepsilon_\mathrm{qsp}\in(0,1]$}
\end{equation}
where $\varepsilon_\mathrm{qsp}$ is the precision to which $R_l$ approximates the optimal filter operator. The QSP subscript stands for ``quantum signal processing.''

The circuit for the eigenstate filtering step is depicted in \cref{fig:QSVT}. To implement it, one has to classically pre-compute the corresponding QSP angles $\{\phi_1,\cdots,\phi_d\}$, which is best done by the methods of \cite{Dong21} (see also \cite{gilyen2019} and \cite{Haah2019product}). The query complexity to the block encoding $U[1]$ is given by $d$, the additional gate overhead is as in \cref{fig:QSVT}, and the total number of qubits is $1+4+\ell$. Finally, using the overall quantum circuit $U$ from \cref{lem:adiabatic-discrete} with constant approximation parameter $\varepsilon_1=\sqrt{2-\sqrt{2}}$ therein (to produce an input state to the quantum circuit of \cref{fig:QSVT}), gives the overall quantum circuit of the QLSS from \cref{lem:adiabatic-discrete-filtered}, which then solves the QLSP to error $\varepsilon_2 = \varepsilon_\mathrm{qsp}$.

The tomography routine also requires the ability to perform controlled versions of the above circuits as described in \cref{eq:controlled-U}, and illustrated in \cref{fig:c-U(s)-circuit} (which replaces \cref{fig:U(s)-circuit}). The controlled circuits can be accomplished by rather simple modifications to the circuits in \cref{fig:U(s)-circuit,fig:QSVT} as follows.

Any QSVT circuit can be made controlled by simply controlling the application of the $z$ rotation gates, since the rest of the circuit contains only symmetric applications of unitary gates and their inverses. Thus, we can create a controlled version of \cref{fig:QSVT} by simply performing controlled-$\sigma_z$ rotations, which requires two CNOT gates and an extra single qubit $\sigma_z$ rotation gate. 

Controlling the linear system portion is not enough to implement \cref{eq:controlled-U}. One must also follow this with a controlled state-preparation routine, controlled on the value of the qubit $c$ being in the $\ket{1}$ state. The full resource analysis for controlled state-preparation was reported in Ref.~\cite{clader2022quantum}, and we refer the reader there for further details. We report the resource counts here in \cref{tab:sp_min_T_depth}.



\section[Implementation and PO resource estimate]{IPM implementation and resource estimates for PO}\label{sec:implementation}

The previous section reviewed the ingredients needed to implement the QIPM, namely, QLSS, block-encoding, and tomography. Here, we combine those ingredients to describe how the QIPM is actually implemented, making several observations that go beyond prior literature. We also perform a full resource analysis of the entire protocol and report resources needed to run the algorithm.

\subsection{Main IPM loop and full pseudo-code}\label{sec:implementation-pseudo}

A QIPM is formed from an IPM by performing the step of solving a linear system with a quantum algorithm; the rest of the steps are classical. 
In Algorithm~\ref{algo:IPM}, we present pseudocode for the interior point method where the single
quantum subroutine---approximately solving a linear system---appears in blue text. The input to Algorithm~\ref{algo:IPM} is an SOCP instance with $N$ variables, $K$ linear constraints, and $r$ second-order cone constraints, along with a tolerance parameter $\epsilon$. Here we note that $K = \bigo{N}$ in the case of the formulation of the PO problem we simulate in \cref{sec:numerics}. The output of the QIPM is a vector $\ve{x}$ that is $\bigo{\epsilon}$ close to feasible, and $\bigo{\epsilon}$ close to optimal. 

The structure of the QIPM is in essence the same as that proposed by Ref.~\cite{Kerenidis2021quantumalgorithms}, but we 
give a more complete specification of the algorithm and make several new observations:
\begin{itemize}
    \item \textbf{Classical costs}: The IPM requires $\bigo{\sqrt{r}\log(1/\epsilon)}$ iterations. In the classical case, when solving the PO problem via SOCP with an IPM, the cost of an iteration is dominated by the time needed to solve a linear system of size $L \times L$, which is $\bigo{N^3}$ if done via Gaussian elimination, since $L\sim \bigo{N}$ in the PO problem. In the quantum case, this step is performed quantumly. However, even in the quantum case, some classical costs are incurred: one must \emph{classically} compute the left-hand and right-hand sides of the Newton system in {\cref{eq:Newton-system-1}} and {\cref{eq:Newton-system-2}} to be able to load this classical data into quantum circuits that perform the QLSS and tomography required to gain a classical estimate of the solution to the linear system. In particular, constructing the linear system requires classical matrix-vector multiplication to compute the residuals on the right-hand-side of the Newton system in \cref{eq:Newton-system-1}. If the SOCP constraint matrix $A$ is $\bigo{N} \times N$ and the number of cones $r = \bigo{N}$, then this classical matrix-vector multiplication takes $\bigo{N^2}$ time in each of the $\bigo{\sqrt{N}}$ iterations. Thus, the QIPM requires at least $\bigo{N^{2.5}}$ classical time. Additionally, in our resource counts we use the minimal depth block-encoding circuits from Ref.~\cite{clader2022quantum}, which require $N^2\polylog(1/\varepsilon)$ classical time per iteration (although this can be parallelized) to compute angles and corresponding gate sequences to precision $\varepsilon$. These classical costs limit the maximum possible speedup of the QIPM over the classical IPM, but if the quantum subroutine is sufficiently fast that classical matrix-vector multiplication and angle computation is the bottleneck step, then this is a good signal for the utility of the QIPM. 
    \item \textbf{Preconditioning}: Since the runtime of the QLSS depends on the condition number of the matrix $G$ that appears in the linear system $G\ve{u} = \ve{h}$, it is worth examining preconditioning techniques \cite{clader2013preconditioned} for reducing the condition number. In the implementation we propose, we perform a very simple form of preconditioning. Let $D$ be a diagonal matrix where entry $D_{ii}$ is equal to the norm of row $i$ of the matrix $G$. Instead of solving the linear system $G\ve{u} = \ve{h}$, we solve the equivalent system $(D^{-1}G)\ve{u} = D^{-1}\ve{h}$. Note that $D^{-1}G$ and $D^{-1}\ve{h}$ can each be classically computed in $\bigo{N^2}$ time, roughly equal to the time required to compute $\ve{h}$ in the first place (see previous bullet), so this step is unlikely to be a bottleneck in the algorithm. In our numerical experiments, we observe that the condition number of $D^{-1}G$ is typically more than an order of magnitude smaller than $G$, and sometimes several orders of magnitude (see \cref{fig:condition_number_scaling} in \cref{sec:numerics}). 
    \item \textbf{Norm of linear system and step length}: As discussed in \cref{sec:qipm-linear-solver}, QLSSs produce a normalized state $\ket{\ve{u}}$, where $\ve{u}$ is the solution to $G\ve{u} = \ve{h}$, and quantum state tomography on $\ket{\ve{u}}$ can only reveal the direction of the solution $\ve{u}$ and not its norm. The norm can be estimated separately with a comparable amount of resources, but we observe that in the context of QIPMs, \emph{it is not necessary to learn the norm of the solution}. If the direction of the solution is known, the amount by which to update the vector in that direction can be determined classically in $\bigo{N}$ time as follows. If $(\ve{\Delta x}; \ve{\Delta y}; \Delta \tau; \Delta \theta; \ve{\Delta s}; \Delta \varkappa)$ is the normalized solution to the Newton linear system in \cref{eq:Newton-system-1,eq:Newton-system-2}, then the amount to step in that direction is equal to
    \begin{equation}
    \frac{\mu(\ve{x},\tau,\ve{s},\varkappa)(1-\sigma)(r+1)}{-(\ve{\Delta x})^\intercal \ve{s} - (\ve{\Delta s})^\intercal \ve{x}-(\Delta \varkappa) \tau - (\Delta \tau)\varkappa}\,.
    \end{equation}
    This expression is chosen such that the duality gap of the new point is exactly a factor of $\sigma$ smaller than the old point, up to deviations that are second order in the step length. Note that if the old point is feasible and the solution to the linear system is exact, the second and higher order contributions vanish anyway.
    \item \textbf{Adaptive tomographic precision and neighborhood detection}: In Ref.~\cite{Kerenidis2021quantumalgorithms}, the choice of tomography precision parameter $\xi$ was determined by a formula that aimed to guarantee staying within the neighborhood of the central path under a worst-case outcome. We observe that, since determining whether a point is within the neighborhood of the central path can be done in classical $\bigo{N}$ time (see \cref{sec:neighborhood_central_path}), the precision parameter can instead be determined adaptively for optimal results: start with $\xi = 1/2$, solve the linear system to precision $\xi$ and check if the resulting point is within the neighborhood of the central path. If yes, continue to the next iteration; if no, repeat the tomography with $\xi \gets \xi/2$. Since the complexity of tomography is $\bigo{1/\xi^2}$, the cost of this adaptive scheme is proportional to a geometric series $4+16+64+\ldots + \bigo{1/\xi^2}$ of which the final term will make up most of the cost (accordingly, for simplicity, in our resource calculation we only account for the final term). This cost could be much lower than the theoretical value if the typical errors are not as adverse for the IPM as a worst-case error of the same size. 
\end{itemize}

\SetKwInput{Input}{Input}
\SetKwInput{Output}{Output}
\SetKwFunction{ApprSolve}{ApprSolve}
\SetStartEndCondition{ }{}{}%
\SetKwProg{Fn}{def}{\string:}{}
\SetKwFunction{Range}{range}
\SetKw{KwTo}{in}\SetKwFor{For}{for}{\string:}{}%
\SetKwIF{If}{ElseIf}{Else}{if}{:}{elif}{else:}{}%
\SetKwFor{While}{while}{:}{fintq}%
\newcommand{\forcond}{$i$ \KwTo\Range{$n$}}
\AlgoDontDisplayBlockMarkers\SetAlgoNoEnd\SetAlgoNoLine%

\begin{algorithm*}
\caption{Quantum Interior Point Method \label{algo:IPM}}
\DontPrintSemicolon
\SetAlgoLined
\Input{SOCP instance $(A,\ve{b}, \ve{c})$, list of cone sizes $(N_1,\ldots,N_r)$ and tolerance $\epsilon$} 
\Output{Vector $\ve{x}$  that optimizes objective function (\cref{eq:SOCP}) to precision $\epsilon$} 
 \tcc{For portfolio optimization, $A$, $\ve{b}$, $\ve{c}$ are given in \cref{eq:SOCP-PO}. First $n$ entries of $\ve{x}$ give optimal stock weights.}
\BlankLine
    $(\ve{x}; \ve{y}; \tau; \theta; \ve{s};\varkappa) \gets (\ve{e};\ve{0}; 1;1;\ve{e};1)$ \tcc*[f]{initialize on central path} \;
    $\mu \gets 1$, \,
    $\sigma \gets 1-\frac{1}{20\sqrt{2}}\frac{1}{\sqrt{r}}$, \,
    $\gamma \gets 1/10$ \tcc*[f]{set parameters}\;
    \While(\tcc*[f]{Follow central path until duality gap less than $\epsilon$}){$\mu \geq \epsilon$}
    {
        $G \gets \begin{pmatrix}
            0                       & A^\intercal            & -\ve{c} &  \ve{\bar{c}}  & I      & 0 \\
            -A                      & 0                      & \ve{b}  & -\ve{\bar{b}}  & 0      & 0 \\
            \ve{c}^\intercal        & -\ve{b}^\intercal      & 0       & -\bar{z}       & \ve{0} & 1 \\
            -\ve{\bar{c}}^\intercal & \ve{\bar{b}}^\intercal & \bar{z} &  0             & \ve{0} & 0 \\
            S                       & 0                      & 0       & 0              & X      & 0 \\
            0                       & 0                      & \varkappa  & 0           & 0      & \tau \\
        \end{pmatrix}$ \tcc*[f]{from \cref{eq:Newton-system-1,eq:Newton-system-2}}\;
        $\ve{h} \gets \begin{pmatrix}
        -A^\intercal \ve{y} + \ve{c}\tau - \ve{\bar{c}}\theta -\ve{s} \\
        A \ve{x} - \ve{b}\tau + \ve{\bar{b}} \theta \\
        -\ve{c}^\intercal \ve{x} + \ve{b}^\intercal \ve{y} + \bar{z}\theta \\
        \ve{\bar{c}}^\intercal \ve{x} -\ve{\bar{b}}^\intercal \ve{y} - \bar{z}\tau \\
        \sigma \mu \ve{e} - \tilde{X}\tilde{S}\ve{e} \\
        \sigma \mu - \varkappa \tau
        \end{pmatrix}$ \tcc*[f]{mat.-vec. mult. performed classically}\;
        \For(\tcc*[f]{preconditioning via row normalization}){$j=1,\ldots,L$}
        {
            $g \gets \sqrt{\sum_{k}|G_{jk}|^2}$ \tcc*[f]{norm of $j$th row of $G$}\;
            $h_j \gets h_j/g$\;
            \For{$k = 1,\ldots,L$}
                {
                $G_{jk} \gets G_{jk}/g$
                }
        }
        Classically compute $L^2$ angles and gate decompositions necessary to perform block-encoding of $G$ and state-preparation of $\ket{\ve{h}}$ (see Ref.~\cite{clader2022quantum})\;
        $\xi \gets 1$\;
        \Repeat(\tcc*[f]{try smaller and smaller $\xi$ until central path is found}){$(\ve{x}'; \ve{y}'; \tau'; \theta'; \ve{s}';\varkappa') \in \mathcal{N}(\gamma)$}
        {
            $\xi \gets \xi/2$\;
            $(\ve{\Delta x}; \ve{\Delta y}; \Delta \tau; \Delta \theta; \ve{\Delta s}; \Delta \varkappa) \gets \textcolor{cyan}{\ApprSolve}(G,\ve{h},\xi)$\;
            $\text{(step length)} \gets \frac{\mu(\sigma-1)(r+1)}{(\ve{\Delta x})^\intercal \ve{s} + (\ve{\Delta s})^\intercal \ve{x}+(\Delta \varkappa) \tau + (\Delta \tau)\varkappa}$\;
            $(\ve{x}'; \ve{y}'; \tau'; \theta'; \ve{s}';\varkappa') \gets (\ve{x}; \ve{y}; \tau;\theta; \ve{s};\varkappa) + \text{(step length)}\cdot (\ve{\Delta x};\ve{\Delta y}; \Delta \tau; \Delta \theta; \ve{\Delta s}; \Delta \varkappa) $\;
        }
        $(\ve{x}; \ve{y}; \tau; \theta; \ve{s};\varkappa) \gets (\ve{x}'; \ve{y}'; \tau'; \theta'; \ve{s}';\varkappa') $\;
        $\mu \gets \sigma \mu$\;
        }

    \Return{$\ve{x}/\tau$}
\BlankLine
\BlankLine
\BlankLine
\Fn{\textcolor{cyan}{\ApprSolve}$(G,\ve{h},\xi)$}
{
$L \gets 2N+K+3$\;
$\delta \gets 0.1$\;
$\varepsilon \gets 0.9\xi$\;
$k \gets 57.5L\ln(6L/\delta)/(\varepsilon^2(1-\varepsilon^2/4))$\;
Run tomography as described in \cref{sec:qipm-tomography} using $k$ applications and $k$ controlled-applications of the QLSS algorithm on the system $(G,\ve{h})$\;
\Return{
Vector $\ve{\tilde{v}}'$ for which $\lVert \ve{\tilde{v}}'\rVert =1$ and $\lVert \ve{\tilde{v}}'-\ve{v} \rVert\leq \xi$ with probability at least $1-\delta$, where $\ve{v} \propto G^{-1}\ve{h}$
}
}

\end{algorithm*}

The pseudocode in Algorithm~\ref{algo:IPM} illustrates the infeasible version of the algorithm (II-QIPM from \cref{tab:QIPM_choices}). To implement the feasible versions (IF-QIPM and IF-QIPM-QR), minor modifications are made to reflect the process described in \cref{sec:socp}.


\subsection{End-to-end quantum resource estimates}\label{sec:implementation-estimate}

The QIPM described in the pseudocode takes $20\sqrt{2}\sqrt{r}\ln(\epsilon^{-1})$ iterations to reduce the duality gap to $\epsilon$, where $r$ is the number of second-order cone constraints. In the case of the portfolio optimization problem we study, $r=3n+1$, where $n$ is the number of stocks in the portfolio. Choosing the constant pre-factor to be $20\sqrt{2}$ allows us to utilize theoretical guarantees of convergence (modulo the issue of infeasibility discussed in \cref{sec:solve_newton_system}); however, it would not be surprising if additional optimization of the parameters or heuristic changes to the implementation of the algorithm (e.g.~adaptive step size during each iteration) would lead to constant-factor speedups in the number of iterations. Since the number of iterations would be the same for both the quantum and classical IPM, these sorts of improvements would not impact the performance of the QIPM relative to its classical counterpart. 

\subsubsection{Quantum circuit compilation and resource estimate for quantum circuits appearing within QIPM}

The QIPM consists of repeatedly performing a quantum circuit associated with the QLSS and measuring in the computational basis. Here we account for all the costs of each of these individual quantum circuits. There are two kinds of circuits that are needed: first, the circuit that creates the output of the QLSS subroutine, given by the state in \cref{eq:qls-state}, and second, the circuit that creates the state needed to determine the signs of the amplitudes during the tomography subroutine corresponding to a controlled-QLSS subroutine, given in \cref{eq:controlled-U}.

To simplify the analysis, we first compile the circuits from the previous section into a primitive gateset that consists of Toffoli gates (and multi-controlled versions of them), rotation gates, block-encoding unitaries, state-preparation and controlled state-preparation unitaries. This compilation allows us to combine our previous in-depth resource analysis for these primitive routines \cite{clader2022quantum} with the additional circuits shown here.

From left to right in the $U[s]$ circuit shown in \cref{fig:U(s)-circuit}, we show the circuits for $U_{Q_{\ve{h}}}$, $CR^{0}(s)$ (and equivalently $CR^{1}(s)$), and $V_G$ in \cref{fig:UQh,fig:CR0,fig:VG}, respectively. In addition to these circuits, we must also perform controlled versions of them within the tomography routine to estimate the sign of the amplitudes. The controlled-$U[s]$ gate is given in \cref{fig:c-U(s)-circuit}. The implementation of the controlled versions of $CR^{0}(s)$ (and equivalently $CR^{1}(s)$), and $V_G$ are also depicted in \cref{fig:CR0,fig:VG}, respectively.

With these decompositions in place, we now report in \cref{tab:qls} the resources required to perform each of the two kinds of quantum circuits involved in the QIPM (which are each performed many times over the course of the whole algorithm). The resource quantities are reported in terms of the number of calls $Q$ to the block-encoding (which scales linearly with the condition number), as well as the controlled-block-encoding and state-preparation resources given previously in \cref{tab:min_T_depth,tab:sp_min_T_depth}. The expressions also depend on various error parameters which must be specified to obtain a concrete numerical value. In \cref{sec:numerics}, after observing empirical scaling of certain algorithmic parameters, we make choices for all error parameters and arrive at a concrete number for a specific problem size.

\begin{figure*}[htp!]
    \centering
\begin{minipage}[b]{\textwidth}
\begin{displaymath}
\Qcircuit @C=0.8em @R=1.2em {
\lstick{a_4} & \ctrl{1}               & \qw &   & & \qw                & \ctrl{1}         & \qw        & \qw &   & & \qw                        & \qw      & \ctrl{1}  & \qw      & \qw                & \qw & \\
\lstick{a_3} & \multigate{2}{U_{Q_{\ve{h}}}} & \qw &   & & \qw                & \ctrl{1}         & \qw        & \qw &   & & \qw                        & \qw      & \ctrl{1}  & \qw      & \qw                & \qw & \\
\lstick{a_1} & \ghost{U_{Q_{\ve{h}}}}       & \qw & = & & \qw                & \multigate{1}{W'} & \qw        & \qw & = & & \qw                        & \qw      & \ctrl{1}  & \qw      & \qw                & \qw & \\
\lstick{L}   & \ghost{U_{Q_{\ve{h}}}}        & \qw &   & & \gate{U_{\ve{h}}^\dagger} & \ghost{W'}        & \gate{U_{\ve{h}}} & \qw &   & & \gate{U_{\ve{h}}^\dagger} & \gate{X} & \ctrl{-1} & \gate{X} & \gate{U_{\ve{h}}} & \qw & \\
}
\end{displaymath}
\caption{\label{fig:UQh}Decomposition of the $U_{Q_{\ve{h}}}$ gate (shown, e.g., in \cref{fig:U(s)-circuit}) into a state-preparation unitary $U_{\ve{h}}$ and multi-controlled-Toffoli gates. The reflection operator $W$ is given by $W':=I_{a_1L}-2\ket{1}\bra{1}_{a_1}\otimes\ket{0}\bra{0}_L$. Not pictured are additional ancillas that begin and end in $\ket{0}$ and are utilized to implement the unitary $U_{\ve{h}}$ in shallower depth.}
\end{minipage}

\begin{minipage}[b]{\textwidth}
\begin{displaymath}
\Qcircuit @C=0.8em @R=1.1em {
\lstick{a_4}    & \multigate{1}{CR^0(s)} & \qw && \raisebox{-0.9cm}{=} &&& \qw                   & \ctrlo{1} & \qw                  & \qw                & \ctrl{1} & \qw               & \qw & \\
\lstick{a_2}    & \ghost{CR^0(s)}        & \qw &&                      &&
& \gate{R_y(-\theta/2)} & \targ     & \gate{R_y(\theta/2)} & \gate{R_y(-\pi/4)} & \targ    & \gate{R_y(\pi/4)} & \qw & \\
}
\end{displaymath}
\vspace{8 pt}
\begin{displaymath}
\Qcircuit @C=0.8em @R=1.1em {
\lstick{c}      & \ctrl{1}               & \qw & &                      & & & \qw                   & \ctrl{1}  & \qw                  & \qw                & \ctrl{1} & \qw               & \qw & \\
\lstick{a_4}    & \multigate{1}{\mathrm{CR}^0(s)} & \qw & & \raisebox{-0.5cm}{=} & & & \qw                   & \ctrlo{1} & \qw                  & \qw                & \ctrl{1} & \qw               & \qw & \\
\lstick{a_2}    & \ghost{\mathrm{CR}^0(s)}        & \qw & &                      & & & \gate{R_y(-\theta/2)} & \targ     & \gate{R_y(\theta/2)} & \gate{R_y(-\pi/4)} & \targ    & \gate{R_y(\pi/4)} & \qw & \\
}
\end{displaymath}
\caption{\label{fig:CR0}Decomposition of the $\mathrm{CR}^{0}(s)$ gate (top) and controlled-$\mathrm{CR}^{0}(s)$ gate (bottom), as defined in \cref{eq:CR0}, into single qubit rotation gates and CNOTs (top) or Toffolis (bottom). The gate $R_y(\phi)$ is defined to map $\ket{0} \mapsto \cos(\phi/2)\ket{0} + \sin(\phi/2)\ket{1}$ and $\ket{1}\mapsto -\sin(\phi/2)\ket{0} + \cos(\phi/2)\ket{1}$. The rotation angle $\theta = 2\arctan(\frac{1-f(s)}{f(s)})$, where $f(s)$ given in \cref{eq:f(s)}. The $\mathrm{CR}^{1}(s)$ gate is identical but with the control bit sign flipped. Note that the $R_y(\pm \pi/4)$ gates are Clifford conjugate to a single $T$ or $T^{\dagger}$ gate.}
\end{minipage}

\begin{minipage}[b]{\textwidth}
\begin{displaymath}
\Qcircuit @C=0.8em @R=1.1em {
\lstick{a_2}    & \multigate{4}{V_G} & \qw & &   & &         & \qw & \ctrlo{1} & \ctrl{1}  & \qw                        & \ctrl{1}  & \qw      & \ctrl{1}  & \qw                & \ctrl{1}  & \qw & \\
\lstick{a_1}    & \ghost{V_G}        & \qw & &   & &         & \qw & \gate{Z}  & \ctrlo{1} & \qw                        & \ctrlo{1} & \gate{X} & \ctrlo{1} & \qw                & \ctrlo{1} & \qw & \\
                &                    &     & & = & & \ket{0} &     & \qw       & \targ     & \ctrl{1}                   & \targ     & \qw      & \targ     & \ctrl{1}           & \targ     & \qw   & \\
\lstick{L}      & \ghost{V_G}        & \qw & &   & &         & \qw & \qw       & \qw       & \multigate{1}{U_G^\dagger} & \qw       & \qw      & \qw       & \multigate{1}{U_G} & \qw       & \qw & \\
\lstick{\ell_G} & \ghost{V_G}        & \qw & &   & &         & \qw & \qw       & \qw       & \ghost{U_G^\dagger}        & \qw       & \qw      & \qw       & \ghost{U_G}        & \qw       & \qw & \\
}
\end{displaymath}
\vspace{8 pt}
\begin{displaymath}
\Qcircuit @C=0.8em @R=1.1em {
\lstick{c}      & \ctrl{1}           & \qw & &   & &         & \qw & \ctrl{1}  & \qw       & \qw                        & \qw       & \ctrl{2} & \qw       & \qw                & \qw       & \qw & \\
\lstick{a_2}    & \multigate{4}{V_G} & \qw & &   & &         & \qw & \ctrlo{1} & \ctrl{1}  & \qw                        & \ctrl{1}  & \qw      & \ctrl{1}  & \qw                & \ctrl{1}  & \qw & \\
\lstick{a_1}    & \ghost{V_G}        & \qw & &   & &         & \qw & \gate{Z}  & \ctrlo{1} & \qw                        & \ctrlo{1} & \targ    & \ctrlo{1} & \qw                & \ctrlo{1} & \qw & \\
                &                    &     & & = & & \ket{0} &     & \qw       & \targ     & \ctrl{1}                   & \targ     & \qw      & \targ     & \ctrl{1}           & \targ     & \qw & \\
\lstick{L}      & \ghost{V_G}        & \qw & &   & &         & \qw & \qw       & \qw       & \multigate{1}{U_G^\dagger} & \qw       & \qw      & \qw       & \multigate{1}{U_G} & \qw       & \qw & \\
\lstick{\ell_G} & \ghost{V_G}        & \qw & &   & &         & \qw & \qw       & \qw       & \ghost{U_G^\dagger}        & \qw       & \qw      & \qw       & \ghost{U_G}        & \qw       & \qw & \\
}   
\end{displaymath}
\caption{\label{fig:VG}Decomposition of the $V_G$ unitary (top) and controlled-$V_G$ unitary (bottom), as defined in \cref{eq:VG}, into calls to a standard block-encoding unitary $U_G$ \cite{clader2022quantum} and other elementary gates, using a single ancilla qubit initialized to the $\ket{0}$ state. Not pictured are additional ancillas that begin and end in $\ket{0}$ and are utilized to implement the unitary $U_{G}$ in shallower depth.}
\end{minipage}

\end{figure*}


\begin{table*}[ht]
\caption{\label{tab:qls}Quantum resources required to create the state output by the quantum linear system solver, given in \cref{eq:qls-state} (QLSS, left) or the state needed to compute the signs during the tomography subroutine, given in \cref{eq:controlled-U} (Controlled QLSS, right) for a square linear system of size $L = 2^\ell$.  Note that these resource quantities do not yet account for the $k$ classical repetitions needed in order to perform tomography on the output state. The parameters $Q$ and $d$ each scale linearly with the condition number of the linear system, as defined in \cref{lem:adiabatic-discrete-filtered}. The symbols $N_{Qcbe}$, $T_{Dcbe}$, and $T_{Ccbe}$ denote the number of logical qubits, the $T$-depth, and the $T$-count, respectively, for performing a \emph{controlled}-block-encoding, as reported in \cref{tab:min_T_depth}. The symbols $T_{Dsp}$ and $T_{Csp}$ are analogous quantities for state-preparation, as reported in \cref{tab:sp_min_T_depth}. The parameters $\varepsilon_\mathrm{ar}$, $\varepsilon_\mathrm{tsp}$ and $\varepsilon_z \in (0,1]$ are error parameters corresponding to the gate synthesis precision required for the $\mathrm{CR}^0(s)$ and $\mathrm{CR}^1(s)$ rotations, controlled state-preparation step required by tomography, and the QSVT phases, respectively. }
\begin{ruledtabular}
\renewcommand{\arraystretch}{1.2}
\begin{tabular}{L{2cm}C{7cm}C{7cm}}
\textbf{Resource} & \textbf{QLSS} & \textbf{Controlled QLSS}\\
\hline
\textbf{\# Qubits}  & $N_{Qcbe} + 5$                  & $N_{Qcbe} + 6$ \\ 
\\
\textbf{$T$-depth}  & $\begin{matrix*}[l] 12Q\log_2(1/\varepsilon_\mathrm{ar}) + 2(Q+d)T_{Dcbe} + 4(Q+d)T_{Dsp} \\ \quad + Q(24\ell + 31) + 3d\log_2(1/\varepsilon_z) + d(32\ell-2)
\end{matrix*}$ & $\begin{matrix*}[l]12Q\log_2(1/\varepsilon_\mathrm{ar}) + 2(Q+d)T_{Dcbe} + 4(Q+d)T_{Dsp} \\ \quad + Q(24\ell + 36)+ 6 d \log_2(1/\varepsilon_z) + d(32\ell - 2) \\ \quad + 12\log_2(1/\varepsilon_\mathrm{tsp}) +3(\ell -1) \end{matrix*}$ \\
\\
\textbf{$T$-count}  & $\begin{matrix*}[l] 12Q\log_2(1/\varepsilon_\mathrm{ar}) + 2(Q+d)T_{Ccbe} + 4(Q+d)T_{Csp} \\ \quad + Q(24\ell + 31) + 3d\log_2(1/\varepsilon_z) + d(32\ell-2)
\end{matrix*}$ & $\begin{matrix*}[l]12Q\log_2(1/\varepsilon_\mathrm{ar}) + 2(Q+d)T_{Ccbe} + 4(Q+d)T_{Csp} \\ \quad + Q(24\ell + 51) + 6 d \log_2(1/\varepsilon_z) + d(32\ell-2) \\ \quad + 12 (L-1)\log_2(1/\varepsilon_\mathrm{tsp}) + 16(L-\ell - 1)\end{matrix*}$ \\
\end{tabular}
\end{ruledtabular}
\end{table*}

\subsubsection{Resource estimate for producing classical approximation to linear system solution}

The resource estimates described above capture the quantum resources required for a single coherent quantum circuit that appears during the algorithm. The output of this quantum circuit is a quantum state, but the QIPM requires a classical estimate of the amplitudes of this quantum state. This classical estimate is produced through tomography, as described in \cref{sec:qipm-tomography}, by performing $k = 57.5L\ln(6L/\delta)/(\varepsilon^2(1-\varepsilon^2/4))$ repetitions each of the QLSS and controlled-QLSS circuits, where $\varepsilon$ is the desired tomography precision and $\delta$ is the probability that the tomography succeeds. In the implementation given in Algorithm~\ref{algo:IPM}, we fix $\delta = 0.1$. Thus, to estimate the quantum resources of a single \textit{iteration} of the QIPM, the previous resource estimates reported in \cref{tab:qls} should each be multiplied by $k$. We note with $P$ processors large enough to prepare the output of the QLSS, these $k$ copies could be prepared in $k/P$ parallel steps, saving a factor of $P$ in the runtime at the expense of a factor of $P$ additional space.  Our resources and scaling estimates do not account for any parallelization, and we assume completely serial execution and runtime.

After multiplication by $k$, these expressions give the quantum resources required to perform the single quantum line of the QIPM, \ApprSolve. This subroutine has both classical input and output and can thus be compared to classical approaches for approximately solving linear systems.

\subsubsection{Estimate for end-to-end portfolio optimization problem}

Recall that the full QIPM algorithm is an iterative algorithm, where each iteration involves approximately solving a linear system by preparing many copies of the same quantum states. The duality gap $\mu$, which measures the proximity of the current interior point to the optimal point, begins at 1 and decreases by a constant factor $\sigma$ with each iteration. Thus, the required number of iterations to reach a final duality gap $\epsilon$ is given by 
\begin{equation}\label{eq:iterations}
\begin{split}
N_{it} & = \lceil\ln(\epsilon)/\ln(\sigma)\rceil = \left\lceil\frac{\ln(\epsilon)}{\ln(1-\frac{1}{20\sqrt{2r}})}\right\rceil \\
& \approx \left \lceil 20\sqrt{2}\ln(\epsilon^{-1})\sqrt{r}\right\rceil.
\end{split}
\end{equation}
Recall from the discussion in \cref{sec:self-dual} that the output of the QIPM will achieve an $\bigo{\epsilon}$ approximation to the optimal value of the objective function.

Pulling this all together, we now estimate the resources to perform the full QIPM algorithm, including the multiplicative factors needed to perform tomography as well as the number of iterations to converge to the optimal solution. Note that the relevant condition number $\kappa_F(G)$ and required linear-system precision $\xi$ will vary from iteration-to-iteration as the Newton matrix $G$ changes. The overall runtime can be upper bounded using the maximum observed value of $\kappa_F(G)$, which we denote by $\kappa_F$, and minimum observed value of $\xi$ across all iterations.  At each iteration, to achieve overall precision $\xi$, the tomography precision $\varepsilon$ is chosen to be just smaller than $\xi$ (we choose $\varepsilon = 0.9 \xi$), while all other error parameters ($\varepsilon_\mathrm{ar}$, $\varepsilon_\mathrm{tsp}$, $\varepsilon_z$, etc.) are chosen to be small constant fractions of $\xi$, such that a total error budget of $\xi$ is not exceeded. As the non-tomographic error parameters all appear underneath logarithms, these small constant factors will drop out of a leading order analysis, and it suffices to replace all of these error parameters with $\xi$.

We may then express the overall runtime in terms of $\kappa_F$, $\xi$, $L$ (the size of the Newton system), and $r$ (the number of second-order cone constraints) up to leading order and including all constant factors, which we report in \cref{tab:resource_full_QIPM}. Recall that for the infeasible version of the QIPM acting on the self-dual embedding, we have $L = 2N + K + 3$, where $N$ is the number of SOCP variables and $K$ is the number of linear constraints. Note that in our leading order expression, we have assumed that the contributions proportional to $Q = \bigo{\kappa_F}$ dominate over terms proportional to $d = \bigo{\kappa_F\log(1/\xi)}$ at practical choices of $\xi$ due to the large constant prefactor in the definition of $Q$ (see \cref{lem:adiabatic-discrete-filtered} and surrounding discussion).  The left column of \cref{tab:summary_scaling} from the introduction is formed using the expressions in \cref{tab:resource_full_QIPM}, and substituting the corresponding relations between $L$ and $n$, where $n$ is the number of stocks in the portfolio optimization problem given in \cref{eq:SOCP-PO}. That is, we substitute $r = 3n+1$ and $L = 2N + K + 3 = 8n + 3m + 6 = 14n +6$ when we take $m = 2n$, where $N$ is the number of SOCP variables, $K$ is the number of SOCP constraints, $n$ is the number of stocks, and $m$ is the number of time epochs used to create the matrix $M$ as described in \cref{sec:portfolio}. 


\begingroup
\squeezetable
\begin{table}[ht]
\caption{\label{tab:resource_full_QIPM}Leading order contribution to the logical qubit count, $T$-depth, and $T$-count for the entire QIPM, including constant factors. The parameter $L$ denotes the size of the Newton linear system and $r$ denotes the number of second-order cone constraints, while $\epsilon$ denotes the final duality gap that determines when the algorithm is terminated. For the infeasible QIPM running on an $n$-asset instance of portfolio optimization, as given in \cref{eq:SOCP-PO}, we have $L = 14n+6$ and $r = 3n+1$; these substitutions yield the results in \cref{tab:summary_scaling}. The parameter $\kappa_F$ denotes the maximum observed Frobenius condition number and $\xi$ denotes the minimum observed tomographic precision parameter across all iterations.}
\begin{ruledtabular}
\renewcommand{\arraystretch}{1.4}
\begin{tabular}{L{1.4
cm}L{7.0cm}}
\textbf{Resource} & \textbf{QIPM complexity} \\
\hline
\textbf{\# Qubits}  & $4L^2 $  \\ 
\\                                        
\textbf{$T$-depth}  & $(5 \times 10^8) \kappa_F L \sqrt{r}\xi^{-2} \log_2(\epsilon^{-1}) \log_2(L) \log_2(\kappa_F L^{14/27}\xi^{-1})$  \\
\\
\textbf{$T$-count}  & $(1 \times 10^8)\kappa_F L^3 \sqrt{r}\xi^{-2} \log_2(\epsilon^{-1}) \log_2(L) \log_2(\kappa_F\xi^{-1})$ \\
\end{tabular}
\end{ruledtabular}
\end{table}
\endgroup




\section[Numerical experiments]{Numerical experiments with historical stock data}\label{sec:numerics}

The resource expressions in \cref{tab:resource_full_QIPM} include constant factors but leave parameters $\kappa_F$ and $\xi$ unspecified. These parameters depend on the specific SOCP being solved. As a final step, we use numerical simulations of small PO problems to study the size of these parameters for different PO problem sizes. This information enables us to give concrete estimates for the resources needed to solve realistic PO problems with our implementation of the QIPM and sheds light on whether there could be an asymptotic quantum advantage. 

Our numerical experiments simulate the entirety of Algorithm~\ref{algo:IPM}. The only quantum part of the algorithm is to carry out the subroutine \ApprSolve$(G,\ve{h},\xi)$. We simulate the quantum algorithm for this subroutine by solving the linear system exactly using a classical solver and then adding noise to the resulting estimated values to simulate the output of tomography. Since the tomography scheme illustrated in \cref{sec:qipm-tomography} repeatedly prepares the same state and draws $k$ samples from measurements in the computational basis, the result is a sample from the multinomial distribution. In our numerical simulation, we draw samples from this same multinomial distribution, thus capturing tomographic noise in a more precise way than by simply adding uniform Gaussian noise, as was done in Ref.~\cite{kerenidis2019quantum}. For simplicity, we assume that the part of the tomography protocol that calculates the signs of each amplitude correctly computes each sign. To numerically estimate resource counts, we must understand ultimately what level of precision $\xi$ is required to stay close enough to the central path throughout the algorithm, as well as how large the Frobenius condition number $\kappa_F$ of the Newton system is. Importantly, we would like to know how these quantities scale with system size and duality gap $\mu$, which decreases by a constant factor with each iteration of the QIPM. 

In \cref{sec:solve_newton_system}, we discussed three formulations of the QIPM (see \cref{tab:QIPM_choices}). The first (II-QIPM) is closely related to the original formulation from Ref.~\cite{Kerenidis2021quantumalgorithms}, which does not guarantee that the intermediate points generated by the IPM are feasible. The other two are instantiations of the inexact-feasible formulation proposed in Ref.~\cite{augustino2021inexact}, which requires pre-computing a basis for the null-space of the SOCP constraint matrix. The first of these computes a valid basis by hand (IF-QIPM), while the second uses a QR decomposition to find the basis (IF-QIPM-QR). We simulated all three versions and found that the II-QIPM was always able to stay close to the central path, despite the lack of a theoretical guarantee that this would be the case. Here we present the results of the II-QIPM. For comparison, in \cref{app:feasible_qipm}, we present some numerical results for the feasible QIPMs, which do benefit from a theoretical convergence guarantee, but have other drawbacks. 

As discussed in \cref{sec:implementation-pseudo}, we also implemented a very simple preconditioner that we find reduces the condition number by at least an order of magnitude with negligible additional classical cost. In all cases, we report resources estimates assuming a preconditioned matrix.

\subsection{Example instance}\label{sec:numerics-example}

\begin{figure}[t]
\centering
\includegraphics{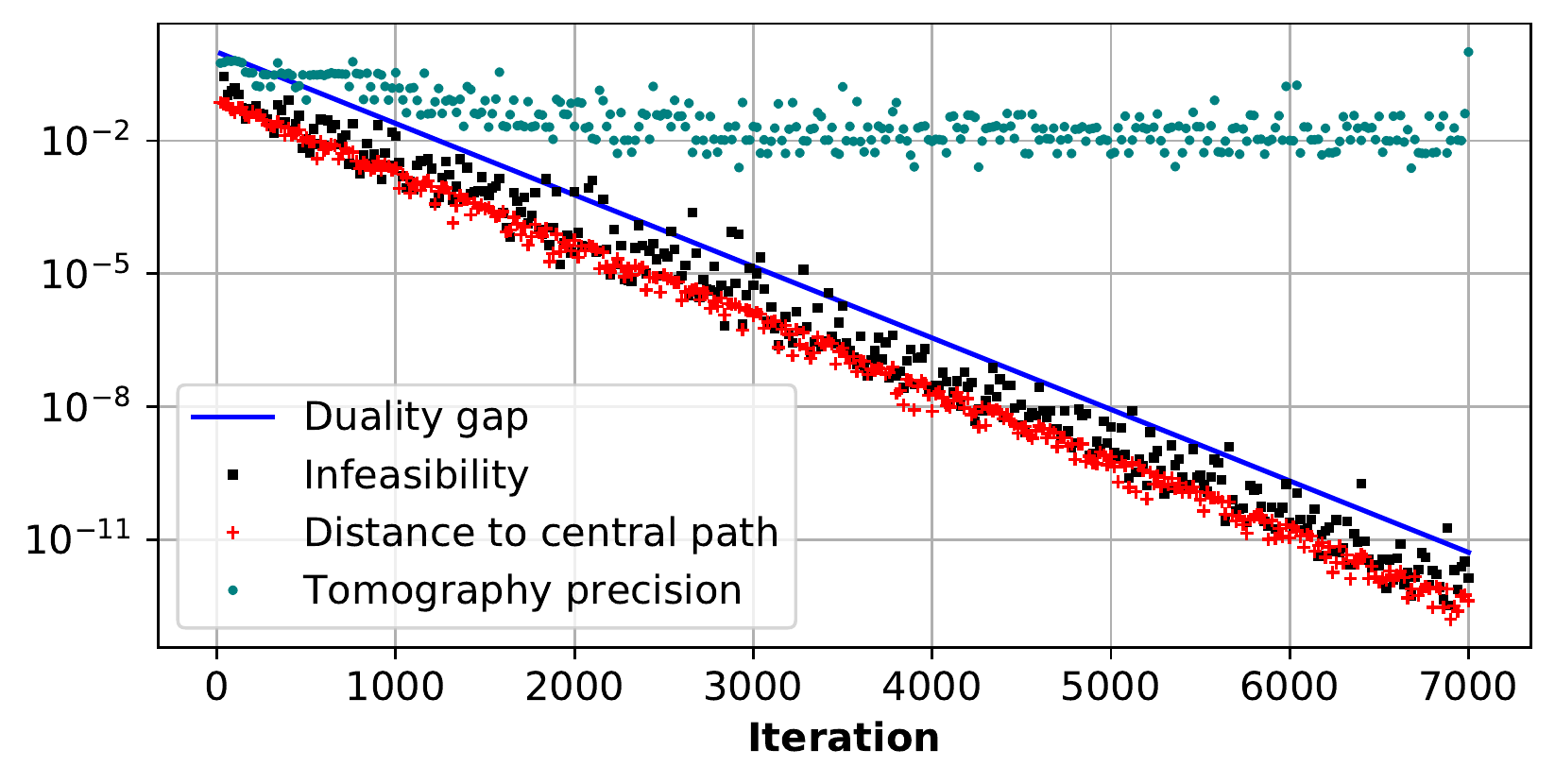}
\caption{\label{fig:example_instance} Simulation of the QIPM on an SOCP instance corresponding to portfolio optimization on $n=30$ randomly chosen stocks using $m=60$ time epochs. The duality gap $\mu$ (defined in \cref{eq:duality_gap_self_dual}), the distance to the central path $d_F$ (defined in \cref{eq:distance_to_cp}), and the infeasibility (defined as the norm of the residual on the right-hand-side in \cref{eq:Newton-system-1}) each decrease exponentially with the number of iterations. The tomography precision $\xi$  required to stay near the central path (defined adaptively as outlined in Algorithm~\ref{algo:IPM}) initially decreases and then plateaus at about $10^{-2}$.}
\end{figure}

In \cref{fig:example_instance}, we present as an example the results of one of our simulations.  We construct a portfolio optimization instance of \cref{eq:PO} by randomly choosing $n=30$ stocks from the Dow Jones U.S.~Total Stock Market Index (DWCF). We (arbitrarily) set parameters $q=1$, $\ve{\zeta} = 0.05 \cdot \ve{1}$, and we assume our previous portfolio $\ve{\bar{w}}$ allocates weight to each stock in proportion to its market capitalization. The returns of the 30 stocks on the first $m=2n=60$ days in our dataset were used to construct an average return vector $\ve{\hat{u}}$ and an $m \times n$ matrix $M$ for which $M^\intercal M = \Sigma$, the covariance matrix for the stock returns, as described in \cref{sec:socp-portfolio}. 

We simulate the infeasible QIPM acting on the corresponding SOCP in \cref{eq:SOCP-PO}. The figure illustrates how the simulation successfully follows the central path to the optimal solution after many iterations. The duality gap decreases with each step, and, crucially, the infeasibility and distance to the central path also decrease (exponentially) with iteration. Also plotted is the tomography precision $\xi$ that was required to ensure that each iteration stayed sufficiently close to the central path (determined adaptively as described in the pseudocode in Algorithm~\ref{algo:IPM}). The plot exemplifies how, despite the lack of theoretical convergence guarantees, our simulations suggest that in practice the II-QIPM acting on the PO SOCP will yield valid solutions.  

Remarkably, for this instance, we also observe that both the Frobenius condition number $\kappa_F$ and the inverse tomography precision $\xi^{-1}$ initially increase but ultimately plateau with the iteration number, even as the duality gap gets arbitrarily small (see \cref{fig:condition_number_scaling_duality_gap} for data on $\kappa_F$). This scaling behavior was a generic feature of our simulations across all the instances we simulated. This contrasts with the worst-case expectation that the condition number can increase as $\kappa_F = \bigo{1/\mu}$ or $\kappa_F = \bigo{1/\mu^2}$ (depending on the formulation of the Newton system) \cite{Kerenidis2021quantumalgorithms,augustino2021inexact}. Prior literature does not say much about whether the quantity $\xi^{-1}$ should be expected to diverge. One might expect that, since the neighborhood of the central path gets smaller as $\mu$ gets smaller (e.g., radius is proportional to $\mu$ in \cref{eq:neighborhood}), the precision requirement to stay close to the central path would get more stringent in proportion to $\mu$. However, it is important to recall that the step size from one iteration to the next also shrinks with $\mu$, and that $\xi$ represents the size of the error on the \emph{normalized} Newton system solution; thus the neighborhood does not shrink \emph{relative} to the distance to the optimum and the length of the next step, and there is no immediate reason that $\xi^{-1}$, as we have defined it, must diverge as $\mu \rightarrow 0$. However, one does expect that in the worst case, if the condition number $\kappa$ diverges, then $\xi^{-1}$ should also diverge, as errors of constant size $\xi$ on the estimate of $\ve{u}/\lVert \ve{u} \rVert$ can lead to residual errors of divergent size $\kappa \xi$ on the normalized product $G\ve{u}/\lVert G \ve{u} \rVert$. We hope that future work can better elucidate why $\kappa_F$ and $\xi^{-1}$ do not diverge on these instances.\footnote{It is perhaps related to whether the instance is nondegenerate and obeys strict complementarity. A known consequence of these properties is that the Jacobian matrix is nonsingular (implying a non-divergent condition number) at the optimum, as discussed in Section 6 of Ref.~\cite{alizadeh2003second}.}

\subsection{Scaling of condition number}\label{sec:numerics-condition}
\begin{figure}
\includegraphics[width=8.4cm]{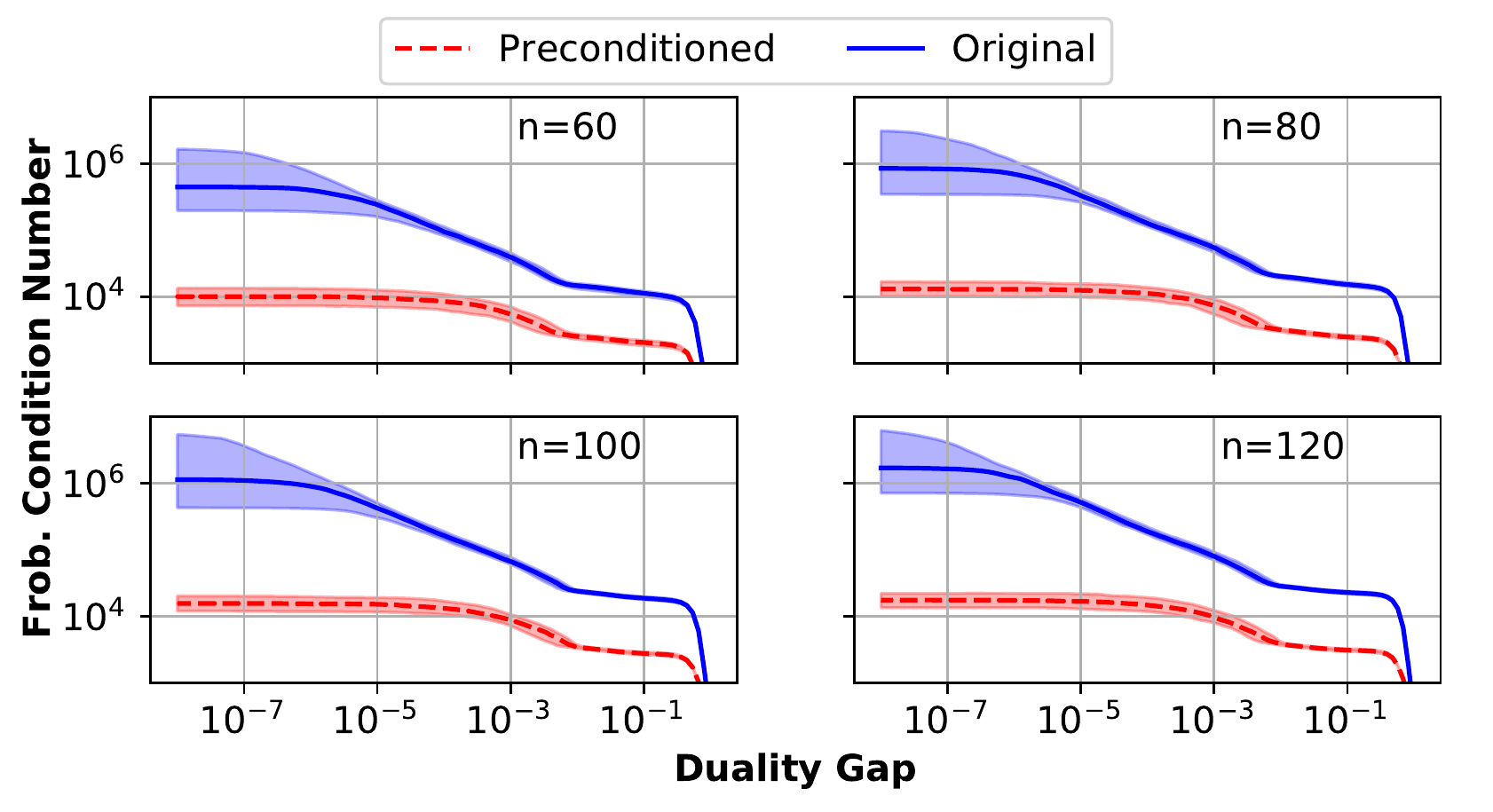}
\caption{\label{fig:condition_number_scaling_duality_gap}Median Frobenius condition $\kappa_F$ number for 128 randomly sampled stock portfolios from the DWCF index as a function of the duality gap for portfolios of size $60$, $80$, $100$, and $120$ stocks. The shaded regions indicate the 16th to 84th percentile. We observe that the condition number appears to plateau at small values of the duality gap.}
\end{figure}

To understand the problem scaling with portfolio size, we generate example problem instances by randomly sampling $n$ stocks from the DWCF, using returns over $m=2n$ time epochs (days) to construct our SOCP as in \cref{eq:SOCP-PO}. Parameters $q$, $\ve{\zeta}$, $\ve{\bar{w}}$, $\ve{\hat{u}}$ and $M$ are all chosen in the same way as described above. 
We plot the Frobenius condition number of the Newton matrix as well as the preconditioned Newton matrix as a function of the duality gap in \cref{fig:condition_number_scaling_duality_gap} for portfolios of size $n \in \{ 60, 80, 100, 120 \}$. Here we confirm our previous remark that the condition number appears to plateau at a certain value of the duality gap, especially for the preconditioned matrix. 

\begin{figure}
\includegraphics[width=8.4cm]{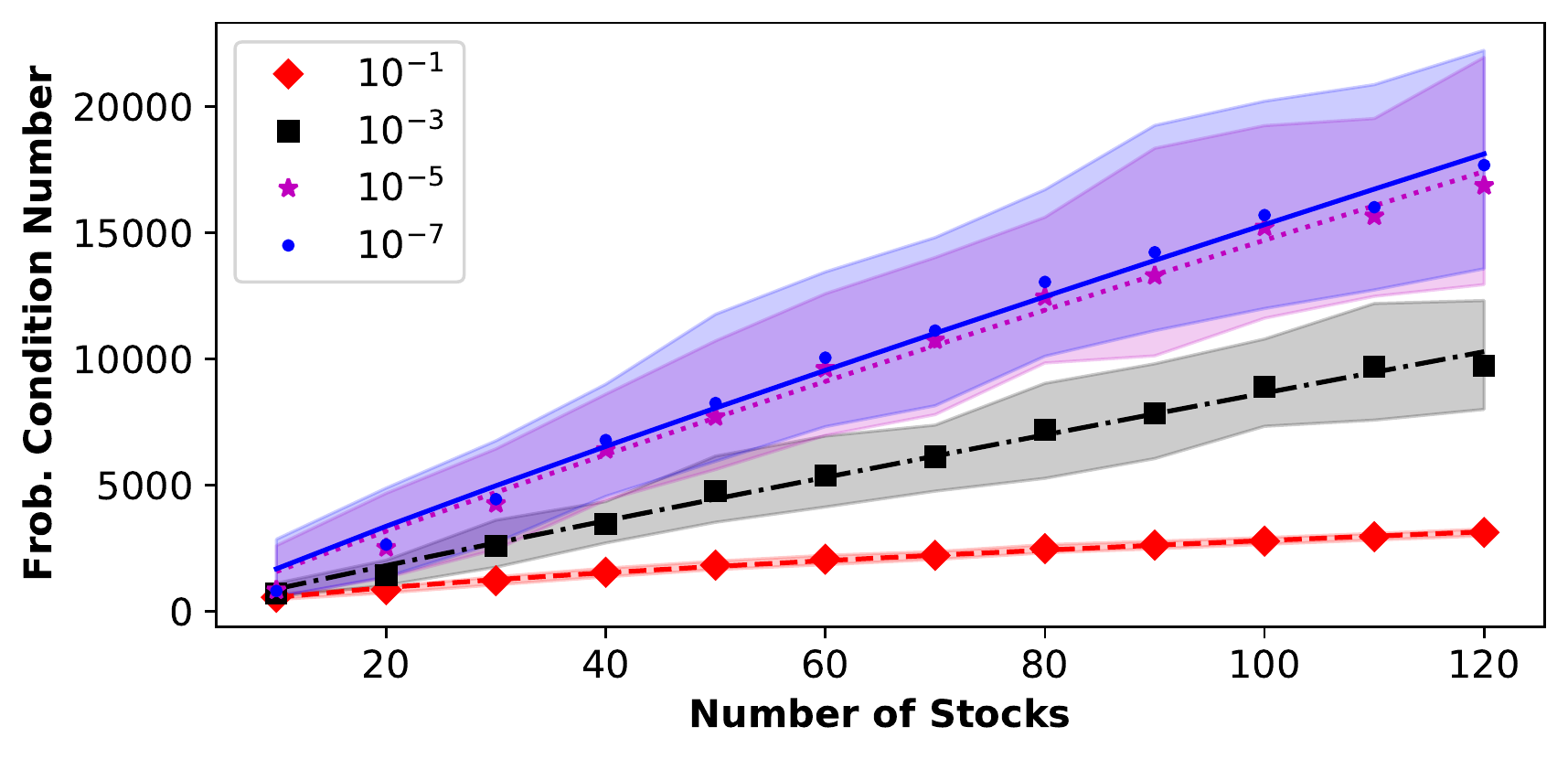}
\caption{\label{fig:condition_number_scaling}Median Frobenius condition number $\kappa_F$ for 128 randomly sampled stock portfolios from the DWCF index as a function of portfolio size for duality gaps of $10^{-1}$, $10^{-3}$, $10^{-5}$, and $10^{-7}$. The shaded regions correspond to the 16th to  84th percentile. The lines represent power-law fits of the form $an^b$, where the values for $b$ are reported in \cref{tab:condnum_slopes}. In all four cases, the exponent is less than 1 and in the latter three cases it is greater than 0.9 suggesting a nearly linear-in-$n$ trend.}
\end{figure}

Key to understanding the asymptotic scaling of the quantum algorithm is to determine how the condition number scales as a function of the number of assets, as the runtime of the QLSS algorithm grows linearly with the condition number. In \cref{fig:condition_number_scaling}, we plot the Frobenius condition number $\kappa_F$ as a function of $n$, the number of stocks, observed at duality gaps $\mu \in \{10^{-1}$, $10^{-3}$, $10^{-5}$, $10^{-7}\}$. At duality gaps of $10^{-5}$ and $10^{-7}$, the condition number $\kappa_F$ has plateaued as observed in \cref{fig:condition_number_scaling_duality_gap}. We perform a non-linear fit to the data using a power law $\kappa_F = an^b$ model, where $a$ and $b$ are fit parameters, and we report the exponents $b$ in \cref{tab:condnum_slopes}. 
All exponents appear to be near or less than unity.

\begin{table}[hb]
\caption{\label{tab:condnum_slopes}Estimated exponent parameters for the Frobenius condition number $\kappa_F$ obtained from the fits that are plotted in \cref{fig:condition_number_scaling}.}
\begin{ruledtabular}
\renewcommand{\arraystretch}{1.4}
\begin{tabular}{C{3cm}C{5.4cm}}
\textbf{Duality Gap} & \textbf{Condition Number Scaling} \\
\hline
$10^{-1}$ & $\bigo{n^{0.60 \pm 0.02}}$ \\ 
$10^{-3}$ & $\bigo{n^{0.94 \pm 0.04}}$  \\
$10^{-5}$ & $\bigo{n^{0.92 \pm 0.04}}$  \\
$10^{-7}$ & $\bigo{n^{0.91 \pm 0.05}}$ \\
\end{tabular}
\end{ruledtabular}
\end{table}

\subsection{Scaling of tomography precision}

\begin{figure}
\includegraphics[width=8.4cm]{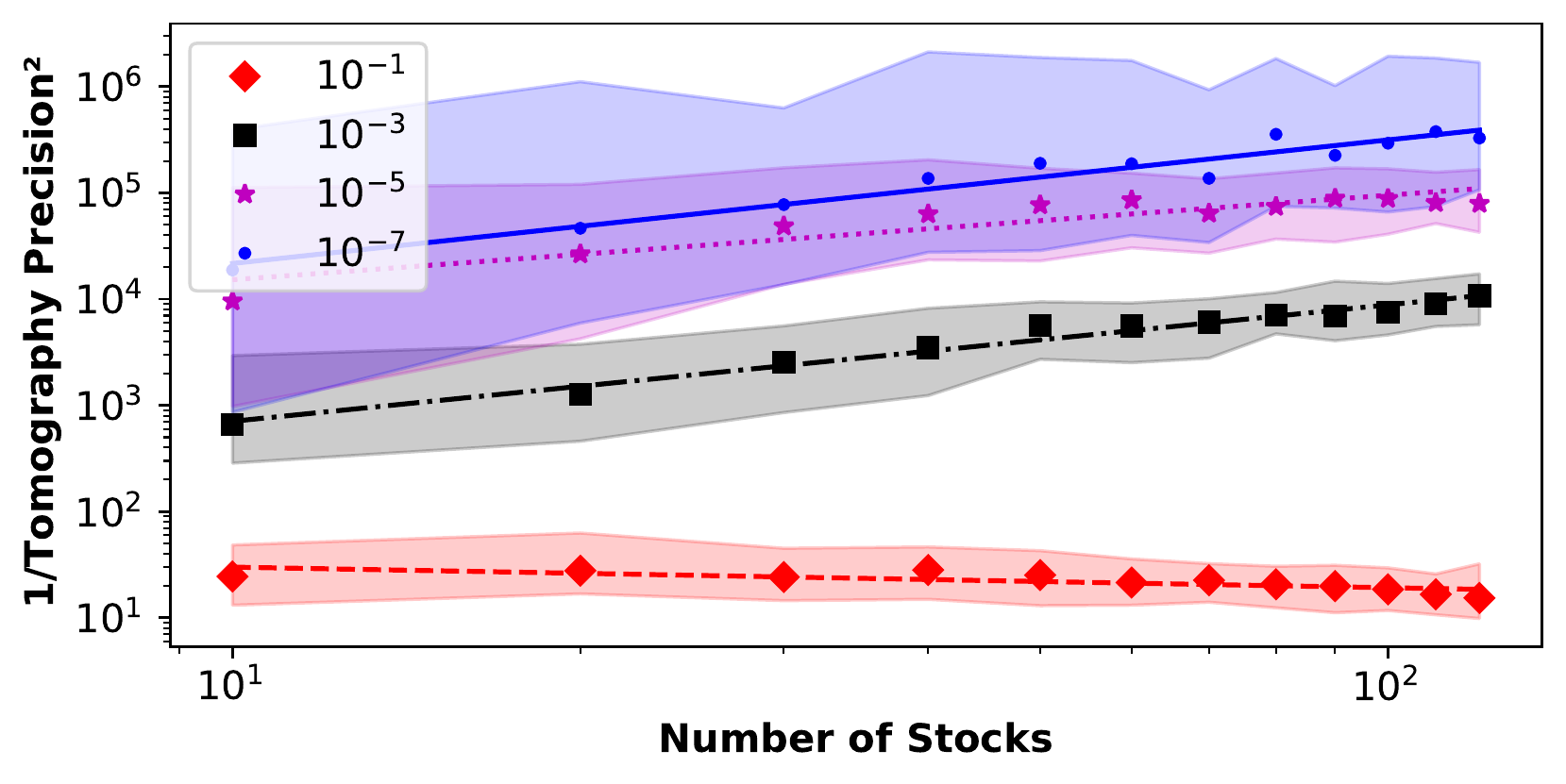}
\caption{\label{fig:invtomography_scaling} Median value of the square of the required inverse tomography precision $\xi^{-2}$ required to remain in the neighborhood of the central path for 128 randomly sampled stock portfolios from the DWCF index as a function of portfolio size for duality gaps of $10^{-1}$, $10^{-3}$, $10^{-5}$, and $10^{-7}$. To reduce iteration-to-iteration variation, an artifact of the adaptive approach to tomography, we average over the observed value of $\xi^{-2}$ at the five iterations for which the duality gap is nearest the indicated value. The shaded regions correspond to the 16th to 84th percentile. Here logarithmic axes are used since (unlike for $\kappa_F$) instance-to-instance variation covers multiple orders of magnitude even for a fixed value of $n$. The dashed lines correspond to a linear fit to the log-log data, where the slope is reported in \cref{tab:invtom_slopes}.}
\end{figure}

While the depth of the individual quantum circuits that compose the QIPM scales only with the Frobenius condition number, the QIPM also requires a number of repetitions of this circuit for tomography that scales as $1/\xi^2$, the inverse of the tomography precision squared. To see how this scales with problem size, we perform a similar analysis for $\xi^{-2}$ that we previously performed for $\kappa_F$. These results are presented in \cref{fig:invtomography_scaling} for the same four duality gaps of $\{10^{-1},10^{-3},10^{-5},10^{-7}\}$. To reduce the iteration-to-iteration variation in the tomography precision (which results from our adaptive approach to tomography in Algorithm~\ref{algo:IPM}), in calculating $\xi^{-2}$ at duality gap $\mu$, we take the average over the value of $\xi^{-2}$ at the five iterations with duality gap nearest to $\mu$.  We fit the median of $\xi^{-2}$ at each value of $n$ to a linear model on a log-log plot, corresponding to a relationship $\xi^{-2} = an^b$, and we report the implied exponent $b$ in \cref{tab:invtom_slopes}.  In this case, it is hard to draw robust conclusions from the fits. The fit suggests that the median of $\xi^{-2}$ is increasing with $n$ on the interval $n \in [10,120]$. However, the most striking feature of the data is that the instance-to-instance variation of $\xi^{-2}$ is significantly larger than that of $\kappa_F$. In fact, at $\mu=10^{-7}$, the 84th percentile of instances at $n=10$, the smallest size we simulated, had a larger value of $\xi^{-2}$ than the 50th percentile of instances at $n=120$, the largest size we simulated.

\begin{table}[ht]
\caption{\label{tab:invtom_slopes}Estimated exponent parameters for $1/\xi^2$ obtained from the fits that are plotted in \cref{fig:invtomography_scaling}.}
\begin{ruledtabular}
\renewcommand{\arraystretch}{1.4}
\begin{tabular}{C{3cm}C{5.4cm}}
\textbf{Duality Gap} & \textbf{Tomography Scaling} \\
\hline
$10^{-1}$ & $\bigo{n^{-0.19 \pm 0.05}}$ \\ 
$10^{-3}$ & $\bigo{n^{1.10 \pm 0.06}}$  \\
$10^{-5}$ & $\bigo{n^{0.79 \pm 0.11}}$  \\
$10^{-7}$ & $\bigo{n^{1.16 \pm 0.10}}$ \\
\end{tabular}
\end{ruledtabular}
\end{table}

\subsection{Asymptotic scaling of overall runtime}\label{sec:numerics-runtime}

\begin{figure}
\includegraphics[width=8.4cm]{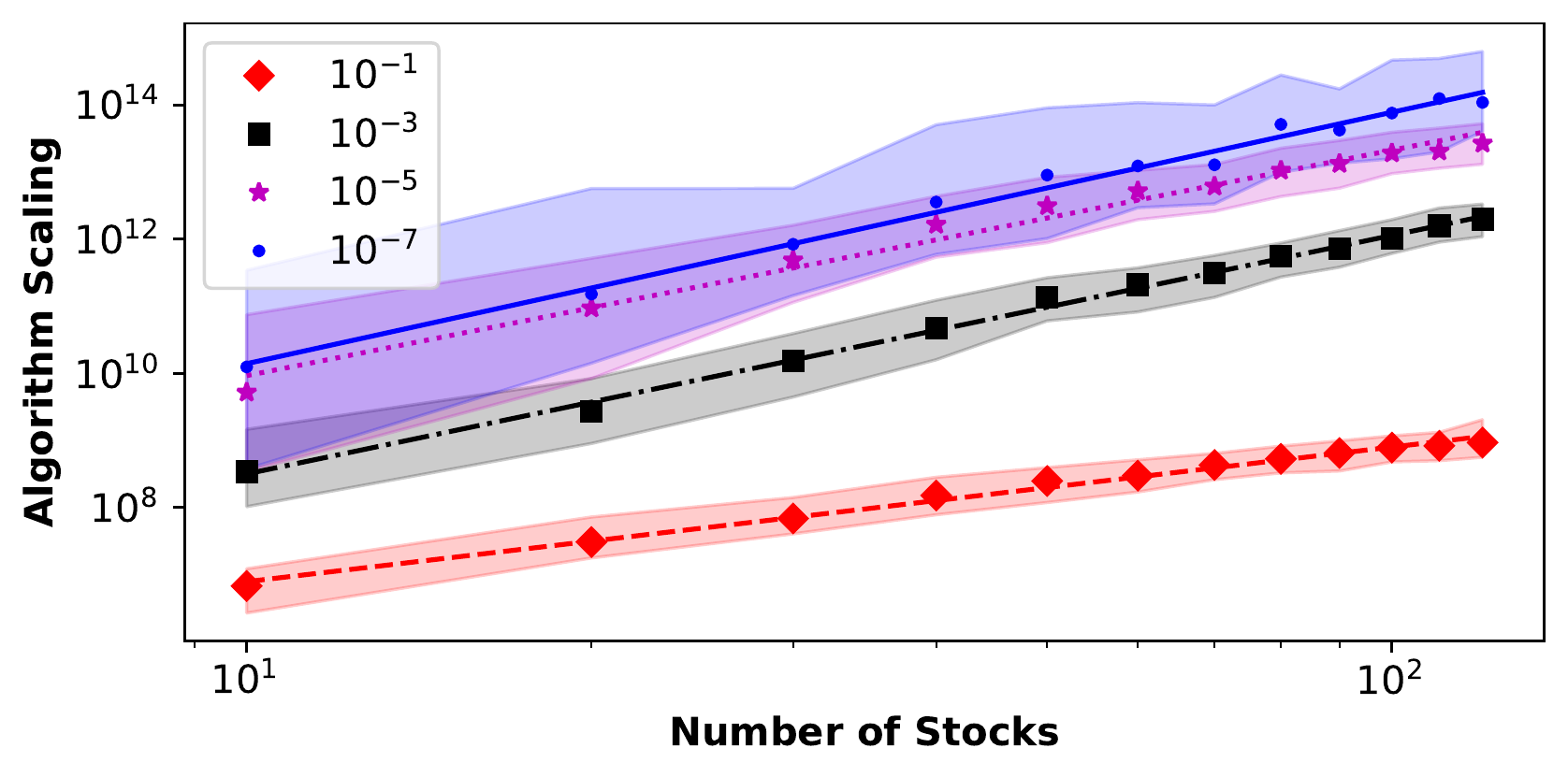}
\caption{\label{fig:algorithm_scaling} Median value of the estimated algorithm scaling factor computed as the median of $n^{1.5} \kappa_F/\xi^2$ for 128 randomly sampled stock portfolios from the DWCF index as a function of portfolio size for duality gaps of $10^{-1}$, $10^{-3}$, $10^{-5}$, and $10^{-7}$. As in Fig.~\ref{fig:invtomography_scaling}, we average over 5 consecutive points to reduce iteration-to-iteration variance deriving from adaptive tomography. Here we also use the actual number of observed samples that were required to achieve sufficient tomographic precision in place of the tomgraphic factor $n/\xi^2$.  The shaded regions correspond to the 16th to 84th percentiles. The lines correspond to a linear fit to the log-log data, where the slope is reported in \cref{tab:scaling_slopes}.}
\end{figure}

Above we provided fits for $\kappa_F$ and $\xi^{-2}$ as a function of $n$ on the range $n \in [10,120]$. Here we study the quantity $n^{1.5}\kappa_F/\xi^2$, which determines the asymptotic scaling of the runtime of the QIPM. In \cref{fig:algorithm_scaling}, we plot this quantity at the same four duality gap values $\mu \in \{10^{-1},10^{-3},10^{-5},10^{-7}\}$. The implied exponents arising from linear fits on a log-log axis are reported in \cref{tab:scaling_slopes}. They are generally consistent with summing the exponents from the previously reported fits. The data inherit from $\xi^{-2}$ the feature that the instance-to-instance variation is orders of magnitude larger than the median. Taken at face value, the fits suggest that the scaling of the median algorithmic runtime on the interval $n \in [10,120]$ is similar to the $n^{3.5}$ scaling of classical IPMs using Gaussian elimination, and worse than the asymptotic $n^{2.87}$ arising from classical IPMs using fast matrix-multiplication techniques to solve linear systems \cite{strassen1969gaussian,alman2021refined} (note that this scaling does not apply until $n$ becomes very large, so it is not a good practical comparator). However, the large variance and imperfect fits do not give us confidence that these trends can be reliably extrapolated to larger $n$.  Accordingly, when we compute actual resource counts in the next subsection, we stick to $n=100$ and do not speculate on precise estimates for larger (more industrially relevant) $n$. Our numerical experiments fail to provide significant evidence for an asymptotic polynomial quantum speedup, but neither do they definitively rule it out. 
Toward that end, note that if the version of tomography we have studied were to be replaced with the more advanced recently proposed tomography scheme of Ref.~\cite{apeldoorn22}, the runtime of the QIPM would instead grow as $n^{1.5}\kappa_F/\xi$, while introducing some additional gate overhead. Our fits from \cref{tab:invtom_slopes} suggest this could reduce the asymptotic exponent, but by no more than about $\bigo{n^{0.6}}$ or so. 

Ultimately, we do not believe it is essential to pin down the asymptotic scaling of the algorithm, because the main finding of our work is that, even if a slight asymptotic polynomial speedup exists, the size of the constant prefactors involved in the algorithm preclude an actual practical speedup, barring significant improvements to multiple aspects of the algorithm. In the next subsection, we elaborate on this point in a more quantitative fashion.

\begin{table}[ht]
\caption{\label{tab:scaling_slopes}Exponent parameter estimates from the fits to the line generated by plotting $n^{1.5} \kappa_F/\xi^2$ in \cref{fig:algorithm_scaling}, which determines the overall scaling of the runtime of the QIPM. For comparison, CIPMs using Gaussian elimination have runtime $\bigo{n^{3.5}}$ and CIPMs using faster methods for solving linear systems have runtime $\bigo{n^{2.87}}$. 
}
\begin{ruledtabular}
\renewcommand{\arraystretch}{1.4}
\begin{tabular}{C{3cm}C{5.4cm}}
\textbf{Duality Gap} & \textbf{Algorithm Scaling} \\
\hline
$10^{-1}$ & $\bigo{n^{2.01 \pm 0.05}}$ \\ 
$10^{-3}$ & $\bigo{n^{3.56 \pm 0.07}}$  \\
$10^{-5}$ & $\bigo{n^{3.36 \pm 0.14}}$  \\
$10^{-7}$ & $\bigo{n^{3.75 \pm 0.12}}$ \\
\end{tabular}
\end{ruledtabular}
\end{table}

\subsection{Numerical resource estimates}\label{sec:numeric-resources}

Rather than examine algorithmic scaling, we now compute actual resource counts for the QIPM applied to PO. Ultimately, it is these resource counts that matter most from a practical perspective. We estimate the total circuit size in terms of the number of qubits, $T$-depth, and $T$-count for a portfolio of 100 assets.
We chose this size because it is small enough that we can simulate the entire quantum algorithm classically. However, at this size, solving the PO problem is not classically hard; generally speaking, the PO problem becomes challenging to solve with classical methods only once $n$ is on the order of $10^3$ to $10^4$. A similar concrete calculation could be performed at larger $n$ by extrapolating trends observed in our numerical simulations, but we are not confident that the fits on $n \in [10,120]$ reported above are reliable predictors for larger $n$.

Recall that the only step in the QIPM performed by a quantum computer is the task of producing a classical estimate to the solution of a linear system to error $\xi$. The complexity of this task as it is performed within the QIPM depends on $\xi$ as well as the Frobenius condition number $\kappa_F$. The first step of our calculation is to fix values for $\xi$ and $\kappa_F$ at $n=100$. We choose them by taking the median over the 128 samples in our numerical simulation at duality gap $\mu = 10^{-7}$.

Once $\kappa_F$ and $\xi$ are fixed, we must now determine concrete values for the various other error parameters that appear in the algorithm such that overall error $\xi$ can be achieved. Tomography dominates the complexity and overall error, but there are a number of other factors that contribute to the error in the final solution. We enumerate and label the sources of error here, for completeness: 

\begin{itemize}
\item $\varepsilon_G$: Error in block-encoding the matrix $G$
\item $\varepsilon_{\ve{h}}$: Error in the unitary that prepares the state $\ket{\ve{h}}$
\item $\varepsilon_\mathrm{ar}$: Gate synthesis error for single-qubit rotations needed by $\mathrm{CR}^0(s)$ and $\mathrm{CR}^1(s)$ (see \cref{fig:CR0})
\item $\varepsilon$: Tomography error
\item $\varepsilon_z$: Gate synthesis error for each single-qubit rotation needed for QSVT eigenstate filtering (see \cref{fig:QSVT})
\item $\varepsilon_\mathrm{qsp}$: Error due to polynomial approximation in eigenstate filtering
\item $\varepsilon_\mathrm{tsp}$: Error in preparing the state $\sum_{i=1}^L\sqrt{p_i}|i\>$ needed for computing the signs in the tomography routine
\end{itemize}

In \cref{sec:qipm}, we described a quantum circuit that prepares a state $\ket{\ve{\tilde{v}}}$ (after postselection) for which $\lVert \ket{\ve{\tilde{v}}} - \ket{\ve{v}} \rVert \leq \varepsilon_\mathrm{QLSP}$. If the block-encoding unitaries, state-preparation unitaries, and single-qubit rotations were perfect, then the only contribution to $\varepsilon_\mathrm{QLSP}$ would be from eigenstate filtering and we would have $\varepsilon_\mathrm{QLSP} \leq \varepsilon_\mathrm{qsp}$. Note the relationship $d = 2 \kappa_F\ln(2/\varepsilon_\mathrm{qsp})$ from \cref{lem:adiabatic-discrete-filtered}. Since the block-encoding unitary $U_G$, the state-preparation unitary $U_{\ve{h}}$, and the single-qubit rotations are implemented imperfectly, there is additional error. In preparing the state, the unitary $U_G$ is called $2Q+2d$ times and the unitary $U_{\ve{h}}$ is called $4Q+4d$ times, where $Q$ is given in \cref{lem:adiabatic-discrete-filtered}. Additionally,  there are $2Q$ combined appearances of $\mathrm{CR}^{0}(s)$ and $\mathrm{CR}^{1}(s)$ gates, where each appearance requires two single-qubit rotations. Note that the appearances of $\mathrm{CR}^{0}(s)$ and $\mathrm{CR}^{1}(s)$ within the eigenstate filtering portion of the circuit do not contribute to the error because at $s=1$ these gates can be implemented exactly. Finally, there are another $d$ single-qubit rotations required to implement the eigenstate filtering step.  Since operator norm errors add sublinearly, we can thus say that
\begin{equation}
    \varepsilon_\mathrm{QLSP} \leq \varepsilon_\mathrm{qsp} + (2Q+2d) \varepsilon_G + (4Q+4d) \varepsilon_{\ve{h}} + 4Q\varepsilon_\mathrm{ar}+2d\varepsilon_z\,.
\end{equation}

Now, the result of \cref{lem:tomography} implies that, in order to assert that the classical estimate $\ve{\tilde{v}}'$ output by tomography satisfies $\lVert \ve{\tilde{v}}' - \ve{v}\rVert \leq \xi$, it suffices to have
\begin{equation}\label{eq:error_budget}
    \begin{split}
    \xi \geq{}&\varepsilon + 1.58 \sqrt{L} \varepsilon_\mathrm{tsp} + 1.58\bigg[\varepsilon_\mathrm{qsp} + (2Q+2d)\\
    &{}  \varepsilon_G + (4Q+4d) \varepsilon_{\ve{h}} + 4Q\varepsilon_\mathrm{ar}+d\varepsilon_z \bigg] \,,
    \end{split}
\end{equation}
where for convenience we recall the definitions (ignoring the $\bigo{\sqrt{\kappa_F}}$ term) of $Q$ and $d$ as
\begin{align}
    Q &= 1.31C\kappa_F  \\
    d &= 2\kappa_F \ln(2/\varepsilon_\mathrm{qsp})
\end{align}
Recalling that the dominant term in the complexity of the algorithm scales as $\varepsilon^{-2}$ but logarithmically in the other error parameters, to minimize the complexity we assign the majority of the error budget to $\varepsilon$: we let $\varepsilon = 0.9 \xi$, and we split the remaining $0.1 \xi$ across the remaining six terms of \cref{eq:error_budget}. There is room for optimizing this error budget allocation, but the savings would be at most a small constant factor in the overall complexity. 

Note that elsewhere in the draft, we have referred to $\xi$ as ``tomography precision'' since $\varepsilon$ will dominate the contribution to $\xi$. Here, the resource calculation requires we differentiate $\varepsilon$ from $\xi$, but when speaking conceptually about the algorithm, we focus on $\xi$ as it is the more fundamental parameter: it represents the precision at which the classical-input-classical-output linear system problem is solved, allowing apples-to-apples comparisons between classical and quantum approaches. 

With values for $\kappa_F$, $\varepsilon_G$, $\varepsilon_{\ve{h}}$, $\varepsilon_\mathrm{qsp}$, $\varepsilon_z$, and $\varepsilon_\mathrm{tsp}$ now fixed, we can proceed to complete the resource count using the expressions in \cref{tab:qls}. Note that for gate synthesis error, we use the formula $R_y = 3\log_2(1/\varepsilon_r)$, where $R_y$ is the number of $T$ gates needed to achieve an $\varepsilon_r$-precise Clifford+$T$ gate decomposition of the rotation gate \cite{ross2016optimal}. Putting this all together yields the resource estimates for a single run of the (uncontrolled) quantum linear system solver in \cref{tab:resources} at $n=100$. We report these estimates both in terms of primitive block-encoding and state-preparation resources, as well as the raw numerical estimates. For the total runtime, we must also estimate the resources required for the controlled state-preparation routine. We have estimated these quantities, but to the precision of the estimates we report, the numbers are the same as the controlled version, so we exclude them for brevity. 

To estimate total runtime, our estimates must be multiplied by the tomography factor $k$ (for controlled and for uncontrolled) as well as the number of iterations $N_{it} = \lceil\ln(\epsilon)/\ln(\sigma)\rceil$, where $\epsilon$ is the target duality gap (which we take to be $\epsilon=10^{-7}$), and $\sigma = 1.0 - 1/(20\sqrt{2r})$. While $k$ will vary from iteration to iteration, in our calculation we assume the total number of repetitions is given by the simple product $(2k) N_{it}$, which, noting that the value of $\xi$ plateaus after a certain number of iterations, will give a roughly accurate estimate. Note that these $2k N_{it}$ repetitions need not be done coherently, in the sense that the entire system is measured and reprepared in between each repetition. One can bound the tomography factor $k$ to be $k \leq 57.5 L \ln (L) / \xi^2$, where $\xi$ is determined empirically. However, our numerical simulations of the algorithm yield an associated value of $k$ needed to generate the estimate to precision $\xi$, so we can use this numerically determined value directly. We find that the observed median value of $k = 3.3 \times 10^8$ from simulation is multiple orders of magnitude smaller than the theoretical bound. Using this substitution for $k$ and $N_{it}$, we find the results shown in the right column of \cref{tab:summary_scaling} in the introduction. 

\begin{figure*}[ht!]
    \begin{tikzpicture}[scale=0.5, transform shape, every node/.style = {scale=1.62, transform shape}]
        \draw[fill=lightgray!50] (1,1) rectangle (28,11) node at (16,2) {Adiabatic evolution: $T_D \approx 3\times 10^{11}$; $T_C \approx 1\times 10^{17}$};
        \draw[fill=green!20!black!30] (1.5,4.5) rectangle (5.5,10.5) node[pos=.5] {$U[0]$};
        \draw[fill=yellow!60!black!30] (5.5,8.5) rectangle (7,10.5) node[pos=.5] {$W$};
        \draw[fill=green!20!black!30] (7,4.5) rectangle (11,10.5) node[pos=.5] {$U[1/Q]$};
        \draw[fill=yellow!60!black!30] (11,8.5) rectangle (12.5,10.5) node[pos=.5] {$W$};
        \draw[fill=green!20!black!30] (12.5,4.5) rectangle (16.5,10.5) node[pos=.5] {$U[2/Q]$};
        \draw[fill=yellow!60!black!30] (16.5,8.5) rectangle (18,10.5) node[pos=.5] {$W$};
        \node at (20,9) {$\cdots$};
        \node at (20,8) {$Q \approx 5\times 10^7$};
        \draw[fill=green!20!black!30] (22,4.5) rectangle (26,10.5) node[pos=.5] {$U[1-1/Q]$};
        \draw[fill=yellow!60!black!30] (26,8.5) rectangle (27.5,10.5) node[pos=.5] {$W$};
        
        \draw[fill=green!20!black!30] (0,13) rectangle (11.5,20) node at (7.75,19.5) {Adiabatic Block-Encode};
        \node at (7.75,18.5) {$T_D \approx 6\times 10^3$};
        \node at (7.75,17.5) {$T_C \approx 2\times 10^{9}$};
        \draw[fill=green!50!black!20] (0,13) rectangle (4,20) node [pos=0.5, text width=6em, align=center] {$V_G$ \\ $T_D \approx 3\times 10^3$ \\ $T_C \approx 2\times 10^{9}$};
        \draw[fill=green!50!black!20] (4,13) rectangle (7.5,16) node [pos=0.5, text width=6em, align=center] {$\mathrm{CR}(s)$ \\ $T_D \approx 5\times 10^2$ \\ $T_C \approx 5\times 10^{2}$};
        \draw[fill=green!50!black!20] (7.5,13) rectangle (11.5,17) node [pos=0.5, text width=6em, align=center] {$U_{Q_{\ve{h}}}$ \\ $T_D \approx 3\times 10^3$ \\ $T_C \approx 3\times 10^{6}$};
        \draw (3,17.2) -- (5,18);
        \draw (5.7,15.8) -- (5.9,17);
        \draw (9.3,16.2) -- (8.5,17.1);
        
        \draw [dashed] (0,13) -- (1.5,10.5);
        \draw [dashed] (11.5,13) -- (5.5,10.5);
        
        \draw[fill=yellow!60!black!30] (15.25,13) rectangle (19.25,17) node [pos=0.5, text width=6em, align=center] {Reflection \\ $T_D \approx 8\times 10^1$ \\ $T_C \approx 8\times 10^{1}$};
        
        \draw [dashed] (16.5,10.5) -- (15.25,13);
        \draw [dashed] (18,10.5) -- (19.25,13);
        
        \draw[fill=red!90!black!10] (28,4.5) rectangle (34,11);
        \draw[fill=red!50!black!30] (28.25,8) rectangle (29.25,10.5);
        \draw[fill=orange!100!black!50] (29.25,9.5) rectangle (29.75,10.5);
        \draw[fill=red!50!black!30] (29.75,8) rectangle (30.75,10.5);
        \draw[fill=orange!100!black!50] (30.75,9.5) rectangle (31.25,10.5);
        \node at (31.65,9) {$\cdots$};
        \draw[fill=red!50!black!30] (32.25,8) rectangle (33.25,10.5);
        \draw[fill=orange!100!black!50] (33.25,9.5) rectangle (33.75,10.5);
        
        \node [text width=10em, align=center] at (31,6) {Eigenstate Filtering \\ $T_D\approx 2\times 10^9$ \\ $T_C\approx 8\times 10^{14}$};
        \draw [decorate, decoration = {calligraphic brace,mirror, amplitude=20pt}] (0.5,0.85) --  (34.5,0.85) node [pos=0.5, below=25pt] {$k \times N_\textrm{it}\approx (3\times10^{8})\times(8\times 10^3)\approx 3\times 10^{12}$ classical repetitions};
        
        \node at (18.1,-1.8) {$+$ same number of repetitions for controlled version};
        
        
    \end{tikzpicture}
    \caption{\label{fig:resource_breakout}Breakdown of the quantum resources required for a single coherent run of the uncontrolled version of the quantum algorithm needed to produce the state \cref{eq:qls-state}. As we did in \cref{tab:resources}, here we take the final duality gap to be $\mu=10^{-7}$ and the number of assets to be $n=100$. Choices for the Frobenius condition number $\kappa_F = 1.6 \times 10^4$ and number of tomographic repetitions $k = 3.3\times 10^8$ are informed by our numerical experiments, as discussed in \cref{sec:numerics}. A similar breakdown for the controlled version needed to produce the state \cref{eq:controlled-U} would be essentially the same. The eigenstate filtering sub-circuit follows a very similar alternating structure to the adiabatic evolution, with the $U[j]$ block-encodings replaced with either $U[1]$ or $U[1]^\dagger$, the reflection operator $W$ replaced with phase rotations, and only $d\ll Q$ total number of iterations (refer to \cref{fig:QSVT} for details.)}
\end{figure*}
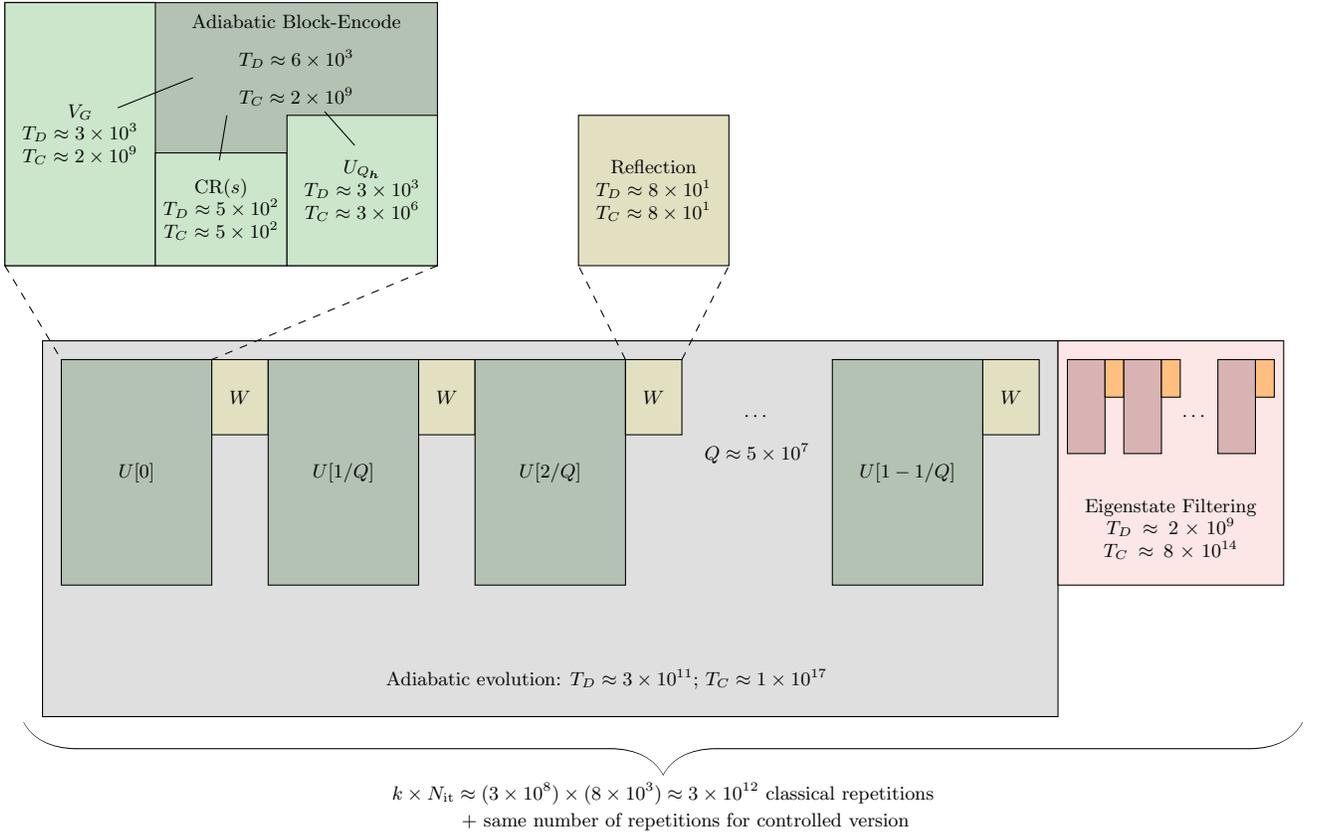

To aid in understanding which portions of the algorithm dominate the complexity, we show a breakdown of the resources in \cref{fig:resource_breakout}. The width of the boxes is representative of the $T$-depth, while the height of the boxes represents the $T$-count. The number of classical repetitions, composed of tomography samples as well as IPM iterations needed to reach a target duality gap, contributes the largest factor to the algorithmic runtime. Of these two, quantum state tomography contributes more than the iterations needed to reach the target duality gap. Our exact calculation confirms that for the individual quantum circuits involved in the QLSS, the discrete adiabatic portion of the algorithm dominates over the eigenstate filtering step in its contribution to the overall quantum circuit $T$-depth. Within the adiabatic subroutine the primary driver of the $T$-depth and $T$-count is the need to apply the block-encoding operator $Q$ times (see e.g.~\cref{eq:walk}), where $Q$ is proportional to the Frobenius condition number. An additional source of a large $T$-count arises from the need to block-encode the linear system, which causes the $T$-count to scale as $\bigo{L^2}$.

\begin{table}[h]
\caption{\label{tab:resources}Estimated number of logical qubits  $N_Q$, $T$-depth $T_D$, and $T$-count $T_C$ required to perform the quantum linear system solver (QLSS) subroutine within the QIPM running on a PO instance with $n=100$ stocks. This calculation uses the empirically observed median value for the condition number at duality gap $\mu = 10^{-7}$, which was $\kappa_F = 1.6 \times 10^{4}$.  The full QIPM repeats this circuit $k = \bigo{n\ln(n)\xi^{-2}}$ times in each iteration to generate a classical estimate of the output of the QLSS, and also performs $N_{it}=\bigo{n^{0.5}}$ iterations, where the linear system being solved changes from iteration to iteration. In the left column, we write the resources as numerical prefactors times the resources required to perform the controlled-block-encoding of the matrix $G$ (denoted by a subscript $cbe$), and the state-preparation of the vector $|\ve{h}\>$ (denoted by a subscript $sp$), defined in \cref{tab:min_T_depth,tab:sp_min_T_depth}. Written this way, one can see the large prefactors occurring from the linear system solver portion of the algorithm. In the right column we compute the exact resources, including those coming from the block-encoding. The notation $AeB$ is short for $A \times 10^B$.}
\begin{ruledtabular}
\renewcommand{\arraystretch}{1.4}
\begin{tabular}{p{5.5cm} p{2.5cm}}
\textbf{QLSS Prefactors} & \textbf{Total} \\
\hline
$N_Q = N_{Qcbe} + 5$                      & $N_Q = 8e6$ \\
$T_D = (1e8)T_{Dcbe} + (2e8) T_{Dsp} + (4e10)$ & $T_D =  3e11$ \\
$T_C = (1e8)  T_{Ccbe} + (2e8)  T_{Csp} + (4e10)$ & $T_C = 1e17$ \\
\end{tabular}
\end{ruledtabular}
\end{table}



\section{Conclusions}\label{sec:discussion}

\subsection{Bottlenecks}

The resource quantities we report are prohibitively large, even for the classically easy problem size of $n=100$ assets in the portfolio optimization instance. Our detailed analysis allows us to see exactly how this large number arises, which is essential for understanding how best to improve it. We outline the several independent factors leading to the large resource estimates.
\begin{itemize}
    \item The block-encoding of the classical data is called many times by the QLSS. This data is arranged in an $L \times L$ matrix (note that for a PO instance of size $n$ with $m=2n$, the Newton linear system has size roughly $L\approx14n$). These block-encodings can be implemented up to error $\varepsilon_G$ in $\bigo{\log(L/\varepsilon_G)}$ $T$-depth using circuits for quantum random access memory (QRAM) as a subroutine \cite{clader2022quantum}. While the asymptotic scaling is favorable, after close examination of the circuits for block-encoding, we find that in practice the $T$-depth can be quite large: at $n=100$ and $\varepsilon_G = 10^{-10}$ (it is necessary to take $\varepsilon_G$ very small since the condition number of $G$ is quite large), block-encoding to precision $\varepsilon_G$ has a $T$-depth of nearly 1000. Importantly, this $T$-depth arises \emph{even after implementing several new ideas to minimize the circuit depth}, presented by a subset of the authors separately in Ref.~\cite{clader2022quantum}.
    
    \item The condition number $\kappa_F$ determines how many calls  to the block-encoding must be made, and we observe that $\kappa_F$ is quite large for the application of portfolio optimization. \emph{Even after an attempt at preconditioning, $\kappa_F$ is on the order of $10^4$ already for small SOCP instances corresponding to $n=100$ stocks}, and empirical trends suggest it grows nearly linearly with $n$.  However, we believe that additional preconditioning could significantly reduce the effective value of $\kappa_F$ in this algorithm.
    
    \item The constant factor in front of the $\bigo{\kappa_F}$ in state-of-the-art QLSSs is also quite large: the theoretical analysis proves an upper bound on the prefactor of $1.2 \times 10^5$. Numerical simulations performed in \cite{costa2021optimal} suggested that, in practice, it can be one order of magnitude smaller than the theoretical value. Following these numerics, we take the constant prefactor to be $1.31\times 2305$ in our numerical estimates, which still contributes significantly to the estimate. Future work should aim to reduce this constant or, alternatively, investigate whether other approaches, such as those based on variable-time amplitude amplification (VTAA) \cite{Ambainis10,chakraborty2019}, could achieve better performance despite being asympototically suboptimal. \footnote{Indeed, after the release of the first version of this paper, Ref.~\cite{jennings2023efficient} proposed a different QLSS implementation based on Ref.~\cite{subasi2019quantumLinearSystemAdiabatic} with slightly suboptimal $\bigo{\kappa \log(\kappa/\epsilon)}$ asymptotic scaling, but with superior actual performance for all $\kappa < 10^{32}$. According to Figure 1 of Ref.~\cite{subasi2019quantumLinearSystemAdiabatic}, substituting their QLSS for the one we have analyzed would reduce the constant prefactor at $\kappa = 10^4$ from our estimated $2 \times 1.31 \times 2305$ (where in this comparison we include the extra factor of 2 arising from the $1/2$ success probability of eigenstate filtering) to about 900, saving about a factor of 6 on our resource estimate. } 
    
    \item Pure state tomography requires preparing many copies of the output $\ket{\ve{v}}$ of the QLSS. We improved the constant prefactors in the theoretical analysis beyond what was known, but even with this improvement, the number of queries needed to produce an estimate $\ve{v}'$ of the amplitudes of $\ket{\ve{v}}$ up to error $\varepsilon$ in $\ell_2$ norm is $115 L \ln(L)/\varepsilon^2$, which for $n = 100$ and $\varepsilon = 10^{-3}$ is on the order of $10^{11}$ (although our simulations suggest $2k = 7 \times 10^8$ suffice in practice). We note that this is another avenue for substantial improvement. For instance, the results of \cite{apeldoorn22} could be used (see \cref{foot:new-tomography} for more details).

    \item QIPMs, like CIPMs, are iterative algorithms; the number of iterations in our implementation is roughly $20\sqrt{2r}\ln(\epsilon^{-1})$, a number chosen to utilize theoretical guarantees of convergence (note that $r \approx 3n$). Taking $n = 100$ and $\epsilon = 10^{-7}$, our implementation would require $8 \times 10^3$ iterations. We suspect that the number of iterations could be significantly decreased if more aggressive choices were made for the step size. For example, similar to our adaptive approach to tomographic precision, one could try longer step sizes first, and shorten the step size when the iteration does not succeed. This sort of optimization would apply equally to CIPMs and QIPMs. 
\end{itemize}
Remarkably, the five factors described above all contribute roughly equally to the overall $T$-depth calculation; the exception being the number of copies needed to do tomography, which is a much larger number than the others. Tomography would be the obvious place to begin to try to reduce the resource depth, perhaps by implementing the scheme recently proposed in Ref.~\cite{apeldoorn22}, and by making modifications to the QIPM that might allow the parameter $\xi$ to be larger in practice, or by using an iterative refinement method \cite{augustino2021inexact}.  Another comment regarding tomography is that, in principle, the $k$ tomographic samples can be taken in parallel rather than in series. Running in parallel leads to a huge overhead in memory: one can reduce the tomographic depth by a multiplicative factor $P$ at the cost of a multiplicative factor $P$ additional qubits. Note that even preparing a single copy requires a daunting number of nearly ten million logical qubits at $n=100$. Moreover, it is unlikely that improvements to tomography alone could make the algorithm practical, as the other four factors still contribute roughly $10^{16}$ to the $T$-depth. 

Besides the rather large constant factors pointed out above for tomography and especially for the QLSS, we also note that the multiplicative ``log factors'' that are typically hidden underneath $\tilde{\mathcal{O}}$ notation in asymptotic analyses contribute meaningfully here. For instance, the entire block-encoding depth is $\bigo{\log(n/\varepsilon_G)}$, which, in practice, is as large as 1000. Moreover, there is an additional $\ln(\epsilon^{-1}) \approx 16$ coming from the iteration count, and a $\ln(L)\approx 7$ from tomography. 

This quantitative analysis of bottlenecks for QIPMs can inform likely bottlenecks in other applications where QLSS, tomography, and QRAM subroutines are required. While some parameters such as $\kappa_F$ and $\xi$ are specific to the application we considered here, other observations such as the numerical size of various constant and logarithmic factors (e.g.~block-encoding depth) would apply more generally in other situations.

\subsection{Resource estimate given dedicated QRAM hardware}

\begin{table*}
\renewcommand{\arraystretch}{1.5}
\caption{\label{tab:solver_complexity}Comparison of time complexities of different approaches for exactly or approximately solving an $L \times L$ linear system with Frobenius condition number $\kappa_F$ to precision $\xi$. The comparison highlights how a quantum advantage only persists when $\kappa_F$ is neither too large nor too small. The constant pre-factor roughly captures the $T$-depth that we found for the quantum case (the same pre-factor from Tab.~\ref{tab:resource_full_QIPM} after discounting the $20\sqrt{2}$ IPM iteration factor) and the number of multiplications in the classical case.}
\begin{ruledtabular}
    \begin{tabular}{cccc}
        \textbf{Solver} & \textbf{Type} & \textbf{Complexity} & \textbf{Pre-factor estimate } \\
        \hline
        QLSS + Tomography & Quantum, Approximate & $L\kappa_F \xi^{-2}\ln(L)\ln(\kappa_F\xi^{-1} L^{14/27})$ &  $4 \times 10^7$ \\
        Gaussian Elimination & Classical, Exact & $L^3$  & $1/3$ \\
        Randomized Kaczmarz \cite{strohmer2009randomized}  & Classical, Approximate & $L\kappa_F^2\ln(\xi^{-1})$  & 8 \\
    \end{tabular} 
\end{ruledtabular}
\end{table*}

The bottlenecks above focused mainly on the $T$-depth and did not take into account the total $T$-count or the number of logical qubits, which are also large. Indeed, our estimate of 8 million logical qubits, as reported in \cref{tab:summary_scaling}, is drastically larger than estimates for other quantum algorithms, such as Shor's algorithm \cite{roetteler2017quantum} and algorithms for quantum chemistry (e.g.~\cite{lee2021chemistry}), both of which can be on the order of $10^3$ logical qubits. By contrast, the current generation of quantum processors have tens to hundreds of \textit{physical} qubits, and no logical qubits; a long way from the resources required for this QIPM.

However, it is important to note that, as for other algorithms requiring repeated access to classical data, the vast majority of the gates and qubits in the QIPM arise in the block-encoding circuits, which are themselves dominated by QRAM-like data-loading subcircuits \cite{clader2022quantum}. These QRAM-like sub-circuits have several special features. Firstly, they are largely composed of controlled-swap gates, each of which can be decomposed into four $T$ gates that can even be performed in a single layer, given one additional ancilla and classical feed-forward capability \cite{jones2013}. 
Furthermore, in some cases, the ancilla qubits can be ``dirty'' \cite{low2018trading,hann2021}, i.e.~initialized to any quantum state, and, if designed correctly, the QRAM circuits can possess a natural noise resilience that may reduce the resources required for error correction \cite{hann2021}. Implementing these circuits with full-blown universal and fault-tolerant hardware could be unnecessary given their special structure. Just as classical computers have dedicated hardware for RAM, quantum computers may have dedicated hardware optimized for performing the QRAM operation. Preliminary work on hardware-based QRAM data structures (as opposed to QRAM implemented via quantum circuits acting on logical qubits) shows promise in this direction~\cite{PRXQuantum.2.030319,jiang2019experimental}.

Our estimates suggest that the size of the QRAM needed to solve an $n = 100$ instance of PO is one megabyte, and the QRAM size for $n=10^4$ (i.e., sufficiently large to potentially be challenging by classical standards) is roughly 10 gigabytes, which is comparable to the size of classical RAM one might find on a modern laptop. These numbers could perhaps be reduced by exploiting the structure of the Newton matrix, as certain blocks are repeated multiple times in the matrix, and many of the entries are zero\footnote{We expect that exploiting the sparsity of the matrix would lead to reduced logical qubit count and $T$-count, but not reduced $T$-depth. In fact, it could lead to non-negligible increases in the $T$-depth, since the shallowest block-encoding constructions from Ref.~\cite{clader2022quantum} are hyper-optimized for low-depth, and are explicitly not compatible with exploiting sparsity.} (see \cref{eq:Newton-system-1,eq:SOCP-PO}). 

With this in mind, we can ask the following hypothetical question: suppose that we had access to a sufficiently large dedicated QRAM element in our quantum computer, and furthermore that the QRAM ran at a 4GHz clock speed (which is comparable to modern classical RAM); would the algorithm become more practical in this case? Under the crude, conservative simplifying assumption that each block-encoding and state-preparation unitary requires just a \emph{single} call to QRAM and the rest of the gates are free, we can give a rough answer by referring to the expression in \cref{tab:resources}, which states that $3 \times 10^8$ total block-encoding and state-preparation queries are needed.\footnote{Separate from the QLSS, a relatively small number of state preparation queries is needed in tomography to create the state in {\cref{eq:tomography_sign_state}}, but this number does not scale with $\kappa$ and we neglect it in this back-of-the-envelope analysis.} Thus, even if the rest of our estimates stay the same, the number of QRAM calls involved in just a single QLSS circuit for $n=100$ would be $3 \times 10^8$. Accounting for the fact that the QIPM involves an estimated $6 \times 10^{12}$ repetitions of similarly sized circuits, the overall number of QRAM calls needed to solve the PO problem would be larger than $10^{21}$, and the total evaluation time would be on the order of ten thousand years.  Thus, even at 4GHz speed for the QRAM, the problem remains decidedly intractable. Nonetheless, we believe that if the QIPM were to be made practical, it would need to involve specialized QRAM hardware in combination with fundamental improvements to the algorithm itself.

\subsection{Comparison between QIPMs and CIPMs and comments on asymptotic speedup}

The discussion above suggests that the current outlook for practicality with a QIPM is pessimistic, but simultaneously highlights several avenues by which to improve the results. Even with such improvements, if QIPMs are to one day be practical, they need to at least have an asymptotic speedup over CIPMs. Here we comment on this possibility. The core step of both QIPMs and CIPMs is the problem of computing a classical estimate of the solution to a linear system, a task that is also of broad use beyond interior point methods. Thus, we need only compare different approaches to solving linear systems, and our conclusions are relevant in any application where linear systems must be solved. 
Accordingly, in \cref{tab:solver_complexity} we give the asymptotic runtime of several approaches to solving an $L \times L$ linear system to precision $\xi$, including the (QLSS $+$ tomography) approach utilized by QIPMs, as well as two classical approaches. Whereas prior literature (e.g.~\cite{Kerenidis2021quantumalgorithms}) primarily compared against Gaussian elimination (which scales as $\bigo{L^3}$), we also note a comparison against the randomized Kaczmarz method \cite{strohmer2009randomized}, which scales as $\bigo{L\kappa_F^2\ln(\xi^{-1})}$. This scaling comes from the fact that $2\kappa_F^2\ln(\xi^{-1})$ iterations are needed, and each iteration involves computing several inner products at cost $\bigo{L}$. We observe that the worst-case cost of an iteration is $4L$ floating point multiplications, meaning all the constant prefactors involved are more-or-less mild. Thus, the asymptotic quantum advantage of the QIPM is limited to an amount equal to $\bigo{\min(\xi^2\kappa_F,\xi^2 L^2/\kappa_F)}$, which is at most $\bigo{L}$ when $\kappa_F \propto L$ and $\xi = O(1)$. Encouragingly, our numerical results are consistent with $\kappa_F \propto L$. However, our results are not consistent with $\xi = \bigo{1}$, suggesting instead that $\xi$ is decreasing with $L$. 

If $\kappa_F \propto L$ and $\xi = \bigo{1}$, we would find a total QIPM runtime of $\bigo{n^{2.5}}$, improving over classical $\bigo{n^{3.5}}$ for a portfolio with $n$ stocks. This speedup would be a material asymptotic improvement over the classical complexity, but leveraging this speedup for a \emph{practical} advantage might still be difficult. Firstly, the difference in the constant prefactor between the quantum and classical algorithms would likely negate the speedup unless $n$ is taken to be very large. Secondly, the speedup would necessarily be sub-quadratic. In the context of combinatorial optimization, where quadratic speedups can be obtained easily via Grover's algorithm, even a quadratic speedup is unlikely to exhibit actual quantum advantage after factoring in slower quantum clock speeds and error-correction overheads~\cite{babbush2021focus}.

Our results suggest that finding a practical quantum advantage for portfolio optimization might require structural improvements to the QIPM itself. In particular, it may be necessary to explore whether additional components of the IPM can be quantized, and whether the costly contribution of quantum state tomography could be completely circumvented. Naively, circumventing tomography entirely is challenging, as it is vitally important to retrieve a \textit{classical} estimate of the solution to the linear system at each iteration in order to update the interior point and construct the linear system at the next iteration. Nevertheless, tomography represents a formidable bottleneck that must be addressed.

While our results are pessimistic on the question of whether quantum interior point methods will deliver quantum advantage for portfolio optimization (and other applications), it is our hope that by highlighting the precise issues leading to daunting resource counts, our work can inspire innovations that render quantum algorithms for optimization more practical. Finally, we conclude by noting that detailed, end-to-end resource estimations of the kind we performed here are vitally important for commercial viability of quantum algorithms and quantum applications. While it is essential to discover and prove asymptotic speedups of quantum algorithms over classical, an asymptotic speedup alone does not imply practicality. For this, a detailed, end-to-end resource estimate is required, as the quantum algorithm may nevertheless be far from practical to implement. As we have seen, the devil is in the details, and there are many details behind which the devil can hide. 


\section*{Acknowledgements}\label{sec:acknowledgements}

We thank Brandon Augustino, Kyle Booth, Paul Burchard, Connor Hann, Iordanis Kerenidis, Anupam Prakash, D\'aniel Szil\'agyi, and Tamas Terlaky for helpful discussions. We are especially grateful to Earl Campbell for early collaboration during an initial phase of the project. GS, HK, and MS are thankful to Shantu Roy for his leadership, trust, and vision for the IACT team at AWS. We also thank James Tarantino for his support throughout the project.



\bibliography{references}

\appendix


\section{Notation}\label{app:notation}

Here we list the important symbols that appear in our paper for reference.
\begin{itemize}
    \item Symbols related to portfolio optimization
    \begin{itemize}
        \item $n$: number of stocks in the portfolio
        \item $\ve{w}$: length-$n$ vector indicating fraction of porfolio allocated to each stock (the object to be optimized)
        \item $\ve{\bar{w}}$: length-$n$ vector indicating current portfolio allocation
        \item $\ve{\zeta}$: length-$n$ vector indicating maximum allowable change to portfolio
        \item $\ve{\hat{u}}$: length-$n$ vector of average returns 
        \item $\Sigma$: $n \times n$ covariance matrix capturing deviations from average returns
        \item $q$: parameter in objective function that determines relative weight of risk vs.~return (\cref{eq:PO})
        \item $M$: $m \times n$ matrix corresponding to the square-root of $\Sigma$, i.e.~$\Sigma = M^\intercal M$
        \item $m$: number of rows in $M$, often equal to the number of time epochs (\cref{sec:socp-portfolio})
    \end{itemize}
    \item Symbols related to second-order cone programs
    \begin{itemize}
        \item $\mathcal{Q}^k$: second-order cone of dimension $k$ (\cref{eq:second-order-cone})
        \item $\mathcal{Q}$: product set of several second-order cones
        \item $\ve{e}$: identity element for $\mathcal{Q}$ or $\mathcal{Q}^k$ (depending on context)
        \item $N$: total number of variables in the SOCP
        \item $K$: total number of linear constraints in the SOCP
        \item $r$: number of second-order cone constraints in the program
        \item $\ve{x}$: length-$N$ vector; primal variable to be optimized, constrained to $\mathcal{Q}$
        \item $\ve{y}$: length-$K$ vector; dual variable to be optimized
        \item $\ve{s}$: length-$N$ vector, appears in dual program, constrained to $\mathcal{Q}$
        \item $A$: $K \times N$ matrix encoding linear constraints (\cref{eq:SOCP})
        \item $\ve{b}$: length-$K$ vector encoding right-hand side of linear constraints (\cref{eq:SOCP})
        \item $\ve{c}$: length-$N$ vector encoding objective function (\cref{eq:SOCP})
        \item $\mu(\ve{x},\ve{s})$: duality gap of the primal-dual point $(\ve{x},\ve{s})$ (\cref{eq:duality_gap})
        \item $\tau$, $\varkappa$, $\theta$: additional scalar variables introduced to implement self-dual embedding (\cref{sec:self-dual})
        \item $\mu(\ve{x},\tau, \ve{s},\varkappa)$: duality gap of the point $(\ve{x},\tau,\ve{s},\varkappa)$ of the self-dual SOCP (\cref{eq:duality_gap_self_dual})
        \item $X$, $S$: arrowhead matrices for vectors $\ve{x}$ and $\ve{s}$ (\cref{eq:arrowhead})
        \item $B$: basis for null space of self-dual constraint matrix
    \end{itemize}
    \item Symbols related to second-order cone programs for portfolio optimization
    \begin{itemize}
        \item $\ve{\phi}$: length-$n$ variable introduced during reduction from PO to SOCP; part of $\ve{x}$ (\cref{eq:SOCP-PO})
        \item $\ve{\rho}$: length-$n$ variable introduced during reduction from PO to SOCP; part of $\ve{x}$ (\cref{eq:SOCP-PO})
        \item $t$: scalar variable introduced during reduction from PO to SOCP; part of $\ve{x}$ (\cref{eq:SOCP-PO})
        \item $\ve{\eta}$: length-$m$ variable introduced during reduction from PO to SOCP; part of $\ve{x}$ (\cref{eq:SOCP-PO})
    \end{itemize}
    \item Symbols related to interior point methods (IPMs)
        \begin{itemize}
            \item $\nu$: parameterizes central path (\cref{eq:central_path_conditions})
            \item $d_F(\ve{x},\tau,\ve{s},\varkappa)$: distance of the point $(\ve{x},\tau,\ve{s},\varkappa)$ to the central path of the self-dual SOCP (\cref{eq:SOCP-self-dual})
            \item $\mathcal{N}$, $\mathcal{N}_F$: neighborhoods of the ``central path'' (\cref{eq:neighborhood,eq:neighborhood_F})
            \item $\gamma$: radius of neighborhood of central path
            \item $\sigma$: step length parameter
            \item $L$: size of (square) Newton matrix
            \item $\epsilon$: input to IPM specifying error tolerance, algorithm terminates once duality gap falls beneath $\epsilon$
        \end{itemize}
    \item Important relations between parameters
        \begin{itemize}
            \item Self-dual embedding has $2N+K+3$ parameters and $N + K + 2$ linear constraints
            \item Newton matrix has size $L = 2N + K + 3$ for infeasible approach and $L=N+1$ for feasible approach
            \item For PO formulation in \cref{eq:SOCP-PO}, $N = 3n+ m + 1$, $r = 3n + 1$, $K = 2n + m + 1$
            \item In our numerical experiments, we choose $m = 2n$
        \end{itemize}
    \item Symbols related to quantum linear system solvers
        \begin{itemize}
            \item $G$: $L \times L$ matrix encoding linear constraints
            \item $\ve{h}$: length-$L$ vector encoding right-hand-side of linear constraints
            \item $\ve{u}$: solution to linear system $G\ve{u} = \ve{h}$
            \item $\ve{v}$: normalized solution to linear system $\ve{u} / \lVert \ve{u}\rVert$
            \item $\varepsilon_\mathrm{QLSP}$: error in solution to linear system
            \item $\ve{\tilde{v}}$: normalized output of the QLSS, which should satisfy $\lVert \ve{v} - \ve{\tilde{v}} \rVert \leq \varepsilon_\mathrm{QLSP}$
            \item $\ell $: $\lceil \log_2 L \rceil$
            \item $U_G$: block-encoding unitary for $G$
            \item $\ell_G$: number of ancilla qubits used by $U_G$
            \item $U_{\ve{h}}$: state-preparation unitary for $\ket{\ve{h}}$
            \item $\kappa_F(G)$: Frobenius condition number $\lVert G \rVert_F \lVert G^{-1}\rVert$ of $G$
            \item $Q$: number of queries to $U_G$ and $U_{\ve{h}}$ (\cref{lem:adiabatic-discrete})
            \item $C$: constant prefactor of $\kappa_F$ (\cref{lem:adiabatic-discrete})
            \item $d$: the degree of the polynomial used in eigenstate filtering (\cref{lem:adiabatic-discrete-filtered})
        \end{itemize}
    \item Symbols related to block encoding and state preparation
        \begin{itemize}
            \item $\varepsilon_G$: block-encoding error for matrix $G$
            \item $\varepsilon_{\ve{h}}$: state-preparation error for vector $\ve{h}$
			\item $\varepsilon_\mathrm{ar}$: Gate synthesis error for rotations needed by $CR^0(s)$ and $CR^1(s)$ 
			\item $\varepsilon_z$: Gate synthesis error for rotations needed by the QSP phases
			\item $\varepsilon_\mathrm{qsp}$: Error due to polynomial approximation in eigenstate filtering
			\item $\varepsilon_\mathrm{tsp}$: Error in preparing the state $\sum_{i=1}^L\sqrt{p_i}|i\>$ needed for the tomography routine
			\item $N_{Qbe}$, $T_{Dbe}$, and $T_{Cbe}$: number of logical qubits, $T$-depth, and $T$-count required for block-encoding.
			\item $N_{Qcbe}$, $T_{Dcbe}$, and $T_{Ccbe}$: number of logical qubits, $T$-depth, and $T$-count required for \emph{controlled}-block-encoding.
			\item $N_{Qsp}$, $T_{Dsp}$, and $T_{Csp}$: number of logical qubits, $T$-depth, and $T$-count required for state preparation.
			\item $N_{Qcsp}$, $T_{Dcsp}$, and $T_{Ccsp}$: number of logical qubits, $T$-depth, and $T$-count required for \emph{controlled}-state preparation.
        \end{itemize}
    \item Symbols related to tomography
        \begin{itemize}
            \item $k$: number of measurements on independent copies of the state
            \item $\delta$: probability of failure
            \item $\varepsilon$: guaranteed error of tomographic estimate
            \item $\xi$: overall precision of solution to linear system, dominated by tomographic error 
        \end{itemize}
\end{itemize}{}

\section{Deferred proofs}\label{app:proofs}


\subsection{Quantum state tomography}\label{app:tomography}

\begin{proof}[Proof of \cref{lem:bernstein}]
Consider a single coordinate $\alpha_j$ with associated probability $p_j = |\alpha_j|^2$, and suppose we take $k$ samples to find an estimate $\tilde{p}_j$ of $p_j$. By Bernstein's inequality,
\begin{align}
\text{Pr}[|\tilde{p}_j - p_j| > \varepsilon_j] \le 2 \exp \left(-\frac{\varepsilon^2}{2(p_j + \varepsilon/3)}k\right)
\end{align}
and so for a given component-wise target deviation in the probability $\varepsilon_j$, choosing 
\begin{align}
k \ge \frac{2(p_j + \varepsilon/3)}{\varepsilon^2}\ln(2/\delta') = \frac{2(|\alpha_j|^2+\varepsilon/3)} {\varepsilon^2}\ln(2/\delta')
\end{align} 
guarantees that $\text{Pr}[|\tilde{p}_j - p_j| > \varepsilon_j] \le \delta'$.

We now pick $\varepsilon_j = \sqrt{3\gamma}|\alpha_j|\varepsilon + \gamma \varepsilon^2$ for some yet undetermined $\gamma > 0$. With this choice
\begin{align}
& \frac{2(|\alpha_j|^2+\frac{\varepsilon}{3})} {\varepsilon^2}\ln(2/\delta') \nonumber\\
&= \frac{2(|\alpha_j|^2 + \sqrt{\frac{\gamma}{3}}\varepsilon + \frac{\gamma}{3}\varepsilon^2)}{(\sqrt{3\gamma}|\alpha_j|\varepsilon + \gamma\varepsilon^2)^2}  \ln(2/\delta')\nonumber\\
&\le \frac{2(|\alpha_j|^2 + 2\sqrt{\frac{\gamma}{3}}\varepsilon + \frac{\gamma}{3}\varepsilon^2)} {3\gamma\varepsilon^2 ( |\alpha_j| + \sqrt{\frac{\gamma}{3}}\varepsilon)^2} \ln(2/\delta') \nonumber\\ 
& = \frac{2}{3\gamma\varepsilon^2} \ln(2/\delta'),
\end{align}
and hence it suffices to choose $k = \frac{2}{3\gamma\varepsilon^2} \ln(2/\delta')$. Letting $\delta' = \delta / L$, the union bound implies that for $k = \frac{2}{3\gamma\varepsilon^2} \ln(2L/\delta)$, all estimates $\tilde{p}_j$ satisfy $|\tilde{p}_j - p_j| \le \varepsilon_j$. We now bound the distance between $|\tilde{\alpha}_j|$ and $|\alpha_j|$. First,
\begin{align}
|\tilde{\alpha}_j| - |\alpha_j| &\le \sqrt{p_j + \varepsilon} - |\alpha_j| \nonumber\\ 
&= \sqrt{|\alpha_j|^2 + \sqrt{3\gamma}|\alpha_j|\varepsilon + \gamma\varepsilon^2} - |\alpha_j| \nonumber\\
&\le (|\alpha_j| + \sqrt{\gamma}\varepsilon) - |\alpha_j| \nonumber\\
&= \sqrt{\gamma}\varepsilon.
\end{align}
Next, we bound $|\alpha_j| - |\tilde{\alpha}_j|$. If $p_j \le \varepsilon_j$ then
\begin{align}
|\alpha_j|^2 \le \sqrt{3\gamma}|\alpha_j|\varepsilon + \gamma\varepsilon^2 \Leftrightarrow |\alpha_j| \le \frac{(\sqrt{3}+\sqrt{7})\sqrt{\gamma}}{2}\varepsilon,
\end{align}
while if $p_j > \varepsilon_j$,
\begin{align}
|\alpha_j| - |\tilde{\alpha_j}| &\le |\alpha_j| - \sqrt{p_j - \varepsilon_j} \nonumber\\
&= |\alpha_j| - \sqrt{|\alpha_j|^2 - \sqrt{3\gamma}|\alpha_j|\varepsilon - \gamma\varepsilon^2} \nonumber\\
& < \frac{(\sqrt{3}+\sqrt{7})\sqrt{\gamma}}{2}\varepsilon,
\end{align}
which follows because the function $f(x) = x - \sqrt{x^2 - \sqrt{3}x - 1}$ has its maximum at $f(\frac{\sqrt{3} + \sqrt{7}}{2}) = \frac{\sqrt{3}+\sqrt{7}}{2}$. Therefore with the choice $\gamma = \left(\frac{\sqrt{3}+\sqrt{7}}{2}\right)^{-2}$, we can guarantee that $| |\tilde{\alpha}_j| - |\alpha_j| | \le \varepsilon$, which corresponds to
\begin{align}
k = \frac{2}{3\gamma\varepsilon^2} \ln(2L/\delta) = \frac{5+\sqrt{21}}{3\varepsilon^2}\ln(2L/\delta)
\end{align}
measurements.
\end{proof}

\begin{proof}[Proof of \cref{lem:tomography}]
Define $\varepsilon' = \frac{1}{\sqrt{2L}} \varepsilon \sqrt{1 - \varepsilon^2/4}$. Then $k = 28.75 \varepsilon'^{-2} \ln(6L/\delta)$. Consider the following three assertions:
\begin{enumerate}
\item The estimates $p_i$ satisfy $|\sqrt{p_i} - |\tilde{v}_i|\sqrt{p}| \le \varepsilon'/3$ for all $i$.
\item The estimates $p^+_i = k^+_i / k$ satisfy $$\left|\sqrt{p^+_i} - \frac{|\sqrt{p}\tilde{v}_i + \sqrt{p'_i}|}{2}\right| \le \varepsilon'/3,$$ and the estimates $p^-_i = k^-_i / k$ satisfy $$\left|\sqrt{p^-_i} - \frac{|\sqrt{p}\tilde{v}_i - \sqrt{p'_i}|}{2}\right| \le \varepsilon'/3,$$ for all $i$.
\item The actual amplitudes $\sqrt{p_i'}$ of the state created in the second step satisfy $|\sqrt{p_i'}-\sqrt{p_i}| \le \varepsilon_\mathrm{tsp}$.
\end{enumerate}
From \cref{lem:bernstein}, we know that Assertion 1 holds with probability at least $1 - \delta/3$, and Assertion 2 holds with probability at least $1 - 2\delta/3$. Therefore both assertions hold with probability at least $1 - \delta$. Moreover, Assertion 3 holds by assumption. From here on we will assume that all three assertions hold.

Let $a_i$ be the real part and $b_i$ be the imaginary part of the quantity $\sqrt{p}\tilde{v}_i$. Let $r^+_i = |\sqrt{p}\tilde{v}_i + \sqrt{p_i}|$, and $r^-_i = |\sqrt{p}\tilde{v}_i - \sqrt{p_i}|$. Note that $r^+_i$ and $r^-_i$ are proportional to the absolute value of the ideal amplitudes of the state created in {\cref{eq:tomography_sign_state}}. One can show that
\begin{align}
    a_i = \frac{\left(r^{+}_i\right)^{2} - \left(r^{-}_i\right)^2}{4\sqrt{p_i}}.
\end{align}
Define $f_i(x,y)=(x^2-y^2)/\sqrt{p_i}$; then $a_i = f_i(r^+_i/2, r^-_i/2)$. We first note that the estimates  $\sqrt{p_i^\pm}$ give us good approximations of $r^+_i/2$ and $r^-_i/2$:
\begin{align}
    \left \lvert\sqrt{p^\pm_i} - \frac{r^\pm}{2}\right\rvert \le \frac{\varepsilon'}{3}\varepsilon + \frac{\varepsilon_\mathrm{tsp}}{2},\label{eq:ppm_rpm_close}
\end{align}
which follows from Assertions 2 and 3. The amplitudes $\tilde{a}_i$ that define the estimate output by the tomography algorithm are given in \cref{eq:tomography_atilde_i}, which can now be rewritten as
\begin{align}
    \tilde{a}_i =         \begin{cases}
            0, &\!\!\!\!\!\!\!\!\!\sqrt{p_i} \le \frac{2}{3}\varepsilon'+\varepsilon_\mathrm{tsp}\text{; else} \\
            \min(\sqrt{p_i}, f_i(\sqrt{p^+_i},\sqrt{p^-_i})), & \!\!f_i(\sqrt{p^+_i},\sqrt{p^-_i}) \ge 0 \\
            \max(-\sqrt{p_i}, f_i(\sqrt{p^+_i},\sqrt{p^-_i})), & \!\!f_i(\sqrt{p^+_i},\sqrt{p^-_i}) < 0 \\ 
        \end{cases}\,.
\end{align}
We prove that the $\tilde{a}_i$ values approximate the $a_i$ values, specifically
\begin{align}
    |\tilde{a}_i-a_i| \le \varepsilon'+\varepsilon_\mathrm{tsp}+|b_i|.
\end{align}
We will prove the claim above using a case-by-case analysis. Assume that $a_i \ge 0$; the case $a_i < 0$ will proceed similarly.

First, consider the case $\sqrt{p_i} \le \frac{2}{3}\varepsilon'+\varepsilon_\mathrm{tsp}$. In this case $\tilde{a}_i = 0$ and $a_i \le \sqrt{p}|\tilde{v}_i| \le \sqrt{p_i} + \frac{\varepsilon'}{3} \le \varepsilon'+\varepsilon_\mathrm{tsp}$, so $|\tilde{a}_i - a_i| \le \varepsilon+\varepsilon_\mathrm{tsp}$.

Second, consider the case $f_i(\sqrt{p^+_i},\sqrt{p^-_i}) \ge a_i$. From the definition of $\tilde{a}_i$ and Assertion 1, we have $\tilde{a}_i \le \sqrt{p_i} \le \sqrt{p}|\tilde{v}_i|+ \frac{\varepsilon'}{3}$, and thus
\begin{align}
    \tilde{a}_i - a_i &\le \sqrt{p}|\tilde{v}_i| - a_i +\frac{\varepsilon'}{3} \nonumber\\
    &= \sqrt{a_i^2 + b_i^2} - a_i +\frac{\varepsilon'}{3} \le |b_i| +\frac{\varepsilon'}{3}.
\end{align}
We also have (again invoking Assertion 1)
\begin{align}
    a_i - \tilde{a}_i \le a_i - \sqrt{p_i} \le a_i - \sqrt{p}|\tilde{v}_i| + \frac{\varepsilon'}{3} \le \frac{\varepsilon'}{3}\,
\end{align}
and thus, $|a_i - \tilde{a}_i| \leq |b_i| + \frac{\varepsilon'}{3}$.

Finally, consider the case $f_i(\sqrt{p^+_i},\sqrt{p^-_i}) < a_i$. Defining $\tilde{\varepsilon}=\frac{2}{3}\varepsilon'+\varepsilon_\mathrm{tsp}$,we can lower bound $f_i(\sqrt{p^+_i},\sqrt{p^-_i})$:

\begin{align}
    f_i(\sqrt{p^+_i},\sqrt{p^-_i}) &= \frac{(2\sqrt{p^+_i})^2 - (2\sqrt{p^+_i})^2}{4\sqrt{p_i}} \nonumber\\
    &\ge \frac{(r^+_i-\tilde{\varepsilon})^2 - (r^-_i+\tilde{\varepsilon})^2}{4\sqrt{p_i}} \nonumber\\
    &= \frac{(r^+_i)^2-(r^-_i)^2}{4\sqrt{p_i}} - \tilde{\varepsilon}\frac{r^+_i+r^-_i}{2\sqrt{p_i}} \nonumber\\
    &= a_i - \tilde{\varepsilon}\frac{r^+_i+r^-_i}{2\sqrt{p_i}}.
\end{align}
Here in the second line we used \cref{eq:ppm_rpm_close} and the fact that $r^+_i \ge \sqrt{p_i} \ge \frac{2}{3}\varepsilon'+\varepsilon_\mathrm{tsp}$.
We now upper bound $r^+_i+r^-_i$:

\begin{align}
    r^+ + r^- &= \sqrt{(a_i+\sqrt{p_i})^2 + b_i^2} + \sqrt{(a_i-\sqrt{p_i})^2 + b_i^2} \nonumber\\
    &\le |a_i+\sqrt{p_i}|+|a_i-\sqrt{p_i}|+2|b_i| \nonumber\\
    &= 2\max(a_i,\sqrt{p_i})+2|b_i| \nonumber\\
    &\le 2(\sqrt{p_i}+\varepsilon'/3+|b_i|),
\end{align}
where in the fourth line we used $a_i \le \sqrt{a_i^2+b_i^2}=\sqrt{p}|\tilde{v}_i| \le \sqrt{p_i}+\varepsilon'/3$ (Assertion 1). Therefore, 
\begin{align}
    f_i(\sqrt{p^+_i},\sqrt{p^-_i}) 
    &= a_i - \tilde{\varepsilon}\frac{r^+_i+r^-_i}{2\sqrt{p_i}} \nonumber\\
    &\ge a_i - \tilde{\varepsilon}\frac{2(\sqrt{p_i}+\varepsilon'/3+|b_i|)}{2\sqrt{p_i}} \nonumber\\
    &= a_i - \tilde{\varepsilon} - \frac{\tilde{\varepsilon}}{\sqrt{p_i}} (\varepsilon'/3+|b_i|))\nonumber\\
    &\ge a_i - (\varepsilon'+\varepsilon_\mathrm{tsp}+|b_i|),
\end{align}
where in the fourth line we have used $\tilde{\varepsilon}/\sqrt{p_i} \leq 1$. This implies
\begin{align}
    |\tilde{a}_i - a_i| &= a_i - \tilde{a}_i \nonumber\\
    &\le a_i - \min(f_i(\sqrt{p^+_i},\sqrt{p^-_i}),\sqrt{p_i}) \nonumber\\
    &\le \varepsilon'+\varepsilon_\mathrm{tsp}+|b_i|.
\end{align}
Here, we used $a_i - \sqrt{p_i} \le \sqrt{p}|\tilde{v}_i|-\sqrt{p_i} \le \varepsilon'/3$.

We've shown that $|\tilde{a}_i-a_i| \le \varepsilon'+\varepsilon_\mathrm{tsp}+|b_i|$ for all cases. Therefore,
\begin{align}
    &\|\tilde{\ve{a}}-\ve{a}\|^2_2 \le \sum_i \left[(\varepsilon'+\varepsilon_\mathrm{tsp})^2 + 2|b_i|(\varepsilon'+\varepsilon_\mathrm{tsp}) + b_i^2 \right] \nonumber\\
    &\le L(\varepsilon'+\varepsilon_\mathrm{tsp})^2 + 2(\varepsilon'+\varepsilon_\mathrm{tsp})\sqrt{L\sum_i b_i^2} + \sum_i b_i^2 \nonumber\\
    &= \left(\sqrt{L}(\varepsilon'+\varepsilon_\mathrm{tsp}) + \sqrt{\sum_i b_i^2}\right)^2,
\end{align} and hence
\begin{align}
    &\|\tilde{\ve{a}}-\sqrt{p}\ve{v}\|_2 \le \|\tilde{\ve{a}}-\ve{a}\|_2 + \|\ve{a}-\sqrt{p}\ve{v}\|_2 \nonumber\\
    &\le \sqrt{L}(\varepsilon'+\varepsilon_\mathrm{tsp}) + \sqrt{\sum_i b_i^2} + \sqrt{\sum_i (\sqrt{p}v_i-a_i)^2} \nonumber\\
    &\le \sqrt{L}(\varepsilon'+\varepsilon_\mathrm{tsp}) + \sqrt{2p}\varepsilon_\mathrm{QLSP},
\end{align}
where we used $\sum_i((v_i-a_i/\sqrt{p})^2+b_i^2/p) \le \varepsilon_\mathrm{QLSP}^2$.
Since $\ve{\tilde{v}'} \propto \tilde{\ve{a}}$, for some proportionality factor $\lambda$ we have $\|\lambda \tilde{\ve{v'}} - \ve{v} \| \le \sqrt{2L}(\varepsilon'+\varepsilon_\mathrm{tsp}) + \sqrt{2}\varepsilon_\mathrm{QLSP}$, where we used $p \le 1/2$. A bit of geometry will show that if $\|\ve{c} - \ve{d} \|_2 \le \gamma < 1$ and $\|\ve{d}\|_2 = 1$, then $\| \frac{\ve{c}} {\|\ve{c}\|_2} - \ve{d} \|_2 \le g(\gamma) \equiv 2 \sin (\frac{1}{2} \sin^{-1} \gamma) = \sqrt{1+\gamma}-\sqrt{1-\gamma}$. Applying this with $\ve{c} = \lambda \tilde{\ve{v}}'$ and $\ve{d} = \ve{v}$ we obtain 
\begin{align}
&\|\tilde{\ve{v}}' - \ve{v}\|_2 \nonumber\\
&\le g(\sqrt{2L} (\varepsilon'+\varepsilon_\mathrm{tsp}) + \sqrt{2}\varepsilon_\mathrm{QLSP}) \nonumber\\
&< g(\sqrt{2L}\varepsilon')\nonumber\\ &\quad+(\sqrt{2L}\varepsilon_\mathrm{tsp}+\sqrt{2}\varepsilon_\mathrm{QLSP})\left.\frac{dg}{dx}\right|_{x=\sqrt{2L} (\varepsilon'+\varepsilon_\mathrm{tsp})+\sqrt{2}\varepsilon_\mathrm{QLSP}} \nonumber\\
&< \varepsilon + 1.58\sqrt{L}\varepsilon_\mathrm{tsp} + 1.58\varepsilon_\mathrm{QLSP}
\end{align}
as claimed. In the second inequality we used the convexity of $g$; in the third inequality we used the fact that $g(\sqrt{2L}\varepsilon')=\varepsilon$, $\sqrt{2L} (\varepsilon'+\varepsilon_\mathrm{tsp})+\sqrt{2}\varepsilon_\mathrm{QLSP} < \varepsilon+\sqrt{2L}\varepsilon_\mathrm{tsp}+\sqrt{2}\varepsilon_\mathrm{QLSP} \le 1/2$, and $\sqrt{2}g'(1/2) < 1.58$.
\end{proof}


\section{Null space matrix for portfolio optimization}\label{app:null-space}

In \cref{sec:scop-interior-c}, an inexact-feasible interior point method was described that requires as input a matrix $B$ with columns that form a basis for the null space of the feasibility equations for the self-dual SOCP that appear in \cref{eq:Newton-system-1}. A straightforward way to find such a $B$ in general would be to perform a QR decomposition of the constraint matrix, costing classical $O(N^3)$ runtime (or, using techniques for fast matrix multiplication, between $O(N^2)$ and $O(N^3)$ time \cite{knight1995fast,camarero2018simple}). The upshot is that $B$ need only be computed once and does not change with each iteration of the algorithm, but depending on other parameters of the problem, this classical runtime could dominate the overall complexity. Alternatively, in many specific cases including ours, a valid matrix $B$ can be determined by inspection. For example, suppose that we have a $(N-K)\times N$ matrix $Q_A$ with full column rank for which $AQ_A = 0$, a $K \times (K-1)$ matrix $P$ with full column rank for which $\ve{\bar{b}}^{\intercal}P = 0$, and a point $\ve{x}_0$ for which $A \ve{x}_0 = \ve{b}$.  Then, letting $\gamma = \ve{b}^\intercal \ve{\bar{b}} / ||\ve{\bar{b}}||^2$, a valid choice for $B$ is
\begin{widetext}
\begin{equation}\label{eq:Q}
 B \qquad ={}\;\begin{matrix}
 \ve{x} \\
 \ve{y} \\
 \tau \\
 \theta \\
 \ve{s} \\
 \kappa
\end{matrix}
\;
 \begin{pmatrix}
0                  & Q_A                                                                            
& \ve{e}                                                        &  \ve{x}_0 \\
P                  & \ve{\bar{b}} \frac{\ve{\bar{c}}^\intercal Q_A}{||\ve{\bar{b}}||^2}             
& -\frac{(r+1)}{||\ve{\bar{b}}||^2}\ve{\bar{b}}                 &  \frac{\ve{\bar{c}}^\intercal \ve{x}_0 - \bar{z}}{||\ve{\bar{b}}||^2}\ve{\bar{b}}   \\
0                  & 0                                                                              
& 1                                                             &  1  \\
0                  & 0                                                                              
& 1                                                             &  0  \\
-A^\intercal P     & -A^\intercal \ve{\bar{b}}\frac{\ve{\bar{c}}^\intercal Q_A}{||\ve{\bar{b}}||^2} 
& \frac{r+1}{||\ve{\bar{b}}||^2}A^\intercal \ve{\bar{b}}+\ve{e} &  \frac{-\ve{\bar{c}}^\intercal \ve{x}_0 + \bar{z}}{||\ve{\bar{b}}||^2}A^\intercal \ve{\bar{b}}  +\ve{c} \\
\ve{b}^\intercal P & (\gamma-1)\ve{c}^\intercal Q_A - \gamma \ve{e}^\intercal Q_A                            
& 1-\gamma(r+1)                                                      &  -\gamma \bar{z}+(\gamma-1)\ve{c}^\intercal \ve{x}_0-\gamma \ve{e}^\intercal \ve{x}_0\
 \end{pmatrix}
 \end{equation}
 \end{widetext}
The leftmost column in the above block matrix corresponds to $K-1$ basis vectors formed by choosing $\ve{y}$ to be a vector perpendicular to $\ve{\bar{b}}$ and $\ve{x}=\ve{0}$, $\tau = \theta = 0$. The second column corresponds to $N-K$ vectors formed by choosing $\ve{x}$ to be in the null space of $A$, and letting $\tau = \theta = 0$, with $\ve{y} = \frac{\ve{\bar{c}}^\intercal \ve{x}}{||\ve{\bar{b}}||^2}\ve{\bar{b}}$. The third column corresponds to the vector formed by choosing $\ve{x} = \ve{e}$, $\tau = \theta = 1$, and then $\ve{y} = \frac{-(r+1)}{||\ve{\bar{b}}||^2}\ve{\bar{b}}$. The final column corresponds to choosing $\ve{x}=\ve{x}_0$, $\tau = 1$, $\theta = 0$, and $\ve{y} = \frac{\ve{\bar{c}}^\intercal \ve{x}_0 - \bar{z}}{||\ve{\bar{b}}||^2}\ve{\bar{b}}$. In each case, the choices of $\ve{x}$, $\ve{y}$, $\tau$, and $\theta$ uniquely determine the values of $\ve{s}$ and $\kappa$. Note that in practice the second and fourth block rows of $B$ can be ignored because in \cref{eq:Newton-system-2} they are left-multiplied by a matrix whose second and fourth block columns are zero.

What remains is to specify $P$, $Q_A$, and $\ve{x}_0$ for the case of portfolio optimization, given in \cref{eq:SOCP-PO}. Finding a valid matrix $P$ is straightforward. Note that from \cref{eq:SOCP-PO}, we have $\ve{b} = (1; \ve{\bar{w}} + \ve{\zeta}; \ve{\bar{w}} - \ve{\zeta};\ve{0} )$. For $j = 1,\ldots, 2n$, we let $\ve{p}_j$ have a 1 in its first entry, and a $-1/b_{j+1}$ in its $(j+1)$th entry, with zeros elsewhere. For $j = 2n+1, \ldots, 2n+m$, we let $\ve{p}_j$ have a single 1 in its $(j+1)$th entry, and zeros elsewhere. Thus, the $\ve{p}_j$ are independent and $\ve{b}^{\intercal}\ve{p}_j = 0$ for all $j$. We then define the matrix $P$ by $P = (\ve{p}_1,\ldots,\ve{p}_{2n+m})$. Similarly, we can generate the columns of a valid matrix $Q_A$ as follows: given a choice of $\ve{w}$ such that $\ve{1}^\intercal \ve{w} = 0$, we choose $\ve{\phi} = -\ve{w}$, $\ve{\rho} = \ve{w}$, $t = 0$, and $\ve{\eta} = M\ve{w}$. As there are $n-1$ linearly independent choices of $\ve{w}$ (e.g.~the vectors $(1;-1;0;0;\ldots;0)$, $(0;1;-1;0;\ldots;0)$, $(0;0;1;-1;\ldots;0)$, etc.), this leads to $n-1$ linearly independent columns of $Q_A$. A final $n$th column can be formed by choosing $t=1$ and $\ve{w} = \ve{\phi} = \ve{\rho} = \ve{0}$ and $\ve{\eta} = \ve{0}$.   Finally, the point $\ve{x}_0$ can be chosen by letting $\ve{w} = \ve{\bar{w}}$, $\ve{\phi} = \ve{\rho} = \ve{\zeta}$, $t = 0$, and $\ve{\eta} = M\ve{\bar{w}}$. 


\section{Alternative search directions}\label{app:alt_search_directions}

The solution $(\Delta \ve{x};\Delta \ve{y};\Delta \tau;\Delta \theta;\Delta \ve{s}; \Delta \varkappa)$ to the Newton systems in \cref{eq:Newton-system-1,eq:Newton-system-2} is one possible \emph{search direction} for the interior point method. Alternative search directions can be found by applying a scale transformation to the convex set. We follow Ref.~\cite{monteiro2000polynomial} and, for the $k$-dimensional second-order cone $\mathcal{Q}^k$, we define the set
\begin{equation}
\mathcal{G}^k = \left\{\lambda T: \lambda > 0, T^\intercal \begin{pmatrix}1 & 0 \\
0 & -I \end{pmatrix} T = \begin{pmatrix}1 & 0 \\
0 & -I \end{pmatrix} \right\}\,.
\end{equation}
For the product $\mathcal{Q}$ of multiple cones, we let the set $\mathcal{G}$ consist of direct sums of entries from $\mathcal{G}^k$. This definition implies that the matrices $G \in \mathcal{G}$ map the set $\mathcal{Q}$ onto itself. Thus for a fixed choice $G \in \mathcal{G}$, we may consider a change of variables $\ve{x}' = G^\intercal \ve{x}$, $\ve{s}' = G^{-1} \ve{s}$, $\ve{y}' = \ve{y}$. We let $X'$ and $S'$ be the arrowhead matrices for $\ve{x}'$ and $\ve{s}'$, and, following the same logic as above, we arrive at a Newton system
\begin{equation}\label{eq:Newton-system-2-scaled}
\begin{pmatrix}
S'G^\intercal & 0 & 0      & 0 & X'G^{-1} & 0 \\
0 & 0 & \varkappa & 0 & 0 & \tau
\end{pmatrix}
\begin{pmatrix}
\ve{\Delta x} \\
\ve{\Delta y} \\
\Delta \tau \\
\Delta \theta \\
\ve{\Delta s} \\
\Delta \varkappa
\end{pmatrix}
=
\begin{pmatrix}
\sigma \mu \ve{e} - X'S' \ve{e}\\
\sigma \mu - \varkappa\tau
\end{pmatrix}.
\end{equation}
The solution to this linear set of equations (along with the feasibility equations of \cref{eq:Newton-system-1}) will be distinct for different choices of $G$. The choice $G=I$ recovers \cref{eq:Newton-system-2} and is called the Alizadeh-Haeberly-Overton (AHO) direction. Ref.~\cite{monteiro2000polynomial} showed that the IPM can reduce the duality gap by a constant factor after $O(\sqrt{r})$ iterations for any choice of $G$. However, some choices of $G$ can yield additional potentially desirable properties; for example, the Nesterov-Todd search direction scales the cone such that $\ve{x}' = \ve{s}'$. However, in our numerical simulations of the QIPM, we did not observe any obvious benefits of choosing a search direction other than the AHO direction.

\section{Numerical results for feasible QIPMs}\label{app:feasible_qipm}

In \cref{sec:numerics}, we presented numerical results for the ``II-QIPM,'' for which intermediate points could be infeasible. Here we also present some results for two variants of the ``feasible'' QIPM inspired by the work of Ref.~\cite{augustino2021inexact}, denoted by ``IF-QIPM'' and ``IF-QIPM-QR,'' as summarized in \cref{tab:QIPM_choices}. The IF-QIPM uses the null space basis $B$ outlined in \cref{app:null-space}, whereas the IF-QIPM-QR version uses a null space basis $B$ determined using a QR decomposition. In all cases, we simulated the algorithm for enough iterations to reduce the duality gap to $10^{-3}$, whereas for the II-QIPM we simulated down to $10^{-7}$. 

\begin{figure*}[p]
\includegraphics[width=0.95\textwidth]{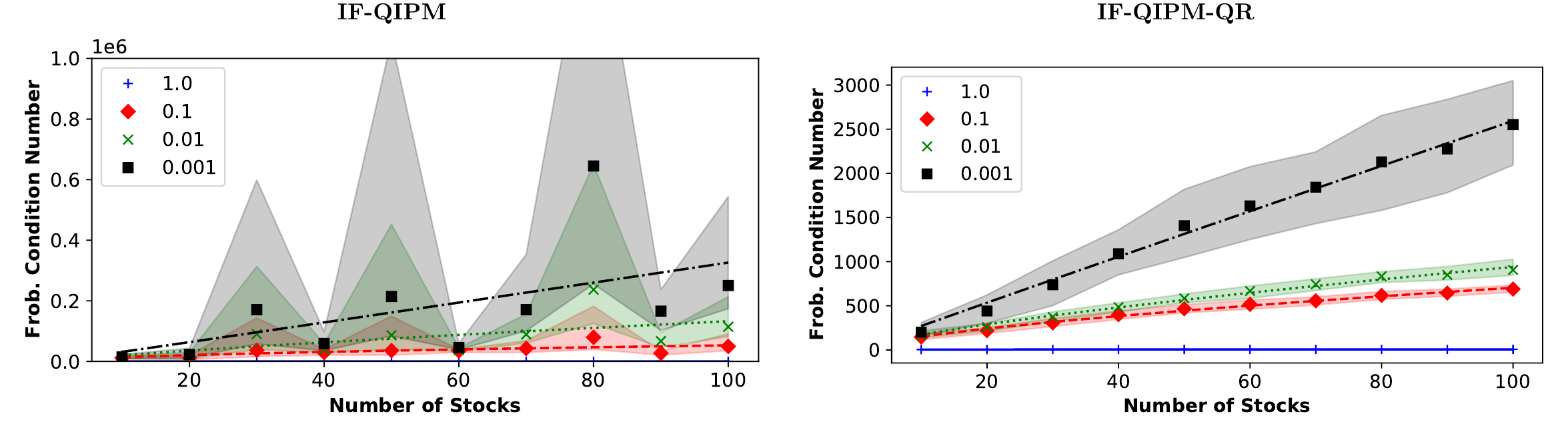}
\caption{\label{fig:condition_number_scaling_appendix}Median Frobenius condition number for 128 randomly sampled stock portfolios from the DWCF index as a function of portfolio size for duality gaps of 1.0, 0.1, 0.01, and 0.001. The error bars show the 68th percentile, which corresponds to one standard deviation if the distribution is Gaussian. We find that a linear trend appears to work quite well for the IF-QIPM-QR case, but that the IF-QIPM is quite noisy. For each duality gap, we also plot a power-law fit of the form $an^b$ and report the values of $b$ in \cref{tab:slopes_appendix}.} 
\end{figure*}

\begin{figure*}[p]
\includegraphics[width=0.95\textwidth]{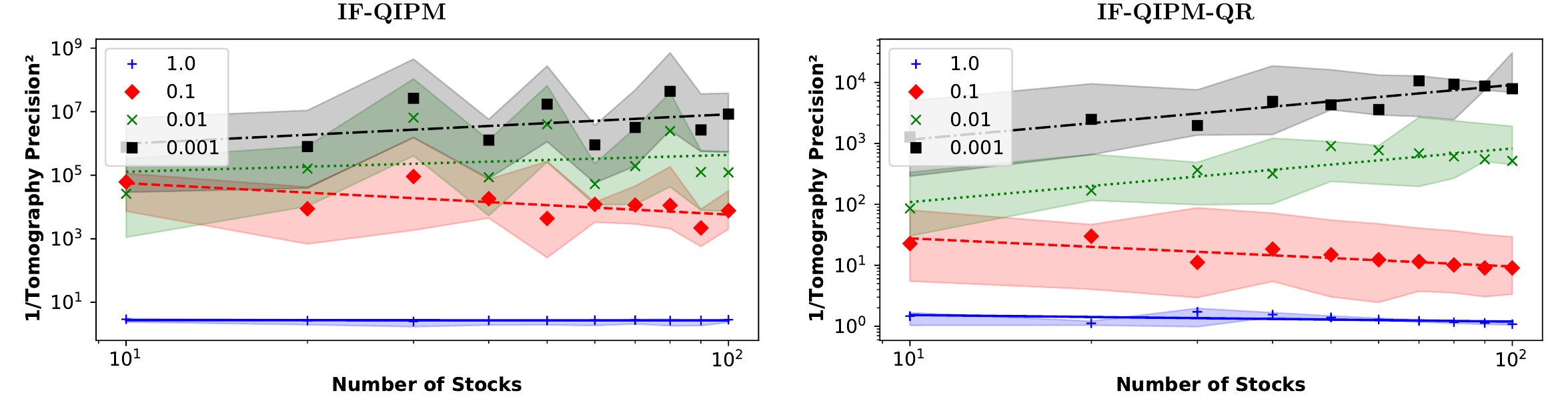}
\caption{\label{fig:invtomography_scaling_appendix} Median value of the square of the required inverse tomography precision required to remain in the neighborhood of the central path for 128 randomly sampled stock portfolios from the DWCF index as a function of portfolio size for duality gaps of 1.0, 0.1, 0.01, and 0.001. The error bars show the 68th percentile, which corresponds to one standard deviation if the distribution is Gaussian. For each duality, gap,  we also plot a linear fit on the log-log data, and report the corresponding slope in \cref{tab:invtom_slopes_appendix}}.
\end{figure*}

\begin{figure*}[p]
\includegraphics[width=0.95\textwidth]{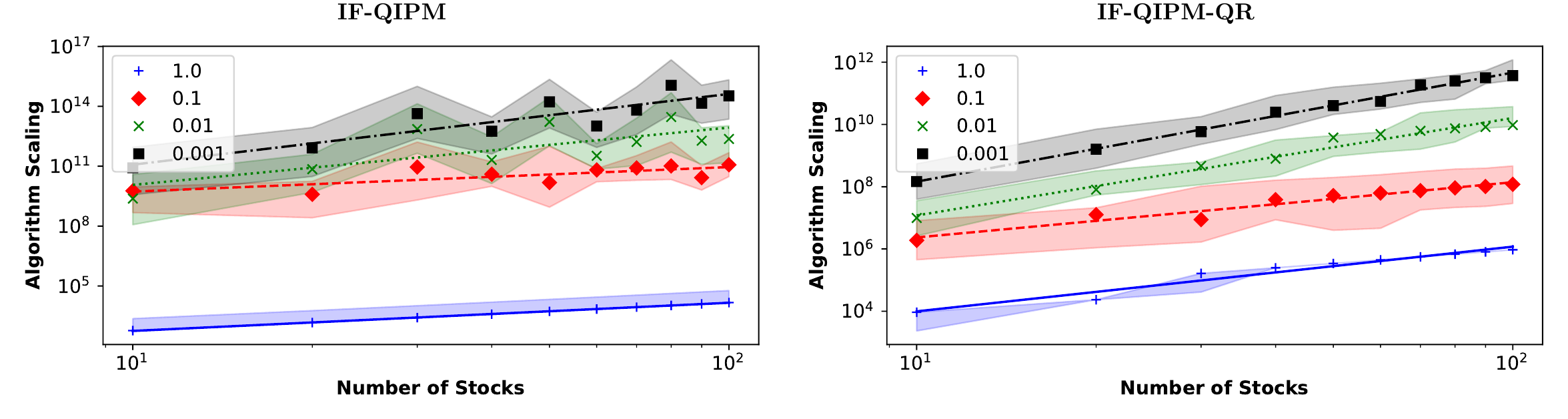}
\caption{\label{fig:algorithm_scaling_appendix} Median value of the estimated algorithm scaling factor computed as the median of $n^{1.5}\kappa_F/\xi^2$ for 128 randomly sampled stock portfolios from the DWCF index as a function of portfolio size for duality gaps of 1.0, 0.1, 0.01, and 0.001. The error bars show the 68th percentile, which corresponds to one standard deviation if the distribution is Gaussian. For each duality, gap,  we also plot a linear fit on the log-log data, and report the corresponding slope in \cref{tab:scaling_slopes_appendix}. 
} 
\end{figure*}

In \cref{fig:condition_number_scaling_appendix,fig:invtomography_scaling_appendix,fig:algorithm_scaling_appendix}, we present the analogous results for the feasible IPMs as were displayed in \cref{fig:condition_number_scaling,fig:invtomography_scaling,fig:algorithm_scaling} for the infeasible case. We find that the IF-QIPM-QR has the best performance, though this must be weighed against the fact that an expensive QR decomposition must be classically pre-computed to implement this method. However, the advantage of the IF-QIPM-QR method is not large enough for any of the qualitative conclusions in \cref{sec:discussion} to change. The IF-QIPM method has the worst performance, which we believe is due to the fact that the null-space basis found by inspection turns out to be a very ill-conditioned matrix (its condition number was observed to be in the vicinity of 1000). Additionally, the IF-QIPM appears to have the largest instance-to-instance variation of any of the methods, leading to lower quality numerical fits.

\begin{table}[h]
\caption{\label{tab:slopes_appendix}Fit parameters for the Frobenius condition number for the four horizontal-axis locations considered on the scaling plot of \cref{fig:condition_number_scaling_appendix}. The uncertainties correspond to one standard deviation errors on the parameter estimates from the fit.
We note that both versions have similar empirical scaling, although the fits are better for IF-QIPM-QR. The constant prefactors are superior for the IF-QIPM-QR version, but} calculating the QR decomposition requires a one-time classical cost proportional to $\bigo{L^3}$.
\begin{ruledtabular}
\renewcommand{\arraystretch}{1.4}
\begin{tabular}{p{0.8cm}p{3.1cm}p{3.1cm}}
\textbf{Dual. Gap} & \textbf{IF-QIPM} & \textbf{IF-QIPM-QR} \\
\hline
1.0   & $\kappa_F(G) \sim n^{0.57 \pm 0.60}$ & $\kappa_F(G) \sim n^{0.228 \pm 0.002}$      \\ 
0.1   & $\kappa_F(G) \sim n^{0.58 \pm 0.28}$ & $\kappa_F(G) \sim n^{0.66 \pm 0.03}$  \\
0.01  & $\kappa_F(G) \sim n^{0.81 \pm 0.53}$ & $\kappa_F(G) \sim n^{0.73 \pm 0.03}$  \\
0.001 & $\kappa_F(G) \sim n^{1.01 \pm 0.77}$ & $\kappa_F(G) \sim n^{0.98 \pm 0.04}$\\
\end{tabular}
\end{ruledtabular}
\end{table}

\renewcommand{\arraystretch}{1}

\begin{table}[t]
\caption{\label{tab:invtom_slopes_appendix}Fit parameters for the square of the inverse of the required tomography precision to stay near the central path, corresponding to \cref{fig:invtomography_scaling_appendix}}. The uncertainties correspond to one standard deviation errors on the parameter estimates from the fit.
\begin{ruledtabular}
\renewcommand{\arraystretch}{1.4}
\begin{tabular}{p{0.8cm}p{3.1cm}p{3.1cm}}
\textbf{Dual. Gap} & \textbf{IF-QIPM} & \textbf{IF-QIPM-QR} \\
\hline
1.0   & $\xi^{-2} \sim \bigo{n^{-0.01 \pm 0.02}}$ & $\xi^{-2} \sim \bigo{n^{-0.11 \pm 0.07}}$ \\ 
0.1   & $\xi^{-2} \sim \bigo{n^{-0.99 \pm 0.41}}$  & $\xi^{-2} \sim \bigo{n^{-0.46 \pm 0.11}}$ \\
0.01  & $\xi^{-2} \sim \bigo{n^{0.53 \pm 0.91}}$    & $\xi^{-2} \sim \bigo{n^{0.89 \pm 0.15}}$ \\
0.001 & $\xi^{-2} \sim \bigo{n^{0.93 \pm 0.66}}$    & $\xi^{-2} \sim \bigo{n^{0.90 \pm 0.15}}$ \\
\end{tabular}
\end{ruledtabular}
\end{table}

\renewcommand{\arraystretch}{1}

\begin{table}[t]
\caption{\label{tab:scaling_slopes_appendix}Estimated scaling of the quantum algorithm as a function of portfolio size for the two feasible versions of the quantum algorithm, corresponding to \cref{fig:algorithm_scaling_appendix}. The uncertainties correspond to one standard deviation errors on the parameter estimates from the fit. 
}
\begin{ruledtabular}
\renewcommand{\arraystretch}{1.4}
\begin{tabular}{p{0.8cm}p{3cm}p{3cm}}
\textbf{Dual. Gap} & \textbf{IF-QIPM} & \textbf{IF-QIPM-QR} \\
\hline
1.0   & $\bigo{n^{1.41 \pm 0.01}}$ & $\bigo{n^{2.07 \pm 0.15}}$ \\ 
0.1   & $\bigo{n^{1.23 \pm 0.40}}$   & $\bigo{n^{1.77 \pm 0.15}}$ \\
0.01  & $\bigo{n^{2.87 \pm 0.91}}$       & $\bigo{n^{3.13 \pm 0.18}}$ \\
0.001 & $\bigo{n^{3.54 \pm 0.64}}$       & $\bigo{n^{3.50 \pm 0.10}}$ \\
\end{tabular}
\end{ruledtabular}
\end{table}
\renewcommand{\arraystretch}{1}
\clearpage

\end{document}